\documentclass[preprint,authoryear,12pt]{elsarticle}

\usepackage{amssymb}
\usepackage{amsmath}

\usepackage{graphicx}
\usepackage{enumerate}
\usepackage{natbib}
\usepackage{url} 
\usepackage{indentfirst}
\usepackage{listings}
\usepackage{xcolor}
\usepackage{enumerate}
\usepackage{amsfonts,amsthm,amssymb}
\usepackage{extarrows}
\usepackage{mathtools}
\usepackage{CJKpunct}
\usepackage{array}
\usepackage{tabu}
\usepackage{bm}
\usepackage{float}
\usepackage{fancyhdr}
\usepackage[bookmarksopen=true]{hyperref}
\usepackage{bookmark}
\usepackage{booktabs}
\usepackage{etoolbox}
\usepackage{algorithm}
\usepackage{algorithmicx}
\usepackage{algpseudocode}
\usepackage{multirow}
\usepackage{comment}
\usepackage{longtable}

\newcommand{\blind}{1}

\vspace{-0.8cm}
\allowdisplaybreaks[4]

\theoremstyle{remark}

\newtheorem{theorem}{\textbf{Theorem}}

\newtheorem{remark}{\textbf{Remark}}

\makeatletter
\newenvironment{breakablealgorithm}
{
		\begin{center}
			\refstepcounter{algorithm}
			\hrule height.8pt depth0pt \kern2pt
			\renewcommand{\caption}[2][\relax]{
				{\raggedright\textbf{\ALG@name~\thealgorithm} ##2\par}%
				\ifx\relax##1\relax 
				\addcontentsline{loa}{algorithm}{\protect\numberline{\thealgorithm}##2}%
				\else 
				\addcontentsline{loa}{algorithm}{\protect\numberline{\thealgorithm}##1}%
				\fi
				\kern2pt\hrule\kern2pt
			}
		}{
		\kern2pt\hrule\relax
	\end{center}
}
\makeatother

\journal{Journal of Econometrics}

\begin{document}
	
	\begin{frontmatter}
		
		
		
		\title{\textbf{Bootstrap Model Averaging}}
		
		
		\author[inst1]{Minghui Song} 
		
		\affiliation[inst1]{organization={School of Mathematical Sciences},
			addressline={Capital Normal University}, 
			city={Beijing},
			postcode={100048}, 
			country={China}}
		
		\author[inst1]{Guohua Zou\corref{cor1}}
		\ead{ghzou@amss.ac.cn}
		\cortext[cor1]{Corresponding author}
		
		\author[inst2]{Alan T.K. Wan}
		
		\affiliation[inst2]{organization={Department of Management Sciences},
			addressline={City University of Hong Kong}, 
			city={Kowloon},
			country={Hong Kong}}
		
		\begin{abstract}
			Model averaging has gained significant attention in recent years due to its ability of fusing information from different models. The critical challenge in frequentist model averaging is the choice of weight vector. The bootstrap method, known for its favorable properties, presents a new solution. In this paper, we propose a bootstrap model averaging approach that selects the weights by minimizing a bootstrap criterion. Our weight selection criterion can also be interpreted as a bootstrap aggregating. We demonstrate that the resultant estimator is asymptotically optimal in the sense that it achieves the lowest possible squared error loss. Furthermore, we establish the convergence rate of bootstrap weights tending to the theoretically optimal weights. Additionally, we derive the limiting distribution for our proposed model averaging estimator. Through simulation studies and empirical applications, we show that our proposed method often has better performance than other commonly used model selection and model averaging methods, and bootstrap variants.
		\end{abstract}
		
		
	\end{frontmatter}
	
		
		
		\section{Introduction}
		\label{sec:intro}
		\indent Model averaging (MA) is an approach that combines candidate models using given weights, which can be regarded as an extension of model selection (MS) from the perspective of estimation and forecasting. Over the past two decades, the MA method has made significant progress in theory and has found wide application in various fields. For instance, \cite{proietti2021nowcasting} used the MA method to forecast monthly GDP, while \cite{chen2020bayesian} applied it to a genetic association study. Due to its ability to utilize information from different models, MA generally achieves a lower mean squared error than MS. The MA approach includes Bayesian model averaging (BMA) and Frequentist model averaging (FMA). \cite{hoeting1999bayesian} provided a comprehensive overview of BMA. In this paper, we focus on the FMA method.\\
		\indent The most significant challenges in frequentist model averaging lie in the selection of weights and the derivation of the asymptotic distribution of the model averaging estimator. Various weight selection criteria have been proposed under different model frameworks. For instance, one common approach for constructing weights is based on information criteria, such as the S-AIC, which is derived by smoothing the AIC. Similarly, if we replace AIC with BIC, we obtain the S-BIC. See, for example, \cite{buckland1997model}. \cite{hansen2007least} introduced the Mallows model averaging (MMA) criterion for weight selection, which has been extended by \cite{wan2010least} for combining non-nested models. \cite{hansen2008least} applied the MMA criterion to forecast combination. \cite{liang2011optimal} proposed an optimal weight choice method. \cite{hansen2012jackknife} presented the jackknife model averaging (JMA) criterion to handle heteroscedasticity. \cite{ando2014model} suggested an MA procedure for high-dimensional linear regression using the JMA method. \cite{gao2016model} developed the leave-subject-out model averaging criterion for longitudinal data and time series data. A corresponding version for averaging predictions from vector autoregression can be found in \cite{liao2019model}. \cite{liao2020corrected} introduced a corrected MMA estimator and established the convergence rates of MMA and JMA weights. \cite{zhao2020average} considered semiparametric MA for high-dimensional longitudinal data. When the number of parameters is large, \cite{zhang2020parsimonious} proposed a criterion to achieve a parsimonious MA coefficient estimator.\\
		\indent In order to utilize MA for statistical inference, a thorough investigation on asymptotic distribution is needed. The asymptotic distribution of averaging estimator was first studied by \cite{hjort2003frequentist} within a local asymptotic framework, where the regression coefficients are assumed to be in a local neighborhood of zero with a rate of $n^{-1/2}$. Under this assumption, \cite{claeskens2007asymptotic} and \cite{zhu2018asymptotic} developed the asymptotic distribution of the model averaging estimator in general semiparametric models. \cite{liu2015distribution} also considered this local-to-zero assumption and derived the distribution theory of the least squares averaging estimator. Departing from the local misspecification framework, \cite{charkhi2018asymptotic} investigated the asymptotic distribution of a post-selection AIC estimator. Furthermore, \cite{zhang2019inference} developed the asymptotic distribution theory for the MMA and JMA estimators with linear regression. \cite{yu2024post} established the asymptotic distributions of optimal model averaging estimators for generalized linear models.\\
		\indent The jackknife method has gained prominence in the field of model averaging. It is widely recognized that bootstrap and jackknife are two important resampling methods in statistics.  \cite{efron1979bootstrap} stated that the bootstrap method provides an alternative perspective on the jackknife. The fundamental concept behind the bootstrap method is relatively straightforward to comprehend. \cite{efron1979bootstrap} demonstrated how resampling can be generated through bootstrapping and compared the bootstrap method with the jackknife. Numerous examples were presented to illustrate situations where the bootstrap outperforms the jackknife in \cite{efron1982jackknife}. \cite{efron1983estimating} found that the ordinary bootstrap gives an estimate of
		error with low variability. Subsequently, the bootstrap method has been extensively developed and employed in both theoretical and practical contexts. \cite{bickel1981some} studied more asymptotic theory related to the bootstrap method. As a sequel to this paper, \cite{freedman1981bootstrapping} developed the asymptotic theory for applications of bootstrap to regression. \cite{bunke1984bootstrap} indicated that the bias corrected bootstrap estimator for regression models is the best unbiased and should be the first choice. More bootstrap methods of regression analysis were considered in \cite{wu1986jackknife}. \cite{rao1997out} proposed a bootstrap aggregating method using the optimal weights to combine estimates from different bootstrap samples.\\
		\indent \cite{shao1996bootstrap} introduced the bootstrap model selection (BMS) and demonstrated the consistency of the BMS estimator from linear regression to nonlinear regression. Notably, the simulations presented in Shao's work indicate that BMS exhibits superior performance compared to Mallows' $C_{p}$ and the BIC. This finding serves as motivation to further extend the application of the bootstrap method from model selection to model averaging.\\
		\indent When employing resampling methods such as the jackknife and bootstrap, it is possible to retain these resampled samples for estimating additional parameters. However, the bootstrap method often outperforms the jackknife method in certain estimation problems. For instance, the bootstrap distribution estimator has shown higher accuracy compared to the jackknife histogram, as demonstrated by \cite{shao2012jackknife}. Moreover, the jackknife method is entirely unsuitable for estimating the median due to the lack of consistency in its median estimator, whereas the bootstrap estimator of the median performs well, as highlighted by \cite{efron1979bootstrap}. In principle, bootstrap methods are more widely applicable and dependable than the jackknife.\\
		\indent Considering the advantages of the bootstrap, we propose a bootstrap model averaging (BTMA) approach in this paper. We determine the weights by minimizing a bootstrap criterion. We anticipate that BTMA will exhibit superior performance compared to JMA. Theoretically, we establish the asymptotic optimality of the BTMA estimator and derive the convergence rate of the BTMA weight vector towards the optimal weight vector. Additionally, we present the asymptotic distribution of our BTMA estimator for statistical inference purposes.\\
		\indent Since the bootstrap method was proposed, many derivative or related methods have been developed, such as the $m$-out-of-$n$ bootstrap, subsampling, and bootstrap aggregating (known as bagging in the machine learning literature). \cite{davison2003recent} provided a review and discussion on these methods. In particular, their article mentions that bootstrap may exhibit inconsistency unless problem-specific regularity conditions hold, while subsampling offers consistency under weaker conditions. A typical condition for the consistency of subsampling is $m/n\rightarrow0$ as $m\rightarrow\infty$ and $n\rightarrow\infty$. The $m$-out-of-$n$ bootstrap can similarly repair the inconsistency when $m\ll n$. Although the $m$-out-of-$n$ bootstrap exhibits efficiency losses when the traditional bootstrap is valid, it may provide higher accuracy than subsampling. Since the choice of $m$ in the $m$-out-of-$n$ bootstrap is a key issue, \cite{bickel2008choice} considered an adaptive rule to pick $m$. Regarding the comparison of bootstrap and subsampling as resampling methods for model selection, \cite{de2016subsampling} conducted numerical studies using real data examples and found that subsampling emerged as a valid alternative to the bootstrap. In the discussion of bootstrap and bagging, \cite{efron2014estimation} proposed a useful formula for the accuracy of bagging after model selection, which provides standard errors for the smoothed estimators. However, Efron's bootstrap smoothing method still performs model selection on each bootstrap replication, resulting in the loss of potentially useful information from other models. In our numerical studies, we also compare our BTMA and these variants of bootstrap mentioned above.\\
		\indent The remainder of this paper is organized as follows. Section 2 presents model set-up and averaging estimators. Section 3 proposes the bootstrap criterion to choose weights. Section 4 establishes the asymptotic optimality of BTMA estimator, and provides the convergence rate of bootstrap weights. Section 5 derives the asymptotic distribution of BTMA estimator under the linear model. Section 6 conducts two simulation studies for prediction and confidence interval, and Section 7 applies our method to a real dataset. Section 8 provides the conclusion and further discussion. Additional simulation, empirical application, and theoretical proofs are included in the online supplemental file.
		\section{Model Framework and Estimation}
		\label{sec:mode}
		Let $(y_{i},x_{i})$, $i=1,\cdots,n$ be a random sample, where $x_{i}=(x_{i1},x_{i2},\cdots)'$ is countably infinite and $y_{i}$ is real-valued. Following \cite{hansen2007least}, we assume that the data-generating
		process is
		\begin{align}
			y_{i}=\mu_{i}+e_{i}, \label{1}
		\end{align}
		where $\mu_{i}=\sum_{j=1}^{\infty}\theta_{j}x_{ij}$ with $\theta_{j}$ being the corresponding coefficients, and $e_{i}$ is the independent random error with $E(e_{i}|x_{i})=0$, $E(e_{i}^{2}|x_{i})=\sigma^{2}$. We also assume $E\mu_{i}^{2}<\infty$ in the sense that $\mu_{i}$ converges in mean square. Consider a sequence of linear approximating models $q=1,\cdots,M$, where the $q$th model uses $k_{q}$ regressors from $x_{i}$ and $k_{q}>0$, i.e., the $q$th approximating model is
		\begin{align}
			y_{i}=\sum\limits_{j=1}^{k_{q}}\theta_{j(q)}x_{ij(q)}+e_{i}, \notag
		\end{align}
		where $\theta_{j(q)}$ are coefficients of the $q$th model, $x_{ij(q)}$ are variables from $x_{i}$, and the approximation error is $b_{i(q)}=\mu_{i}-\sum_{j=1}^{k_{q}}\theta_{j(q)}x_{ij(q)}$. In matrix notation, the model (\ref{1}) can be rewritten as $Y=\mu+e=X_{(q)}\Theta_{q}+b_{q}+e$, where $Y=(y_{1},\cdots,y_{n})'$, $\mu=(\mu_{1},\cdots,\mu_{n})'$, $X_{(q)}=(x_{1(q)},\cdots,x_{n(q)})'$ is an $n\times k_{q}$ design matrix with $x_{i(q)}=(x_{i1(q)},\cdots,x_{ik_{q}(q)})'$, $\Theta_{q}=(\theta_{1(q)},\cdots,\theta_{k_{q}(q)})'$, $b_{q}=(b_{1(q)},\cdots,b_{n(q)})'$ and $e=(e_{1},\cdots,e_{n})'$.\\
		\indent The least squares estimator of $\Theta_{q}$ in the $q$th candidate model is $\hat{\Theta}_{q}=(X_{(q)}'X_{(q)})^{-1}X_{(q)}'Y$. Denote the corresponding estimator of $\mu$ as $\hat{\mu}_{q}=X_{(q)}\hat{\Theta}_{q}=X_{(q)}(X_{(q)}'X_{(q)})^{-1}X_{(q)}'Y=H_{q}Y$, and the residual vector as $\hat{e}_{q}=Y-\hat{\mu}_{q}$. Let $\omega=(\omega_{1},\cdots,\omega_{M})'$ be a weight vector in $\textbf{H}_{n}=\big\{\omega\in[0,1]^{M}:\sum_{q=1}^{M}\omega_{q}=1\big\}$. Then, the model average estimator of $\mu$ is
		\begin{align}
			\hat{\mu}(\omega)=\sum\limits_{q=1}^{M}\omega_{q}X_{(q)}\hat{\Theta}_{q}=\sum\limits_{q=1}^{M}\omega_{q}H_{q}Y=H(\omega)Y, \notag
		\end{align}
		where $H(\omega)=\sum_{q=1}^{M}\omega_{q}H_{q}$. Setting the weight vector $\omega$ to be a unit vector $w_{q}^{0}$, where the $q$th element is 1 and others are 0, the averaging estimator simplifies to a selection estimator $\hat{\mu}_{q}$. \\
		\indent Define the squared error loss as $L_{n}(\omega)=||\hat{\mu}(\omega)-\mu||^{2}$ and the corresponding risk as $R_{n}(\omega)=E(L_{n}(\omega)|X)$, where $X=(x_{1},\cdots,x_{n})'$. It can be shown that $R_{n}(\omega)=||A(\omega)\mu||^{2}+\sigma^{2}\text{tr}\{H^{2}(\omega)\}$, where $A(\omega)=I_{n}-H(\omega)$. Our purpose is to find the optimal weights to minimize the risk.
		
		\section{Bootstrap Weighting}
		\label{sec:boot}
		In this section, we propose a novel criterion for bootstrapping pairs to select the weight vector, and build its similarity to MMA.
		\subsection{Bootstrap criterion for choosing weights}
		Let $\hat{F}_{n}$ be the empirical distribution putting mass $1/n$ on each pair $(y_{i},x_{i})$, $i=1,\cdots,n$. Bootstrapping pairs generates i.i.d bootstrap observations $\{(y_{i}^{*},x_{i}^{*}),\ i=1,\cdots,m\}$ from $\hat{F}_{n}$. Consider a bootstrap resampling matrix denoted by $P$, then the bootstrap design matrix $X_{(q)}^{*}=(x_{1}^{*},\cdots,x_{m}^{*})'$\ $=PX_{q}$, where $P$ is an $m\times n$ matrix. Let $P=(p_{1},\cdots,p_{m})'$, where $p_{i}=(p_{i1},\cdots,p_{in})'$, $i=1,\cdots,m$ is a vector with one of the elements being 1 and the others being 0. The meaning of $p_{ij}=1$ indicates that the $j$th observation in $X$ is chosen as the $i$th observation in the resampled dataset. Then we have $\sum_{j=1}^{n}p_{ij}=1$. Let $\Pi=P'P$. Note that $p_{ij}$ is 0 or 1, then
		\begin{align}
			\Pi&=\begin{pmatrix}p_{1},p_{2},\cdots,p_{m}\end{pmatrix}
			\begin{pmatrix}
				p_{1}'\\
				p_{2}'\\
				\vdots\\
				p_{m}'
			\end{pmatrix}
			=\text{diag}\bigg(\sum\limits_{i=1}^{m}p_{i1},\sum\limits_{i=1}^{m}p_{i2},\cdots,\sum\limits_{i=1}^{m}p_{in}\bigg)\\
			&\triangleq\text{diag}(\pi_{1},\pi_{2}\cdots,\pi_{n}). \notag
		\end{align}
		The matrix $\Pi$ can be regarded as a selection matrix, with each element $\pi_{j}$ representing the number of times that the $j$th observation has been selected. Hence, $\sum_{j=1}^{n}\pi_{j}=m$ and the vector $\pi\triangleq(\pi_{1},\cdots,\pi_{n})'\sim \text{Multinomial}(m,n,1/n,\cdots,1/n)$, such that
		\begin{align}
			E_{*}(\pi_{j})=\dfrac{m}{n},\quad Var_{*}(\pi_{j})=\dfrac{m(n-1)}{n^{2}},\quad Cov_{*}(\pi_{j},\pi_{l})=-\dfrac{m}{n^{2}}. \notag
		\end{align}
		For example, in the least squares estimation with bootstrap sample, let $Y^{*}=(y_{1}^{*},\cdots,y_{n}^{*})'$, then we can write
		\begin{align}
			X_{(q)}^{*'}X_{(q)}^{*}=X_{(q)}'P'PX_{q}=X_{(q)}'\Pi X_{(q)}=\sum\limits_{i=1}^{n}\pi_{i}x_{i(q)}x_{i(q)}' \notag
		\end{align}
		and
		\begin{align}
			X_{(q)}^{*'}Y^{*}=X_{(q)}'P'PY=X_{(q)}'\Pi Y=\sum\limits_{i=1}^{n}\pi_{i}x_{i(q)}y_{i}. \notag
		\end{align}
		\indent After applying the bootstrap pairs procedure, the least squares estimator for $\Theta_{q}$ is given by $\tilde{\Theta}_{q}^{*}=(X_{(q)}^{*'}X_{(q)}^{*})^{-1}X_{(q)}^{*'}Y^{*}$. Similarly, the corresponding estimators for $\mu$ and the residual vector are obtained as $\tilde{\mu}_{q}=X_{(q)}\tilde{\Theta}_{q}^{*}=X_{(q)}(X_{(q)}^{*'}X_{(q)}^{*})^{-1}X_{(q)}^{*'}Y^{*}\triangleq H_{q}^{*}Y^{*}$ and $\tilde{e}_{q}=Y-\tilde{\mu}_{q}$, respectively. Hence, the bootstrap version of the averaging estimator for $\mu$ is
		\begin{align}
			\tilde{\mu}(\omega)=\sum\limits_{q=1}^{M}\omega_{q}X_{(q)}\tilde{\Theta}_{q}^{*}=\sum\limits_{q=1}^{M}\omega_{q}H_{q}^{*}Y^{*}=H(\omega)^{*}Y^{*}, \notag
		\end{align}
		where $H(\omega)^{*}=\sum_{q=1}^{M}\omega_{q}H_{q}^{*}$.\\
		\indent We propose the following bootstrap criterion for choosing weights in the model average estimator $\hat{\mu}(\omega)$:
		\begin{align}
			\Gamma_{n,m}(\omega)=\dfrac{1}{n}E_{*}||Y-\tilde{\mu}(\omega)||^{2}, \notag
		\end{align}
		where $E_{*}$ is the expectation of bootstrap sample. Similarly, we employ $Var_{*}$ and $Cov_{*}$ to denote the variance and covariance of the bootstrap sample, respectively. These measures can also be expressed as $E_{*}(\bm\cdot)=E(\bm\cdot|X,Y)$, $Var_{*}(\bm\cdot)=Var(\bm\cdot|X,Y)$ and $Cov_{*}(\bm\cdot)=Cov(\bm\cdot|X,Y)$. Alternatively, we can represent them as $E(\bm\cdot|\hat{F}_{n})$, $Var(\bm\cdot|\hat{F}_{n})$ and $Cov(\bm\cdot|\hat{F}_{n})$, respectively. The bootstrap method provides an estimate of the true expected squared error by utilizing bootstrap samples.\\
		\indent As \cite{shao1996bootstrap} mentioned, by using the bootstrap, we can approximate the characteristics of the population and obtain different ``future responses" to fit our model. We also follow this approach to construct the criterion. In the subsequent theoretical proofs, it can be observed that the dominant term in $\Gamma_{n,m}(\omega)$ is $L_{n}(\omega)$. Consequently, selecting a weight vector among all $\omega\in\textbf{H}_{n}$ that minimizes $\Gamma_{n,m}(\omega)$ is equivalent to choosing, within this set of weights, the one that yields the best in-sample prediction.\\
		\indent The weight vector of bootstrap criterion is given by
		\begin{align}
			\hat{\omega}=\mathop{\arg\min}_{\omega\in\textbf{H}_{n}}\Gamma_{n,m}(\omega), \notag
		\end{align}
		and the corresponding BTMA estimator is $\hat{\mu}(\hat{\omega})$. If we consider a unit weight vector $\omega_{q}^{0}$, the BTMA criterion will be simplified to the bootstrap model selection criterion in \cite{shao1996bootstrap}. Thus, BTMA can be considered as an extension of bootstrap model selection. The BTMA criterion is a quadratic function of $\omega$ with the constraint $\omega\in\textbf{H}_{n}$, making it convenient to determine the minimizer.
		\subsection{Implementation of bootstrap model averaging}
		\indent In practical applications, there may be situations where $(X^{*'}X^{*})^{-1}$ does not exist. To address this issue, two potential approaches can be considered: repeating the bootstrapping procedure in pairs or replacing $\tilde{\Theta}^{*}_{q}$ with $\hat{\Theta}_{q}$. The latter reduces the computational burden but may yield the same result as the MMA estimator in extreme cases. Hence, we opt for the former and provide an algorithm below for computing the BTMA estimator $\hat{\mu}(\hat{\omega})$.\\
		\begin{breakablealgorithm}
			\caption{Bootstrap model averaging}
			\begin{algorithmic}[1]
				\Require The sample $S=\{(y_{i},x_{i}),i=1,\cdots,n\}$;
				\Ensure The BTMA estimator $\hat{\mu}(\hat{\omega})$;
				\State Set the Monte Carlo size to be $B$, resample size to be $m$ and $\hat{\Gamma}_{n,m}(\omega)=0$;
				\State Denote $X=(x_{1},\cdots,x_{n})'$ and $Y=(y_{1},\cdots,y_{n})'$;
				\For{each $b\in\{1,\cdots,B\}$}
				\Repeat
				\State Let $S^{*}=\{(y^{*}_{i},x^{*}_{i}),i=1,\cdots,m\}$ be a bootstrap sample drawn by simple random sampling with replacement from sample $S$;
				\State Denote $X^{*}=(x^{*}_{1},\cdots,x^{*}_{m})'$ and $Y^{*}=(y^{*}_{1},\cdots,y^{*}_{m})'$;
				\Until{$(X^{*'}X^{*})^{-1}$ exists}
				\State $\tilde{\Theta}^{*}_{q}$ is obtained from bootstrap sample $S^{*}$, $q=1,\cdots,M$;
				\State Compute the residual error $\tilde{e}_{q}=Y-X\tilde{\Theta}^{*}_{q}$;
				\State Compute the residual error matrix $\bar{e}=(\tilde{e}_{1},\cdots,\tilde{e}_{M})$;
				\State Compute $\hat{\Gamma}_{n,m}(\omega)=\hat{\Gamma}_{n,m}(\omega)+\dfrac{1}{B}\omega'\bar{e}'\bar{e}\omega$;
				\EndFor
				\State $\hat{\omega}$ is obtained by minimizing $\hat{\Gamma}_{n,m}(\omega)$;
				\State The BTMA estimator $\hat{\mu}(\hat{\omega})$ is an MA estimator using bootstrap weight $\hat{\omega}$.
			\end{algorithmic}
		\end{breakablealgorithm}
		\vspace{1cm}
		\begin{remark}
			Our weight selection criterion can also be interpreted as a bootstrap aggregating method, as in Algorithm 1. We replace $n\Gamma_{n,m}(\omega)$ with $\sum_{b=1}^{B}||Y_{b}-\tilde{\mu}_{b}(\omega)||^{2}\big/B$ for computation, which serves as a Monte Carlo approximation. Here, $Y_{b}$ is the observation vector of response variable in the $b$th bootstrap replication, $\tilde{\mu}_{b}(\omega)$ is the BTMA estimator based on the $b$th bootstrap resampling observations, and $B$ is the Monte Carlo sample size (also known as the number of bootstrap replications). This is similar to bagging, where a model is selected from each bootstrap replication of the original dataset, and then the corresponding predictors are smoothed. The difference is that our method first calculates the criterion for each bootstrap replication, then smooths the results of these $B$ calculations, and finally derives the weight vector from this smoothed output. This demonstrates that our bootstrap weight vector also has the ability of generalization.
		\end{remark}
		\begin{remark}
			For the $m$-out-of-$n$ approach, whether in bootstrap (sampling with replacement) or subsampling (sampling without replacement), it is often required that the selection of $m$ should satisfy $m/n\rightarrow0$ as $m\rightarrow\infty$ and $n\rightarrow\infty$. This assumption aims to ensure the consistency. In model averaging approach, we usually focus on prediction, and consider to prove the asymptotic optimality. Therefore, we do not need the restriction of $m/n\rightarrow0$ in the proof of our theorem, which means that we can allow $m\propto n$ or to take a higher order than $n$.
		\end{remark}
		\indent\quad\quad To implement our BTMA approach, an important issue is the selection of resample size $m$. In applications, we recommend using the generalized cross validation (GCV) method to select the corresponding $m$. It is also clear that GCV would increase the complexity of the algorithm and the computation time. In Section 6, we find that setting $m=n/2$ also yields good performance in terms of prediction. So, as a simple and quick choice for resample size, we can take $m=n/2$. The theoretical motivation for such a choice comes from the relationship between Mallows' $C_{p}$ criterion and the bootstrap model selection criterion. In fact, if $\sigma^{2}$ is known, then Mallows' $C_{p}$ and the bootstrap model selection criteria can be expressed as $C_{p}=||Y-\hat{\mu}||^{2}/n+(2\sigma^{2}/n)\text{tr}(H_{q})$ and $\Gamma_{n,m}(q)=||Y-\hat{\mu}||^{2}/n+(\sigma^{2}/m)\text{tr}(H_{q})+o_{p}(1/m)$, respectively (see \cite{shao1996bootstrap}). It is seen that when $m=n/2$, these two criteria are asymptotically equivalent. Considering that MMA is an extension of Mallows' $C_{p}$ and our BTMA is a generalization of bootstrap model selection, a natural choice of $m$ is $n/2$.
		\subsection{Relationship with MMA}
		\indent In this subsection, we investigate the relationship between our proposed BTMA and MMA. To this end, we suppose that the following conditions hold almost surely:
		\begin{flalign}
			&\text{\textbf{Condition} (C.1).}\ \max_{1\leq i\leq n}\big\{|x_{ij(q)}|\big\}=O(1)\ \text{uniformly for}\ q\in{1,\cdots,M}.& \notag\\
			&\text{\textbf{Condition} (C.2).}\ C_{1}<\lambda_{\text{min}}\big\{(X_{(q)}'X_{(q)}/n)\big\}\leq\lambda_{\text{max}}\big\{(X_{(q)}'X_{(q)}/n)\big\}<C_{2}\ \text{for all}\ 1\leq q\leq M,\notag\\
			&\text{where}\ C_{1}\ \text{and}\ C_{2}\ \text{are two positive constants},\ \text{and $\lambda_{\text{min}}\{C\}$ and $\lambda_{\text{max}}\{C\}$ are the minimum and } \notag\\
			&\text{maximum singular values of a general real matrix $C$, respectively}.& \notag
		\end{flalign}
		\indent Condition (C.1) means that the elements of design matrix $X$ are bounded. Condition (C.2) is also an assumption about covariates. These two conditions are mild and widely encountered in the literature. See, for example, \cite{shao1993linear} and \cite{zhang2020parsimonious}. The following theorem reveals the relationship between MMA and BTMA:
		\begin{theorem}
			Suppose that $k_{q}$ is fixed and all candidate models are true or over-fitted. Under Conditions (C.1) and (C.2), when $m/n\rightarrow0$, we have
			\begin{align}
				\Gamma_{n,m}(\omega)
				=\dfrac{||Y-\hat{\mu}(\omega)||^{2}}{n}+\dfrac{\sigma^{2}}{m}\text{tr}H^{2}(\omega)+o_{p}(\dfrac{1}{m}). \notag
			\end{align}
		\end{theorem}
		\begin{proof}
			See the online supplemental file.
		\end{proof}
		\indent It is interesting to observe that the form of MMA criterion is $C_{\text{MMA}}=||Y-\hat{\mu}(\omega)||^{2}/n+(2\sigma^{2}/n)\text{tr}H(\omega)$. So essentially, the difference between MMA and BTMA lies in the penalty terms. Especially, from Theorem 1, BTMA and MMA are asymptotically equivalent for model selection, as shown in Section 3.3 or \cite{shao1996bootstrap}.
		\section{Theoretical Properties}
		\label{sec:theo}
		\indent This section demonstrates the asymptotic optimality of the BTMA estimator $\hat{\mu}(\hat{\omega})$ and the consistency of the bootstrap weight vector $\hat{\omega}$. Denote the $M^{*}$th model as the largest candidate model such that $k_{M^{*}}=\max_{1\leq q\leq M}\{k_{q}\}$, and define $h_{ii(q)}$ as the $i$th diagonal element of $H_{q}$. We assume that the following conditions hold almost surely.
		\begin{flalign}
			&\text{\textbf{Condition} (C.3).}\ \lambda_{\text{min}}\big\{X^{*'}_{(q)}X^{*}_{(q)}/m\big\}\geq C_{3}>0\ \text{for all}\ 1\leq q\leq M,\ \text{where}\ C_{3}\ \text{is a positive con}\notag\\
			&\text{-stant}.& \notag\\
			&\text{\textbf{Condition} (C.4).}\ \max_{1\leq i\leq n}\big\{\mu_{i}^{2}\big\}=O(1).& \notag\\
			&\text{\textbf{Condition} (C.5).}\ \max_{1\leq q\leq M}\max_{1\leq i\leq n}\big\{h_{ii(q)}\big\}=O(k_{M^{*}}/n).& \notag\\
			&\text{\textbf{Condition} (C.6).}\ \frac{n(n+m)^{1/2}M^{3/2}k_{M^{*}}^{2}}{m\xi_{n}}\rightarrow0\ \text{and}\ \frac{n^{2}k_{M^{*}}^{3}}{m^{2}\xi_{n}}\rightarrow0, \text{where}\ \xi_{n}=\inf_{\omega\in\textbf{H}_{n}}R_{n}(\omega).& \notag\\
			&\text{\textbf{Condition} (C.7).}\ \max_{1\leq i\leq n}\big\{E(e^{4G}_{i}|x_{i})\big\}=O(1)\ \text{for some fixed integer}\ 1\leq G<\infty.& \notag\\
			&\text{\textbf{Condition} (C.8).}\ Mn\xi_{n}^{-G}\rightarrow 0\ \text{for some fixed integer}\ 2\leq G<\infty.& \notag\\
			&\text{\textbf{Condition} (C.9).}\ M\xi_{n}^{-2G}\sum\limits_{q=1}^{M}\Big(R_{n}(\omega_{q}^{0})\Big)^{G}\rightarrow 0\ \text{for some fixed integer}\ 1\leq G<\infty.& \notag
		\end{flalign}
		\indent Condtion (C.3) makes matrices $X_{(q)}^{*'}X_{(q)}^{*}$, $q=1,\cdots,M$ invertible. If $X_{(q)}^{*'}X_{(q)}^{*}$ is not invertible, the bootstrapping procedure is repeated until an invertible matrix is obtained, as mentioned in Section 3.2.  \cite{shao2012jackknife} assumed the same condition. Condition (C.4) restricts $\mu_{1}^{2},\cdots,\mu_{n}^{2}$ to be bounded, which implies that $\mu'\mu$ is of the order $O(n)$.
		For example, \cite{shibata1981optimal} considered a data-generating process $y_{i}=\mu_{i}+e_{i}=f(z_{i})+e_{i}$, $i=1,\cdots,n$, where $f(z_{i})=\sum_{s=1}^{\infty}\theta_{s}\text{cos}\big((s-1)z_{i}\big)/s$ with $z_{i}=2\pi(i-1)/n$ and $e_{i}$ are i.i.d. $N(0,\sigma^{2})$ errors (See also \cite{wan2010least}). Assuming $\theta_{s}=s^{-\alpha}$ for some $\alpha>0$, since $\big|\text{cos}\big((s-1)z_{i}\big)\big|$ is bounded, we can immediately obtain that $\max_{1\leq i\leq n}|\mu_{i}|$ is bounded by a constant, and so $\max_{1\leq i\leq n}\{\mu_{i}^{2}\}=O(1)$. Condition (C.5) indicates the absence of extremely influential covariates. Similar conditions can be found in other works, such as \cite{li1987asymptotic}, \cite{zhang2013model} and \cite{liao2020corrected}. Condition (C.6) imposes assumptions on the sample size $n$, resample size $m$, dimension $k_{M}^{*}$ and the number of candidate models $M$. We allow the dimension $k_{M^{*}}$ and the number of candidate models $M$ to be divergent when $n$ and $m$ increase, but with certain constraints on the rates of divergence. There are analogous assumptions in \cite{ando2014model} and \cite{gao2016model}, among others. Condition (C.7) places a bound on the conditional moments of $e_{i}$, $i=1,\cdots,n$. Conditions (C.8) and (C.9) imply that $\xi_{n}\rightarrow\infty$ and also require $M\rightarrow\infty$ with a slow rate. Thus, we have to reduce the number of candidate models. Model screening can be performed to determine the set of candidate models before computing the weight vector. These conditions are similar to those in \cite{andrews1991asymptotic}, \cite{wan2010least} and \cite{liao2019model}.
		\begin{theorem}
			Under Conditions (C.1)-(C.8), we have
			\begin{align}
				\dfrac{L_{n}(\hat{\omega})}{\inf_{\omega\in\textbf{H}_{n}}L_{n}(\omega)}\stackrel{p}{\longrightarrow}1. \notag
			\end{align}
		\end{theorem}
		\begin{proof}
			See the online supplemental file.
		\end{proof}
		\indent Theorem 2 demonstrates that the BTMA estimator achieves asymptotic optimality by obtaining the lowest possible squared error. The bootstrap weight vector selected by $\hat{\Gamma}_{n,m}(\omega)$ is asymptotically equivalent to the infeasible best weight vector to minimize $L_{n}(\omega)$.\\
		\indent Now we consider the consistency of the bootstrap weight vector $\hat{\omega}$. Denote $\Lambda=(\hat{\mu}_{1},\cdots,\hat{\mu}_{M})$, $\Omega=(\mu-\hat{\mu}_{1},\cdots,\mu-\hat{\mu}_{M})$ and $\tilde{\Omega}=(\tilde{\mu}_{1}-\hat{\mu}_{1},\cdots,\tilde{\mu}_{M}-\hat{\mu}_{M})$. We make the following regularity conditions.
		\begin{flalign}
			&\text{\textbf{Condition} (C.10).}\ C_{4}<\lambda_{\text{min}}\big\{\Lambda'\Lambda\big/n\}\leq\lambda_{\text{max}}\big\{\Lambda'\Lambda/n\big\}<C_{5}\ \text{in probability tending to 1},\notag\\
			&\text{where}\ C_{4}\ \text{and}\ C_{5}\ \text{are two positive constants}.& \notag\\
			&\text{\textbf{Condition} (C.11).}\ ||X_{(q)}'e/\sqrt{n}||=O_{p}(k_{M^{*}}^{1/2})\ \text{uniformly in}\ q.& \notag\\
			&\text{\textbf{Condition} (C.12).}\ m=C_{6}n^{\gamma},\ \text{where}\ \gamma\geq1\ \text{and}\ C_{6}\ \text{is a positive constant}.& \notag\\
			&\text{\textbf{Condition} (C.13).}\ \frac{n^{1/2}M^{3/2}k_{M^{*}}^{2}}{\xi_{n}n^{\delta}}\rightarrow 0\ \text{and}\ \frac{Mk_{M^{*}}^{3}}{n^{\delta}}\rightarrow 0,\ \text{where}\ \delta\ \text{is a positive constant sa} \notag\\
			&\text{-tisfying}\ 0<\delta<\dfrac{1}{2}.& \notag
		\end{flalign}
		\indent Conditions (C.10) and (C.11) are the same as those in \cite{liao2020corrected}. Condition (C.10) requires that the minimum singular value of $\Lambda'\Lambda$ has a lower bound away from zero, and the maximum singular value of $\Lambda'\Lambda$ is also bounded. Condition (C.11) is a high-level condition that can be derived from certain original conditions. Condition (C.12) pertains to the resample size and is considered mild because we have flexibility in choosing the resample size $m$. Condition (C.13) imposes restrictions on the dimension $k_{M^{*}}$ and the number of candidate models $M$. From Conditions (C.12) and (C.13), we can also find that $\delta+\dfrac{1}{2}<\gamma$.\\
		\indent Denote the optimal weight as $\omega^{0}=\arg\min_{\omega\in\textbf{H}_{n}}E||\mu-\hat{\mu}(\omega)||^{2}$ and assume that it is an interior point of $\textbf{H}_{n}$. The following theorem describes the convergence rate at which $\hat{\omega}$ tends to $\omega^{0}$.
		\begin{theorem}
			Suppose that Conditions (C.1)-(C.8) and (C.10)-(C.13) are satisfied, then there exists a local minimizer $\hat{\omega}$ of the bootstrap criterion $\Gamma_{n,m}(\omega)$, such that
			\begin{align}
				||\hat{\omega}-\omega^{0}||=O_{p}(\xi_{n}^{1/2}n^{-1/2+\delta}). \notag
			\end{align}
		\end{theorem}
		\begin{proof}
			See the online supplemental file.
		\end{proof}
		\indent Theorem 3 provides insights into the convergence rate of the bootstrap weight vector $\hat{\omega}$ towards the optimal weight vector $\omega^{0}$. The result of this theorem exhibits similarity to those obtained through the MMA and JMA approaches, given stricter conditions regarding the dimension $k_{M^{*}}$ and the number of candidate models $M$. The theorem reveals that the convergence rate is determined by two key factors: the sample size $n$ and the minimized squared risk $\xi_{n}$. As the sample size $n\rightarrow\infty$, the rate at which $\hat{\omega}$ converges to $\omega^{0}$ depends on the behavior of $\xi_{n}$. In particular, when $\xi_{n}$ tends to infinity at a slower rate, the convergence of $\hat{\omega}$ to $\omega^{0}$ becomes faster.
		\section{Asymptotic Distribution}
		\label{sec:asym}
		\indent In this section, our focus is on investigating the asymptotic distribution of the BTMA estimator for the linear regression model, aiming to address the issue of statistical inference. Let $\mu_{i}=x_{i}'\Theta=x_{1i}'\beta_{1}+x_{2i}'\beta_{2}$ in model (\ref{1}) and consider a set of $M$ nested candidate models with fixed $k_{q}$, $q=1,\cdots,M$. We denote $X_{1}=(x_{11},\cdots,x_{1n})'$, $X_{2}=(x_{21},\cdots,x_{2n})'$ and $X=(X_{1},X_{2})$. The core regressors which must be included in the model are contained in the $k_{1^{*}}\times1$ vector $x_{1i}$, and the auxiliary regressors are contained in the $k_{2^{*}}\times1$ vector $x_{2i}$. Without loss of generality, we assume that the first $M_{0}$ candidate models are under-fitted, where $0\leq M_{0}<M$. Other candidate models are just-fitted and over-fitted models. A just-fitted model has no omitted variable and no irrelevant variable, and an over-fitted model has no omitted variable but irrelevant variables. Both of these two types of model can be called correct models. Let $k=k_{1^{*}}+k_{2^{*}}$ be the number of regressors in the model (\ref{1}) and $k_{q}$ be the number of regressors in the $q$th model. In applications, we have at most $k_{2^{*}}+1$ candidate models, since the $k_{1^{*}}$ core regressors are contained in the candidate models. Namely, the $q$th candidate model includes all regressors in $X_{1}$ and the first $q-1$ regressors in $X_{2}$. Additionally, we assume that $X$ is full of column rank.\\
		\indent Let $S_{q}$ be a $k_{q}\times k$ selection matrix and denote $S_{q}=(I_{k_{q}}, \textbf{0}_{k_{q}\times(k-k_{q})})$, then $X_{(q)}=XS_{q}'=(X_{1},X_{2})S_{q}'$. The least squares estimator of the $q$th candidate model is $\hat{\beta}_{q}=S_{q}'\hat{\Theta}_{q}=S_{q}'(X_{(q)}'X_{(q)})^{-1}X_{(q)}'Y$. The definition of model averaging estimator in this section is
		\begin{align}
			\hat{\beta}(\omega)=\sum\limits_{q=1}^{M}\omega_{q}
			\Bigg(
			\begin{matrix}
				\hat{\Theta}_{q}\\
				0_{(k-k_{q})\times1}
			\end{matrix}
			\Bigg)
			=\sum\limits_{q=1}^{M}\omega_{q}\hat{\beta}_{q}. \notag
		\end{align}
		We can also rewrite $\beta=(\beta_{1}',\beta_{2}')'=(\Theta'_{t},\textbf{0}')'$ in this section, where $\Theta_{t}$ is the parameter vector of true model. In matrix notation, we denote the model (\ref{1}) as $Y=X\beta+e=X_{1}\beta_{1}+X_{2}\beta_{2}+e=X_{t}\Theta_{t}+e$, where $X_{t}$ is the design matrix corresponding to $\Theta_{t}$. Minimizing the bootstrap criterion $\Gamma_{n,m}(\omega)$, we obtain the BTMA estimator of $\beta$ as $\hat{\beta}(\hat{\omega})=\sum_{q=1}^{M}\hat{\omega}_{q}\hat{\beta}_{q}$. We now list the following conditions:
		\begin{flalign}
			& \text{\textbf{Condition} (C.14).}\ n=C_{7}m,\ \text{where}\ C_{7}>0\ \text{is a positive constant}. &\notag\\
			& \text{\textbf{Condition} (C.15).}\ \text{As}\ n\rightarrow\infty,\ Q_{n}=X'X/n\rightarrow Q\ \text{almost surely and}\ Z_{n}=n^{-1/2}X'e\stackrel{d}{\rightarrow}Z \notag\\
			&\sim N(0,\Xi),\ \text{where}\ Q=E(x_{i}x_{i}')\ \text{and}\ \Xi=E(x_{i}x_{i}'e_{i}^{2})\ \text{are positive definite matrices}. &\notag
		\end{flalign}
		\indent Condition (C.14) establishes the relationship between the sample size $n$ and the resample size $m$. Condition (C.15) is a commonly used condition. As stated in \cite{liu2015distribution} and \cite{zhang2019inference}, this condition serves as a high-level requirement that allows for the application of cross-sectional, panel, and time-series data with finite fourth moment. It holds under appropriate primitive assumptions. The second part of Condition (C.15) is similar to Condition (C.11) when the dimension $k$ is fixed.\\
		\indent Since the bootstrap weight vector is random, it will affect the asymptotic distribution of the BTMA estimator. The following theorem demonstrates the behavior of the BTMA weights assigned to the under-fitted models.
		\begin{theorem}
			Suppose that Conditions (C.1)-(C.7), (C.14) and (C.15) hold, then for any $q\in\{1,\cdots,M_{0}\}$,
			\begin{align}
				\hat{\omega}_{q}=O_{p}(m^{-1}). \notag
			\end{align}
		\end{theorem}
		\begin{proof}
			See the online supplemental file.
		\end{proof}
		\indent Theorem 4 states that the order of the weights of under-fitted models is $O_{p}(m^{-1})$, which is determined by the order of $m$. By Condition (C.14), we have $\hat{\omega}_{q}=O_{p}(n^{-1})$ for any $q\in\{1,\cdots,M_{0}\}$. Generally, this theorem indicates that the weights of under-fitted models converge to zero at a rate of $1/n$, so that we can omit them in the asymptotic properties.\\
		\indent Next, we study the BTMA weights of candidate models with $q\in\{M_{0}+1,\cdots,M\}$. Let $R=M-M_{0}$ be the number of just-fitted and over-fitted models. Since $R$ is not smaller than one, we define a new weight vector $\nu=(\nu_{1},\cdots,\nu_{R})'$ and the new weight set $\mathbf{L}_{n}=\{\nu\in[0,1]^{R}:\sum_{r=1}^{R}\nu_{r}=1\}$. This set excludes the under-fitted models.\\
		\indent Let $\Xi_{r}=S_{M_{0}+r}\Xi S_{M_{0}+r}'$, $Q_{r}=S_{M_{0}+r}QS_{M_{0}+r}'$ and $V_{r}=S_{M_{0}+r}'Q_{r}^{-1}S_{M_{0}+r}$, $r=1,\cdots,R$ be the matrices associated with the new weight vector. Denote the asymptotic BTMA weight vector of ture and over-fitted models as $\tilde{\nu}=(\tilde{\nu}_{1},\cdots,\tilde{\nu}_{R})'=\text{argmin}_{\nu\in\mathbf{L}_{n}}\nu'\Delta\nu$, where $\Delta$ is an $R\times R$ matrix with the $(r,t)$-th element
		\begin{align}
			\Delta_{rt}=\dfrac{n\sigma^{2}}{m}k_{\min\{r,t\}}+Z'\big(Q-V_{\max\{r,t\}}\big)Z=C_{7}\sigma^{2}k_{\min\{r,t\}}+Z'\big(Q-V_{\max\{r,t\}}\big)Z. \notag
		\end{align}
		We present the asymptotic distribution of $\hat{\beta}(\hat{\omega})$ in the following theorem:
		\begin{theorem}
			Suppose that Conditions (C.1)-(C.7), (C.14) and (C.15) hold. Then we have
			\begin{align}
				\sqrt{n}(\hat{\beta}(\hat{\omega})-\beta)\stackrel{d}{\longrightarrow}\sum\limits_{r=1}^{R}\tilde{\nu}_{r}V_{r}Z. \notag
			\end{align}
		\end{theorem}
		\begin{proof}
			See the online supplemental file.
		\end{proof}
		\indent Theorem 5 establishes that the BTMA estimator can be asymptotically expressed as a nonlinear function of the normal random vector $Z$. This implies that the BTMA estimator exhibits a nonstandard limiting distribution in large sample sizes. To derive the asymptotic distribution, it is necessary to include at least one correct model in the set of candidate models. Unlike \cite{hjort2003frequentist}, Theorem 5 does not rely on a local asymptotic framework, such as $\beta_{2}=\epsilon/\sqrt{n}\rightarrow0$, where $\epsilon$ represents an unknown fixed parameter. An important application of Theorem 5 is to construct the confidence interval. We propose an algorithm to obtain the confidence interval of $\beta_{j}$.
		\vspace{0.5cm}
		\begin{breakablealgorithm}
			\caption{Confidence interval of $\beta_{j}$}
			\begin{algorithmic}[1]
				\Require The sample $S=\{(y_{i},x_{i}),i=1,\cdots,n\}$;
				\Ensure The confidence interval of $\beta_{j}$;
				\State Denote $X=(x_{1},\cdots,x_{n})'$ and $Y=(y_{1},\cdots,y_{n})'$;
				\State Let $\hat{e}_{i}=y_{i}-x_{i}'\hat{\beta}_{M}$ be the least squares residual from the full model, where $\hat{\beta}_{M}=(X'X)^{-1}X'Y$;
				\State Obtain the estimators of $\sigma^{2}$, $Q$ and $\Xi$, which are given by $\hat{\sigma}^{2}=\dfrac{1}{n-k}\sum\limits_{i=1}^{n}\hat{e}_{i}^{2}$, $\hat{Q}=\dfrac{1}{n}\sum\limits_{i=1}^{n}x_{i}x_{i}'$ and $\hat{\Xi}=\dfrac{1}{n}\sum\limits_{i=1}^{n}x_{i}x_{i}'\hat{e}_{i}^{2}$, respectively;
				\State Generate a large number of $k\times1$ normal random vectors $Z^{(u)}\sim N(0,\hat{\Xi})$ for $u=1,\cdots,U$, and compute the quantities in the asymptotic distribution of Theorem 5 for each $u$;
				\State Compute $\hat{Q}_{r}=S_{M_{0}+r}\hat{Q}S_{M_{0}+r}'$ and $\hat{V}_{r}=S_{M_{0}+r}'\hat{Q}_{r}^{-1}S_{M_{0}+r}$ for a given $M_{0}$. In practice, $M_{0}$ can be obtained by the procedure of BIC model selection;
				\State Compute $\hat{\Delta}_{rt}^{(u)}=\dfrac{n\hat{\sigma}^{2}}{m}k_{\min\{r,t\}}+Z^{(u)'}\big(\hat{Q}^{-1}-\hat{V}_{\max\{r,t\}}\big)Z^{(u)}$;
				\State Compute the weight vector $\tilde{\nu}^{(u)}=\text{argmin}_{\nu\in\mathbf{L}_{n}}\nu'\hat{\Delta}^{(u)}\nu$, where the $(r,t)$th element of $\hat{\Delta}^{(u)}$ is $\hat{\Delta}_{rt}^{(u)}$. Then we compute $\sum\limits_{r=1}^{R}\tilde{\nu}_{r}^{(u)}V_{r}Z^{(u)}$;
				\State Let $\Upsilon^{(u)}_{\beta,j}$ be the $j$th element of $\sum\limits_{r=1}^{R}\tilde{\nu}_{r}^{(u)}V_{r}Z^{(u)}$. Then the $(\alpha/2)$th and $(1-\alpha/2)$th quantiles are obtained from $\Upsilon^{(u)}_{\beta,j}$ for $u=1,\cdots,U$, which are denoted as $\hat{z}_{\beta,j}(\alpha/2)$ and $\hat{z}_{\beta,j}(1-\alpha/2)$, respectively;
				\State Let $\hat{\beta}_{j}(\hat{\omega})$ be the $j$th element of $\hat{\beta}(\hat{\omega})$, where $\hat{\omega}$ is obtained by Algorithm 1;
				\State The confidence interval of $\beta_{j}$ is $CI_{\beta,n}=[\hat{\beta}_{j}(\hat{\omega})-n^{-1/2}\hat{z}_{\beta,j}(1-\alpha/2),\hat{\beta}_{j}(\hat{\omega})-n^{-1/2}\hat{z}_{\beta,j}(\alpha/2)]$.
			\end{algorithmic}
		\end{breakablealgorithm}
		\section{Simulation Study}
		\label{sec:simu}
		\indent In this section, we conduct three simulation studies to evaluate the performance of our averaging estimator using bootstrap weights. The first simulation is based on the setting of \cite{hansen2007least} and is designed to compare the risks of different methods. The second simulation examines the behavior of confidence intervals constructed with  different estimators. The third simulation, based on the setting of \cite{shao1996bootstrap}, aims to compare the performance of BTMA, BMS, and JMA estimators, and the related results are contained in Section S1 of the online supplemental file.
		\subsection{Simulation 1 based on the setting of Hansen (2007)}
		\indent This simulation setting is the same as that in \cite{hansen2007least}. The data are generated from the model
		\begin{align}
			y_{i}=\sum_{j=1}^{100}\theta_{j}x_{ij}+e_{i}, i=1,\cdots,n, \notag
		\end{align}
		where $x_{i1}=1$ and the remaining $x_{ij}$, $j=2,\cdots$ are from $N(0,1)$; the independent errors $e_{i}$, $i=1,\cdots,n$ are from $N(0,1)$ and independent of $x_{ij}$, $i=1,\cdots,n$; and the regression coefficients are given by $\theta_{j}=c\sqrt{2\alpha}j^{-\alpha-1/2}$, $j=1,\cdots,100$ with $\alpha$ varying in $\{0.5,1.0,1.5\}$, and $c$ being determined by $R^{2}=c^{2}/(1+c^{2})$ which is between 0.1 and 0.9. The sample size is set to be $n\in\{50, 150, 400\}$, and the number of candidate models $M$ is determined by $M=[3n^{1/3}]$. A larger $\alpha$ means that the regression coefficient $\theta_{j}$ decreases with a faster rate.\\
		\indent We compare the following eleven different methods: (\romannumeral1) AIC model selection (AIC), (\romannumeral2) BIC model selection (BIC), (\romannumeral3) Mallows model selection (Mallows), (\romannumeral4) smoothed AIC (S-AIC) model averaging, (\romannumeral5) smoothed BIC (S-BIC) model averaging, (\romannumeral6) Mallows model averaging (MMA), (\romannumeral7) jackknife model averaging (JMA), (\romannumeral8) bootstrap model selection (BMS), (\romannumeral9) subsampling model averaging (Sub), (\romannumeral10) bagging (Bag), and (\romannumeral11) bootstrap model averaging (BTMA). Subsampling model averaging is similar to the BTMA method, but it uses sampling without replacement. For the subsampling, we take the resample sizes of $[0.632n]$ and $[n^{2/3}]$, denoted as Sub1 and Sub2, respectively. These choices are based on \cite{davison2003recent} and \cite{de2016subsampling}. Regarding bagging, we consider the smoothing based on the $C_{p}$ criterion which is outlined by \cite{efron2014estimation}. In order to compare these methods, we compute the simulated risks.\\
		\indent To compute the risk of BTMA estimator, after generating the data $(y_{i},x_{i})$, $i=1,\cdots,n$, we resample it based on bootstrapping pairs $m$ times. If $X_{(q)}^{*'}X_{(q)}^{*}$ is not of full rank, then we discard this bootstrap sample and resample the data again until $X_{(q)}^{*'}X_{(q)}^{*}$ is invertible. The bootstrap criterion is computed by Monte Carlo with size $B=500$. Then the weight vector $\hat{\omega}$ is obtained. If we denote $\bar{e}_{b}=(\tilde{e}_{1,b},\cdots,\tilde{e}_{M,b})$ as an $n\times M$ residual estimation matrix, where $\tilde{e}_{q,b}=Y_{b}-\tilde{\mu}_{q,b}=Y_{b}-X_{(q),b}\tilde{\Theta}^{*}_{q,b}$ is the bootstrap residual vector for the $b$th computation, $b=1,\cdots,B$ under the $q$th candidate model, $q=1,\cdots,M$. Then the bootstrap criterion is equivalent to
		\begin{align}
			\Gamma_{n,m}(\omega)=\dfrac{1}{nB}\sum\limits_{b=1}^{B}\omega'\bar{e}_{b}'\bar{e}_{b}\omega. \notag
		\end{align}
		This alternative criterion derived from the law of large numbers is presented for calculating $\hat{\omega}$. To optimize this alternative criterion, the quadprog package in the R language can be used. The simulation results are displayed in Figures 1-4.\\
		\indent Figures 1-3 with the resample size $m=n/2$ reveal that BTMA achieves the smallest risks in most cases. When $\alpha=0.5$, Figure 1 shows that when the sample size $n$ is small, BTMA has the lowest risk for not large $R^{2}$ (such as $R^{2}\leq0.6$). As $n$ increases, the behavior of BTMA with larger $R^{2}$ becomes better and is similar to those of the best estimators. When $\alpha=1$, Figure 2 presents that BTMA outperforms all the competing methods over the vast majority of $R^{2}$. All of MA methods perform almost equally well as $n$ becomes large. When $\alpha=1.5$, Figure 3 displays that BTMA still dominates the other methods for a wide range of $R^{2}$ with small $n$ ($=50$). It performs as well as other MA methods for moderate $n$ ($=150$). For large $n$ ($=400$), as $R^{2}$ increases, the risk of BTMA gradually approaches that of MMA, JMA, and S-AIC, while still performs better than other methods. The subsampling methods exhibit critical flaws under the settings in Figures 1-3. It shows some competitiveness only when $\alpha=0.5$. Additionally, subsampling performs poorly for small values of $R^{2}$ when $m=[0.632n]$, and large values of $R^{2}$ when $m=[n^{2/3}]$. Bagging also shows unacceptable results in these settings.\\
		\indent In Figure 4, we compare the risks of BTMA with different resample sizes $m$. When $m$ has the same order as $n$, the corresponding risk is lower than that when $m$ has a higher or lower order than $n$. Among the various resample sizes $m$ that are of the same order as $n$, both $m=n/2$ and GCV consistently perform well.\\
		\indent In addition, we further compare our BTMA with BMS and JMA based on the setting of \cite{shao1996bootstrap}. The simulation results show better performance of our method in the vast majority of cases. The details can be found in Section S1 of the online supplemental file.
		\begin{figure}[!ht] 
			\begin{center}
				\begin{minipage}[htbp]{0.32\linewidth}
					\includegraphics[scale=0.3]{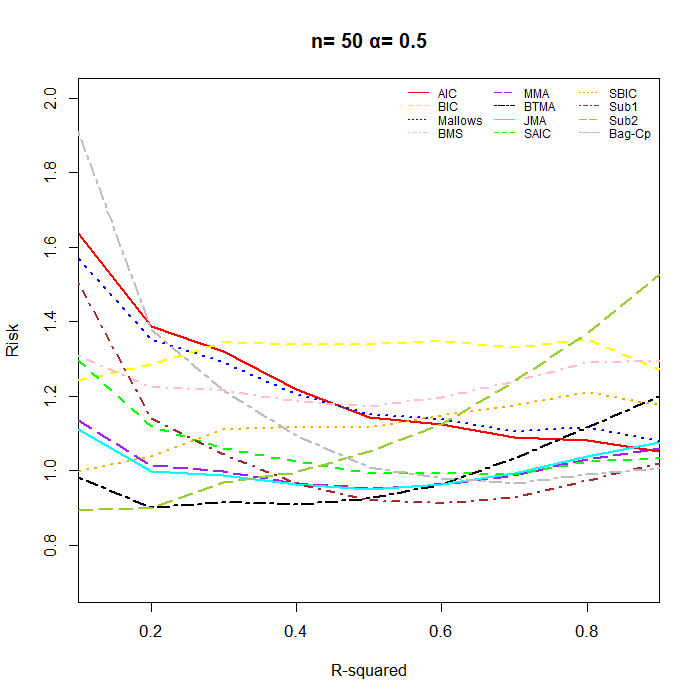}
				\end{minipage}%
				\begin{minipage}[htbp]{0.32\linewidth}
					\includegraphics[scale=0.3]{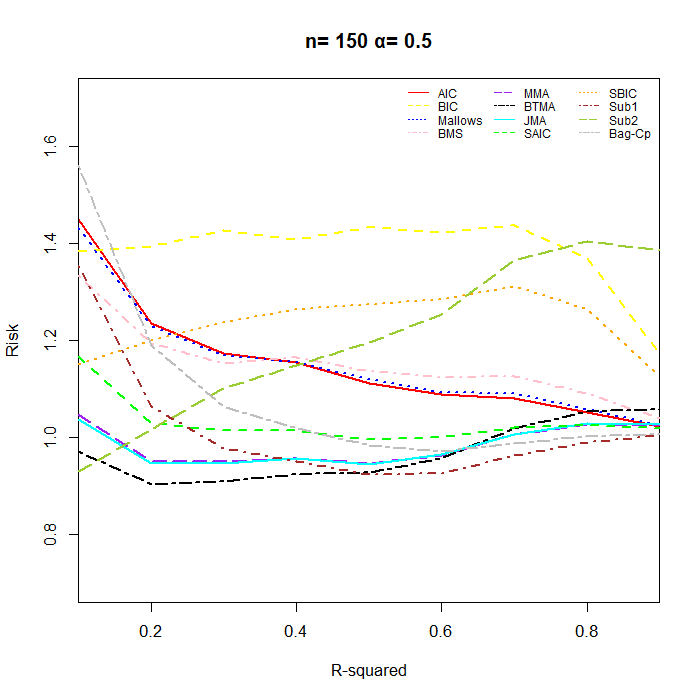}
				\end{minipage}%
				\begin{minipage}[htbp]{0.32\linewidth}
					\includegraphics[scale=0.3]{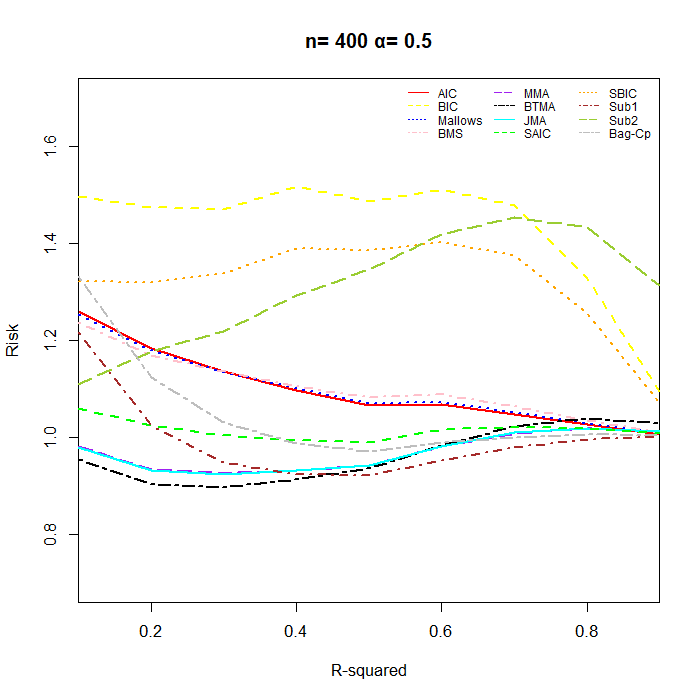}
				\end{minipage}%
			\end{center}
			\caption{Risks of various methods under different sample size $n$ with $\alpha=0.5$ \label{fig:first}}
		\end{figure}%
		\begin{figure}[!ht]
			\begin{center}
				\begin{minipage}[htbp]{0.32\linewidth}
					\includegraphics[scale=0.3]{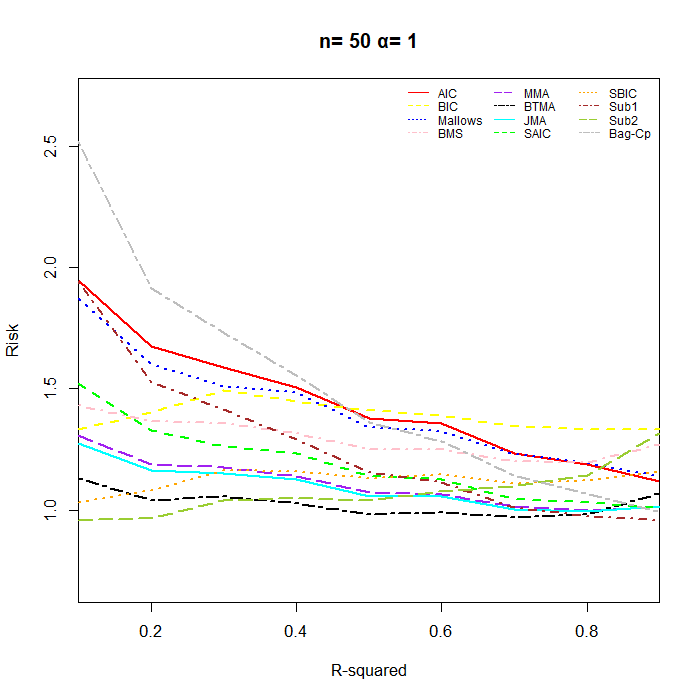}
				\end{minipage}%
				\begin{minipage}[htbp]{0.32\linewidth}
					\includegraphics[scale=0.3]{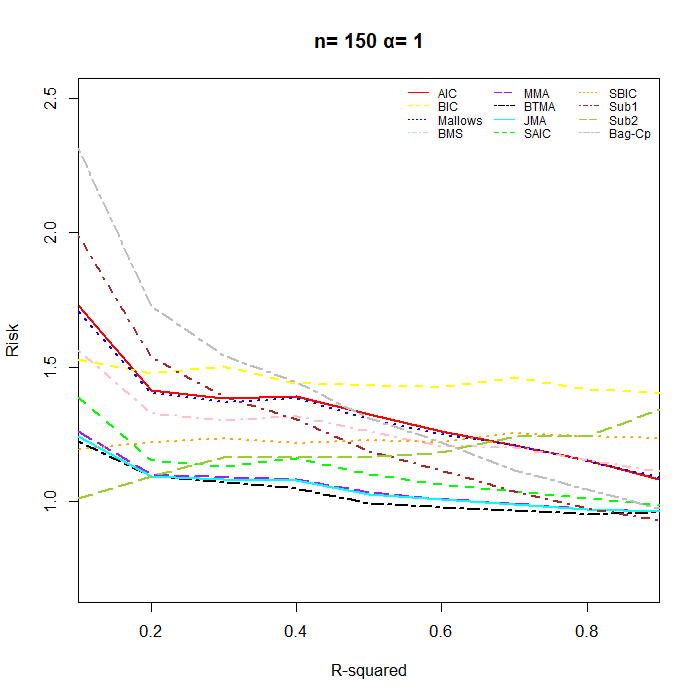}
				\end{minipage}%
				\begin{minipage}[htbp]{0.32\linewidth}
					\includegraphics[scale=0.3]{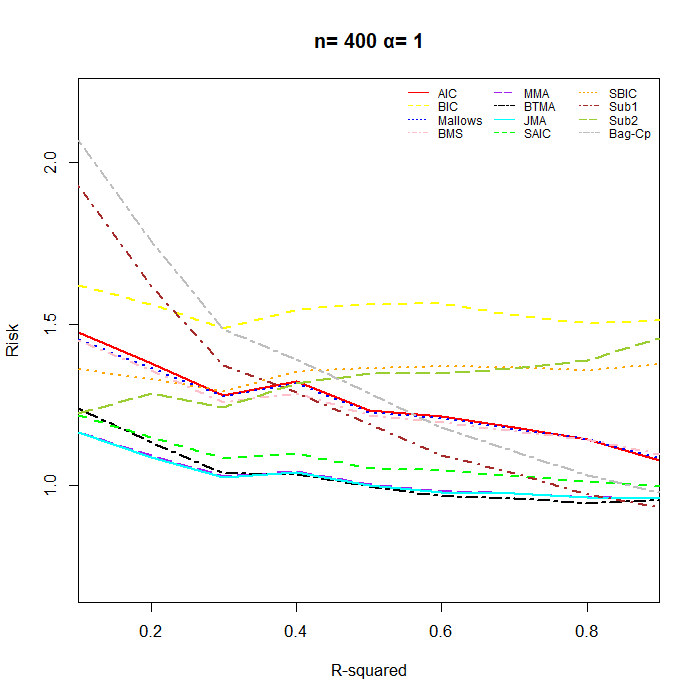}
				\end{minipage}%
			\end{center}
			\caption{Risks of various methods under different sample size $n$ with $\alpha=1$ \label{fig:second}}
		\end{figure}%
		\begin{figure}[!ht]
			\begin{center}
				\begin{minipage}[htbp]{0.32\linewidth}
					\includegraphics[scale=0.3]{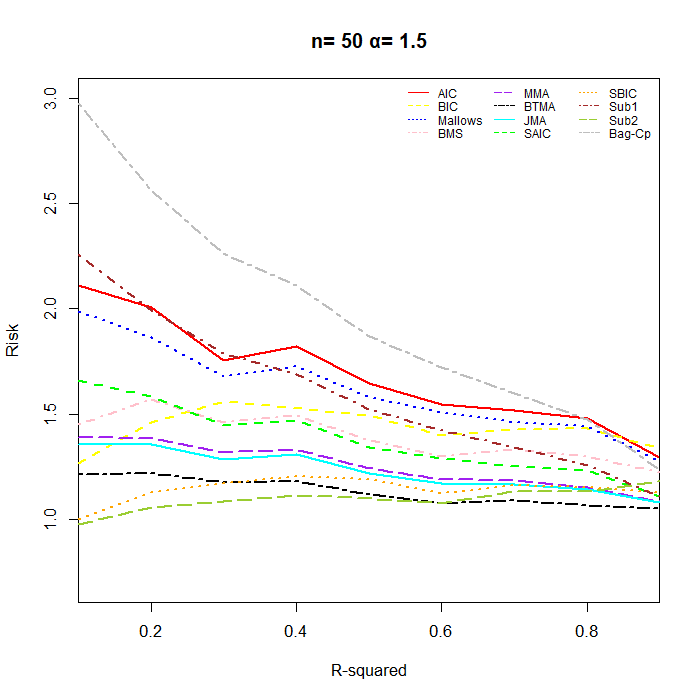}
				\end{minipage}%
				\begin{minipage}[htbp]{0.32\linewidth}
					\includegraphics[scale=0.3]{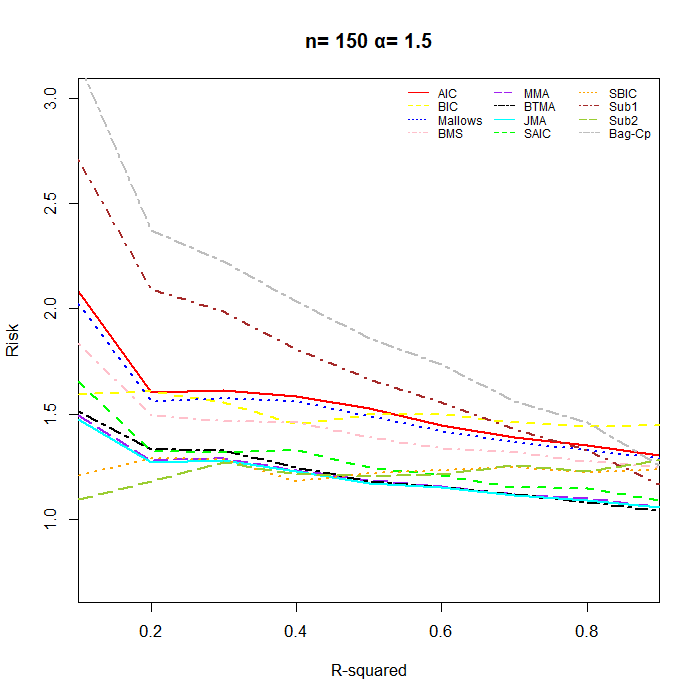}
				\end{minipage}%
				\begin{minipage}[htbp]{0.32\linewidth}
					\includegraphics[scale=0.3]{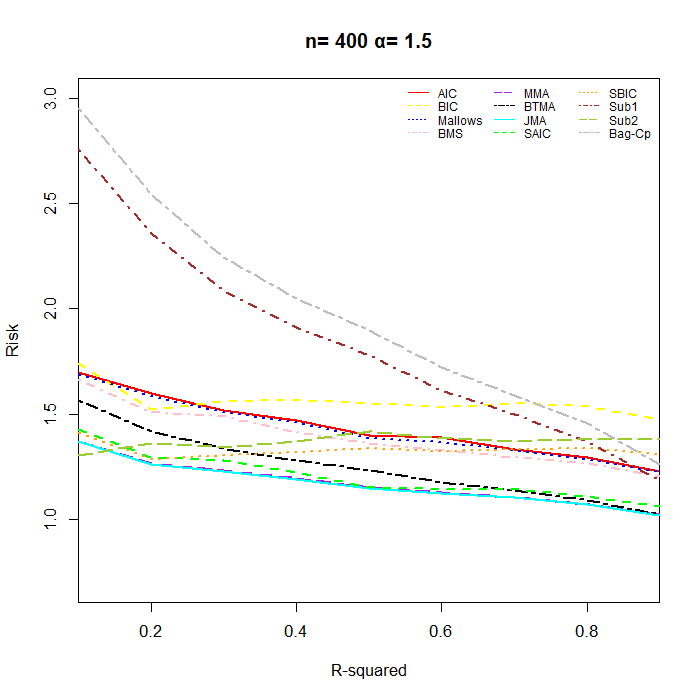}
				\end{minipage}%
			\end{center}
			\caption{Risks of various methods under different sample size $n$ with $\alpha=1.5$ \label{fig:third}}
			\vspace{1cm}
		\end{figure}%
		\begin{figure}[!ht]
			\begin{center}
				\begin{minipage}[htbp]{0.32\linewidth}
					\includegraphics[scale=0.3]{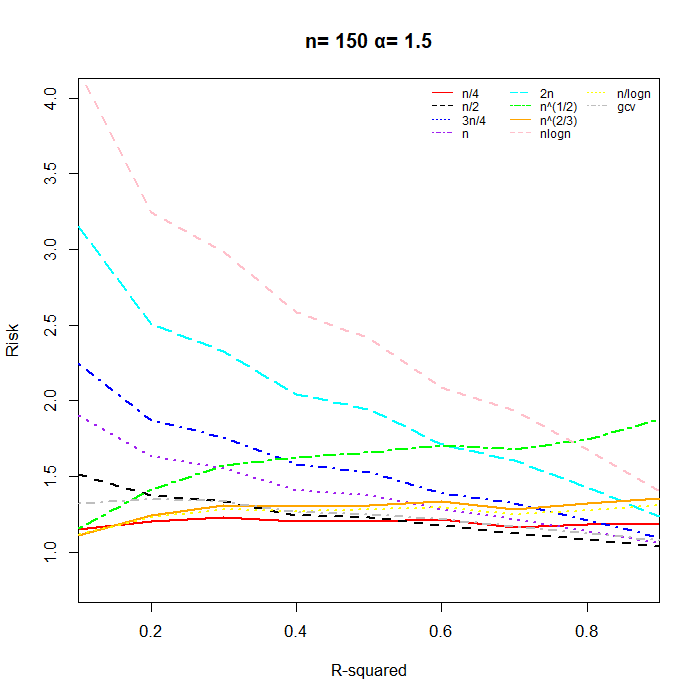}
				\end{minipage}%
				\begin{minipage}[htbp]{0.32\linewidth}
					\includegraphics[scale=0.3]{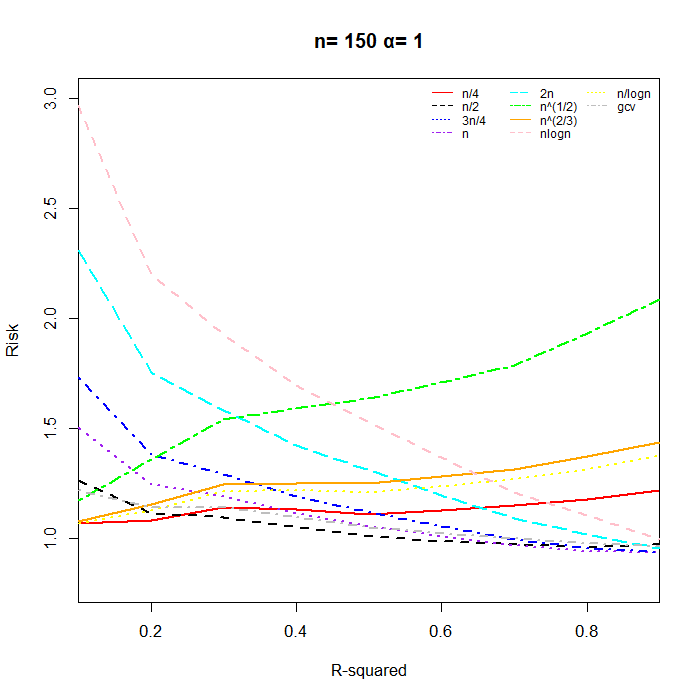}
				\end{minipage}%
				\begin{minipage}[htbp]{0.32\linewidth}
					\includegraphics[scale=0.3]{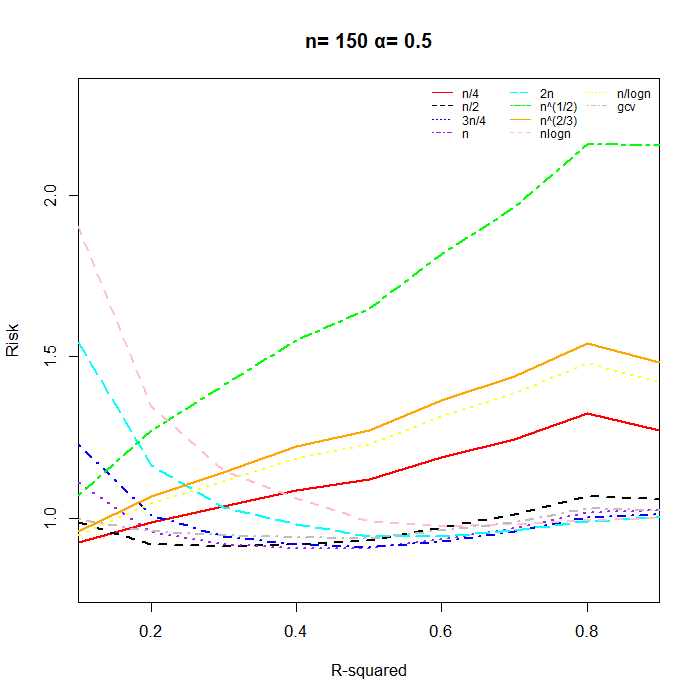}
				\end{minipage}%
			\end{center}
			\caption{Risks of various resample sizes $m$ under different values of $\alpha$ with $n=150$ \label{fig:fourth}}
			\vspace{1cm}
		\end{figure}%
		\subsection{Simulation 2 on confidence interval}
		\indent In this subsection, we compare the confidence intervals based on eight different estimators: (\romannumeral1) LSE for the just-fitted model (JUST), (\romannumeral2) LSE for the largest model (FULL), (\romannumeral3) AIC model selection (AIC), (\romannumeral4) BIC model selection (BIC), (\romannumeral5) Bootstrap model selection (BMS), (\romannumeral6) Mallows model averaging (MMA), (\romannumeral7) jackknife model averaging (JMA), and (\romannumeral8) bootstrap model averaging (BTMA). We consider a linear model with finite regressors. The data generating process is as follows:
		\begin{align}
			y_{i}=\sum_{j=1}^{k}\beta_{j}x_{ij}+e_{i}, i=1,\cdots,n, \notag
		\end{align}
		where $x_{i1}=1$ and $(x_{i2},\cdots,x_{ik})'$, $i=1,\cdots,n$ are independent normal vectors from $N(0, \Sigma_{x})$. The diagonal elements of the covariance matrix $\Sigma_{x}$ are $\rho$ and off-diagonal elements are $\rho^{2}$, where $\rho$ is set to be 0.7. The random error $e_{i}=\eta\sigma_{i}$, where $\sigma_{i}$, $i=1,\cdots,n$ are independent and generated from a standard normal distribution $N(0,1)$, and $\eta$ is varied at 0.5, 1 and 1.5. Let $\{x_{i1}, x_{i2}\}$ be the core regressors and others be the auxiliary regressors. We consider a nested candidate model set and let $k=10$. Two cases of the regression coefficients are studied
		\begin{flalign}
			& \text{\textbf{Case 1.}}\ \beta=(1, 1, c, c^{2}, c^{3}, c^{4}, 0, 0, 0, 0)', &\notag\\
			& \text{\textbf{Case 2.}}\ \beta=(1, 1, c^{4}, c^{3}, c^{2}, c, 0, 0, 0, 0)', &\notag
		\end{flalign}
		where $c=0.5$. Since we consider a nested set of candidate models, the number of under-fitted models in both cases is $M_{0}=4$. In Case 1, $\{\beta_{3}, \beta_{4}, \beta_{5},\beta_{6}\}$ is decreasing, while in Case 2, $\{\beta_{3}, \beta_{4}, \beta_{5}, \beta_{6}\}$ is increasing. We set the sample size $n=20$, 100 and 500, then we compare the confidence intervals based on different methods for parameters $\beta_{3}$ and $\beta_{4}$, where $(\beta_{3}, \beta_{4})=(c, c^{2})$ in Case 1 and $(\beta_{3}, \beta_{4})=(c^{4}, c^{3})$ in Case 2.\\
		\indent Here we briefly introduce how to construct the confidence intervals based on the above seven methods (\romannumeral1)-(\romannumeral7). The JUST estimator is the LSE for the $(M_{0}+1)$th model(the just-fitted model) and the FULL estimator is the LSE for the $M$th model(the largest model). Thus, the confidence intervals based on the JUST and FULL are constructed as
		\begin{align}
			CI_{\beta,n}^{JF}=[\hat{\beta}_{j}(q)-z_{1-\alpha/2}s(\hat{\beta}_{j}(q)),\hat{\beta}_{j}(q)+z_{1-\alpha/2}s(\hat{\beta}_{j}(q))] \notag
		\end{align}
		with $q=M_{0}+1$ and $M$, respectively, where $z_{1-\alpha/2}$ is the $1-\alpha/2$ quantile of the standard normal distribution and $s(\hat{\beta}_{j}(q))$ is the standard error of $\hat{\beta}_{j}$ calculated under the $q$th model (See \cite{zhang2019inference}).\\
		\indent To construct the confidence intervals based on AIC and BIC, we first use the AIC and BIC criteria to select the model respectively, and then proceed with the following construction,
		\begin{align}
			CI_{\beta,n}^{AB}=[\hat{\beta}_{j}(\hat{q})-z_{1-\alpha/2}s(\hat{\beta}_{j}(\hat{q})),\hat{\beta}_{j}(\hat{q})+z_{1-\alpha/2}s(\hat{\beta}_{j}(\hat{q}))], \notag
		\end{align}
		here $\hat{q}$ is the model selected by the AIC or BIC criterion and $s(\hat{\beta}_{j}(\hat{q}))$ is the standard error of $\hat{\beta}_{j}$ calculated under the $\hat{q}$th model (See also \cite{zhang2019inference}).\\
		\indent The estimator of BMS is defined as $\tilde{\beta}^{*}(q)=S_{q}'(X^{*'}_{(q)}X^{*}_{(q)})^{-1}X^{*'}_{(q)}Y^{*}$, where $X^{*}_{(q)}$ and $Y^{*}$ are the bootstrap samples defined in Section 3, for $q=1.\cdots,M$. Let $\tilde{\beta}^{*}_{j}(q)$ and $\hat{z}^{*}_{j}(\alpha)$ be the $j$th component of $\tilde{\beta}^{*}(q)$ and the $\alpha$th quantile of the bootstrap distribution of $|\sqrt{n}\big(\tilde{\beta}^{*}_{j}({q})-\hat{\beta}_{j}(q)\big)|$, respectively. Then the confidence interval based on BMS estimator is constructed as
		\begin{align}
			CI_{\beta,n}^{*}=[\tilde{\beta}^{*}_{j}(\hat{q})-n^{-1/2}\hat{z}^{*}_{j}(\alpha),\tilde{\beta}^{*}_{j}(\hat{q})+n^{-1/2}\hat{z}^{*}_{j}(\alpha)], \notag
		\end{align}
		where $\hat{q}$ is chosen by the BMS criterion (See (15) of \cite{shao1996bootstrap}).\\
		\indent \cite{zhang2019inference} proposed the simulation-based confidence intervals for MMA and JMA. We replace $\hat{\Delta}_{rt}^{(u)}$ in Algorithm 2 with $\hat{\Delta}_{\text{MMA},rt}^{(u)}=\hat{\sigma}^{2}(k_{M_{0}+r}+k_{M_{0}+t})-Z^{(u)'}\hat{V}_{\max{r,t}}Z^{(u)}$ and $\hat{\Delta}_{\text{JMA},rt}^{(u)}=\text{tr}\{\hat{Q}_{r}^{-1}\hat{\Omega}_{r}\}+\text{tr}\{\hat{Q}_{t}^{-1}\hat{\Omega}_{t}\}-Z^{(u)'}\hat{V}_{\max{r,t}}Z^{(u)}$ respectively. Then we obtain the matrices $\hat{\Delta}_{\text{MMA}}^{(u)}$ and $\hat{\Delta}_{\text{JMA}}^{(u)}$ whose $(r,t)$th elements are $\hat{\Delta}_{\text{MMA},rt}^{(u)}$ and $\hat{\Delta}_{\text{JMA},rt}^{(u)}$ respectively. Next, we compute the weight vectors $\tilde{\nu}_{\text{MMA}}^{(u)}=\text{argmin}_{\nu\in\mathbf{L}_{n}}\nu'\hat{\Delta}_{\text{MMA}}^{(u)}\nu$ and $\tilde{\nu}_{\text{JMA}}^{(u)}=\text{argmin}_{\nu\in\mathbf{L}_{n}}\nu'\hat{\Delta}_{\text{JMA}}^{(u)}\nu$, where $\mathbf{L}_{n}=\{\nu\in[0,1]^{R}:\sum_{r=1}^{R}\nu_{r}=1\}$. Finally, we can use the same algorithm as Algorithm 2 to compute the quantiles of the asymptotic distributions of MMA and JMA estimators and obtain the confidence intervals by replacing $\hat{\omega}$ with $\hat{\omega}_{\text{MMA}}$ and $\hat{\omega}_{\text{JMA}}$, respectively.\\
		\indent In all cases, we set $U=500$, $B=500$, and use GCV for resample size $m$ to construct the confidence intervals and repeat the Monte Carlo experiment 500 times. We focus on the coverage probability of a nominal 95\% confidence interval (CP) for $\beta_{3}$ and $\beta_{4}$, and also report average length (Len) in Tables 1-3.\\
		\indent From Tables 1-3, we draw the following conclusions. First, except for a few cases, the converge probability of BTMA yields a good agreement to the nominal level 95\% and this supports the conclusion of Theorem 5. Second, in most cases, AIC, BIC and BMS totally fail, indicating that their performance in coverage probability is far inferior to that of the MA methods. Third, in comparison to FULL, the MA methods have shorter average lengths of interval. The FULL and MA methods are close to the nominal values in the cases of $n=100$ and $n=500$. However, when $n=20$, the MA methods show better coverage probability than FULL. Finally, three MA methods generally have coverage probabilities close to the nominal value. Their average lengths of intervals are also similar. Overall, BTMA performs better than JMA and as well as MMA in terms of coverage probability. Especially, under three different values of $\eta$, when $n=500$, BTMA achieves much better coverage probability for $\beta_{4}$ in Case 1 compared to MMA and JMA.\\
		\newpage
		\small
		\begin{longtable}{cccccccccc}
			\caption{Simulation results in Case 1 and Case 2 for $\eta=0.5$ \label{tab:fifth}}\\
			\hline
			&  & \multicolumn{4}{c}{Case1} & \multicolumn{4}{c}{Case2}\\
			\cmidrule(r){3-6}\cmidrule(r){7-10}
			&  & \multicolumn{2}{c}{$\beta_{3}$} & \multicolumn{2}{c}{$\beta_{4}$} & \multicolumn{2}{c}{$\beta_{3}$} & \multicolumn{2}{c}{$\beta_{4}$}\\
			\cmidrule(r){3-4}\cmidrule(r){5-6}\cmidrule(r){7-8}\cmidrule(r){9-10}
			$n$ &  Method & CP & Len  & CP & Len  & CP & Len  & CP & Len\\
			\hline
			\endfirsthead
			\multicolumn{10}{c}{Table 1 (continued):\ Simulation results in Case 1 and Case 2 for $\eta=0.5$ \label{tab:fifthcontinued}}\\
			\hline
			&  & \multicolumn{4}{c}{Case1} & \multicolumn{4}{c}{Case2}\\
			\cmidrule(r){3-6}\cmidrule(r){7-10}
			&  & \multicolumn{2}{c}{$\beta_{3}$} & \multicolumn{2}{c}{$\beta_{4}$} & \multicolumn{2}{c}{$\beta_{3}$} & \multicolumn{2}{c}{$\beta_{4}$}\\
			\cmidrule(r){3-4}\cmidrule(r){5-6}\cmidrule(r){7-8}\cmidrule(r){9-10}
			$n$ &  Method & CP & Len  & CP & Len  & CP & Len  & CP & Len\\
			\hline
			\endhead
			\endfoot
			\endlastfoot
			\multirow{8}*{20} & JUST& 0.935 & 1.032 & 0.939 & 1.036 & 0.942 & 1.033 & 0.940 & 1.027\\
			& FULL& 0.918 & 1.277 & 0.926 & 1.275 & 0.937 & 1.280 & 0.932 & 1.263\\
			& AIC & 0.857 & 0.901 & 0.638 & 0.940 & 0.832 & 0.997 & 0.797 & 1.002\\
			& BIC & 0.827 & 0.838 & 0.437 & 0.861 & 0.688 & 0.930 & 0.628 & 0.943\\
			& BMS & 0.883 & 1.022 & 0.569 & 1.044 & 0.821 & 1.146 & 0.763 & 1.152\\
			& MMA & 0.931 & 1.070 & 0.953 & 1.073 & 0.924 & 1.075 & 0.936 & 1.065\\
			& JMA & 0.937 & 1.070 & 0.955 & 1.073 & 0.932 & 1.075 & 0.944 & 1.065\\
			& BTMA& 0.936 & 1.091 & 0.950 & 1.094 & 0.932 & 1.100 & 0.936 & 1.089\\
			\hline
			& JUST& 0.943 & 0.400 & 0.953 & 0.400 & 0.955 & 0.397 & 0.947 & 0.397 \\
			& FULL& 0.950 & 0.427 & 0.949 & 0.428 & 0.948 & 0.424 & 0.945 & 0.426 \\
			& AIC & 0.916 & 0.384 & 0.916 & 0.384 & 0.949 & 0.399 & 0.946 & 0.400 \\
			\multirow{2}*{100}& BIC & 0.869 & 0.371 & 0.821 & 0.373 & 0.954 & 0.397 & 0.950 & 0.397 \\
			& BMS & 0.965 & 0.492 & 0.956 & 0.488 & 0.967 & 0.469 & 0.963 & 0.471 \\
			& MMA & 0.935 & 0.402 & 0.933 & 0.401 & 0.947 & 0.399 & 0.953 & 0.400 \\
			& JMA & 0.937 & 0.402 & 0.935 & 0.401 & 0.946 & 0.399 & 0.954 & 0.400 \\
			& BTMA& 0.942 & 0.404 & 0.938 & 0.403 & 0.946 & 0.402 & 0.949 & 0.403 \\
			\hline
			\multirow{8}*{500} & JUST& 0.938 & 0.174 & 0.954 & 0.174 & 0.938 & 0.174 & 0.954 & 0.174 \\
			& FULL& 0.939 & 0.183 & 0.954 & 0.183 & 0.936 & 0.182 & 0.950 & 0.183 \\
			& AIC & 0.926 & 0.172 & 0.926 & 0.173 & 0.933 & 0.175 & 0.955 & 0.175 \\
			& BIC & 0.899 & 0.168 & 0.740 & 0.170 & 0.939 & 0.174 & 0.953 & 0.174 \\
			& BMS & 0.887 & 0.164 & 0.892 & 0.164 & 0.856 & 0.152 & 0.861 & 0.152 \\
			& MMA & 0.927 & 0.175 & 0.915 & 0.174 & 0.939 & 0.174 & 0.950 & 0.174 \\
			& JMA & 0.927 & 0.175 & 0.919 & 0.174 & 0.938 & 0.174 & 0.951 & 0.174 \\
			& BTMA& 0.930 & 0.176 & 0.934 & 0.176 & 0.934 & 0.175 & 0.946 & 0.175 \\
			\hline
		\end{longtable}
		\small
		\begin{longtable}{cccccccccccccc}
			\caption{Simulation results in Case 1 and Case 2 for $\eta=1.0$ \label{tab:sixth}}\\
			\hline
			&  & \multicolumn{4}{c}{Case1} & \multicolumn{4}{c}{Case2}\\
			\cmidrule(r){3-6}\cmidrule(r){7-10}
			&  & \multicolumn{2}{c}{$\beta_{3}$} & \multicolumn{2}{c}{$\beta_{4}$} & \multicolumn{2}{c}{$\beta_{3}$} & \multicolumn{2}{c}{$\beta_{4}$}\\
			\cmidrule(r){3-4}\cmidrule(r){5-6}\cmidrule(r){7-8}\cmidrule(r){9-10}
			$n$ &  Method & CP & Len  & CP & Len  & CP & Len  & CP & Len\\
			\hline
			\endfirsthead
			\multicolumn{10}{c}{Table 2 (continued):\ Simulation results in Case 1 and Case 2 for $\eta=1.0$ \label{tab:sixthcontinued}}\\
			\hline
			&  & \multicolumn{4}{c}{Case1} & \multicolumn{4}{c}{Case2}\\
			\cmidrule(r){3-6}\cmidrule(r){7-10}
			&  & \multicolumn{2}{c}{$\beta_{3}$} & \multicolumn{2}{c}{$\beta_{4}$} & \multicolumn{2}{c}{$\beta_{3}$} & \multicolumn{2}{c}{$\beta_{4}$}\\
			\cmidrule(r){3-4}\cmidrule(r){5-6}\cmidrule(r){7-8}\cmidrule(r){9-10}
			$n$ &  Method & CP & Len  & CP & Len  & CP & Len  & CP & Len\\
			\hline
			\endhead
			\endfoot
			\endlastfoot
			\multirow{8}*{20} & JUST& 0.927 & 2.057 & 0.915 & 2.067 & 0.923 & 2.069 & 0.925 & 2.040 \\
			& FULL& 0.933 & 2.539 & 0.924 & 2.586 & 0.912 & 2.553 & 0.920 & 2.527 \\
			& AIC & 0.696 & 1.741 & 0.431 & 1.842 & 0.577 & 1.832 & 0.493 & 1.860 \\
			& BIC & 0.536 & 1.572 & 0.215 & 1.624 & 0.329 & 1.658 & 0.228 & 1.691 \\
			& BMS & 0.699 & 1.950 & 0.348 & 1.998 & 0.553 & 2.062 & 0.434 & 2.082 \\
			& MMA & 0.957 & 2.139 & 0.955 & 2.159 & 0.945 & 2.140 & 0.967 & 2.110 \\
			& JMA & 0.967 & 2.139 & 0.968 & 2.159 & 0.945 & 2.140 & 0.964 & 2.110 \\
			& BTMA& 0.960 & 2.171 & 0.955 & 2.192 & 0.941 & 2.173 & 0.965 & 2.143 \\
			\hline
			\multirow{8}*{100} & JUST& 0.949 & 0.797 & 0.957 & 0.798 & 0.952 & 0.799 & 0.946 & 0.802 \\
			& FULL& 0.947 & 0.852 & 0.943 & 0.854 & 0.947 & 0.854 & 0.951 & 0.856 \\
			& AIC & 0.898 & 0.732 & 0.652 & 0.754 & 0.907 & 0.793 & 0.886 & 0.800 \\
			& BIC & 0.814 & 0.691 & 0.294 & 0.716 & 0.557 & 0.749 & 0.472 & 0.776 \\
			& BMS & 0.974 & 1.019 & 0.669 & 1.031 & 0.954 & 1.029 & 0.906 & 1.039 \\
			& MMA & 0.939 & 0.803 & 0.980 & 0.804 & 0.950 & 0.804 & 0.955 & 0.807 \\
			& JMA & 0.939 & 0.803 & 0.980 & 0.804 & 0.950 & 0.804 & 0.955 & 0.807 \\
			& BTMA& 0.942 & 0.805 & 0.985 & 0.805 & 0.961 & 0.807 & 0.965 & 0.810 \\
			\hline
			\multirow{8}*{500} & JUST& 0.951 & 0.348 & 0.954 & 0.348 & 0.953 & 0.348 & 0.946 & 0.348 \\
			& FULL& 0.943 & 0.366 & 0.952 & 0.366 & 0.948 & 0.366 & 0.948 & 0.366 \\
			& AIC & 0.928 & 0.337 & 0.930 & 0.338 & 0.947 & 0.350 & 0.944 & 0.350 \\
			& BIC & 0.871 & 0.323 & 0.817 & 0.327 & 0.953 & 0.348 & 0.945 & 0.348 \\
			& BMS & 0.970 & 0.410 & 0.967 & 0.411 & 0.956 & 0.400 & 0.965 & 0.401 \\
			& MMA & 0.944 & 0.349 & 0.933 & 0.348 & 0.942 & 0.348 & 0.948 & 0.349 \\
			& JMA & 0.941 & 0.349 & 0.933 & 0.348 & 0.943 & 0.348 & 0.949 & 0.349 \\
			& BTMA& 0.944 & 0.350 & 0.940 & 0.350 & 0.947 & 0.350 & 0.951 & 0.351 \\
			\hline
		\end{longtable}
		\small
		\begin{longtable}{cccccccccccccc}
			\caption{Simulation results in Case 1 and Case 2 for $\eta=1.5$ \label{tab:seventh}}\\
			\hline
			&  & \multicolumn{4}{c}{Case1} & \multicolumn{4}{c}{Case2}\\
			\cmidrule(r){3-6}\cmidrule(r){7-10}
			&  & \multicolumn{2}{c}{$\beta_{3}$} & \multicolumn{2}{c}{$\beta_{4}$} & \multicolumn{2}{c}{$\beta_{3}$} & \multicolumn{2}{c}{$\beta_{4}$}\\
			\cmidrule(r){3-4}\cmidrule(r){5-6}\cmidrule(r){7-8}\cmidrule(r){9-10}
			$n$ &  Method & CP & Len  & CP & Len  & CP & Len  & CP & Len\\
			\hline
			\endfirsthead
			\multicolumn{10}{c}{Table 3 (continued):\ Simulation results in Case 1 and Case 2 for $\eta=1.5$ \label{tab:seventhcontinued}}\\
			\hline
			&  & \multicolumn{4}{c}{Case1} & \multicolumn{4}{c}{Case2}\\
			\cmidrule(r){3-6}\cmidrule(r){7-10}
			&  & \multicolumn{2}{c}{$\beta_{3}$} & \multicolumn{2}{c}{$\beta_{4}$} & \multicolumn{2}{c}{$\beta_{3}$} & \multicolumn{2}{c}{$\beta_{4}$}\\
			\cmidrule(r){3-4}\cmidrule(r){5-6}\cmidrule(r){7-8}\cmidrule(r){9-10}
			$n$ &  Method & CP & Len  & CP & Len  & CP & Len  & CP & Len\\
			\hline
			\endhead
			\endfoot
			\endlastfoot
			& JUST& 0.944 & 3.064 & 0.944 & 3.023 & 0.919 & 3.069 & 0.925 & 3.059 \\
			& FULL& 0.921 & 3.795 & 0.933 & 3.760 & 0.923 & 3.823 & 0.905 & 3.809 \\
			& AIC & 0.576 & 2.585 & 0.386 & 2.695 & 0.469 & 2.678 & 0.387 & 2.740 \\
			\multirow{2}*{20} & BIC & 0.337 & 2.297 & 0.138 & 2.319 & 0.229 & 2.376 & 0.129 & 2.398 \\
			& BMS & 0.564 & 2.864 & 0.297 & 2.920 & 0.459 & 2.937 & 0.305 & 3.018 \\
			& MMA & 0.958 & 3.175 & 0.976 & 3.142 & 0.949 & 3.199 & 0.950 & 3.192 \\
			& JMA & 0.961 & 3.175 & 0.974 & 3.142 & 0.957 & 3.199 & 0.963 & 3.192 \\
			& BTMA& 0.962 & 3.224 & 0.969 & 3.187 & 0.949 & 3.250 & 0.950 & 3.241 \\
			\hline
			\multirow{8}*{100} & JUST& 0.934 & 1.191& 0.937 & 1.190& 0.942 & 1.198& 0.939 & 1.198 \\
			& FULL& 0.931 & 1.274 & 0.933 & 1.274 & 0.938 & 1.279 & 0.939 & 1.278 \\
			& AIC & 0.836 & 1.073 & 0.464 & 1.126 & 0.768 & 1.154 & 0.674 & 1.173 \\
			& BIC & 0.615 & 1.011 & 0.112 & 1.074 & 0.220 & 1.067 & 0.145 & 1.113 \\
			& BMS & 0.904 & 1.487 & 0.473 & 1.551 & 0.809 & 1.564 & 0.689 & 1.584 \\
			& MMA & 0.953 & 1.200 & 0.977 & 1.195 & 0.954 & 1.207 & 0.974 & 1.202 \\
			& JMA & 0.951 & 1.200 & 0.979 & 1.195 & 0.957 & 1.207 & 0.972 & 1.202 \\
			& BTMA& 0.958 & 1.202 & 0.981 & 1.197 & 0.960 & 1.210 & 0.978 & 1.204 \\
			\hline
			\multirow{8}*{500} & JUST& 0.952 & 0.521 & 0.946 & 0.522 & 0.943 & 0.521 & 0.953 & 0.521 \\
			& FULL& 0.948 & 0.548 & 0.950 & 0.548 & 0.949 & 0.548 & 0.951 & 0.548 \\
			& AIC & 0.903 & 0.494 & 0.819 & 0.502 & 0.937 & 0.525 & 0.945 & 0.525 \\
			& BIC & 0.806 & 0.464 & 0.429 & 0.485 & 0.852 & 0.513 & 0.835 & 0.517 \\
			& BMS & 0.976 & 0.651 & 0.881 & 0.659 & 0.970 & 0.633 & 0.976 & 0.630 \\
			& MMA & 0.950 & 0.523 & 0.937 & 0.523 & 0.940 & 0.522 & 0.963 & 0.522 \\
			& JMA & 0.952 & 0.523 & 0.937 & 0.523 & 0.938 & 0.522 & 0.964 & 0.522 \\
			& BTMA& 0.949 & 0.524 & 0.950 & 0.524 & 0.941 & 0.524 & 0.965 & 0.525 \\
			\hline
		\end{longtable}
		\normalsize
		\section{Empirical Analysis}
		\label{sec:empi}
		\indent In this section, we apply our proposed BTMA method to analyze the U.S. crime dataset. We also consider the Motor Trend Car data from the dataset package in R, and the related results can be found in Section S2 of the online supplementary file.\\
		\indent The U.S. crime dataset is from R package named MASS, which includes aggregate data on 47 states of the USA in 1960. It is of interest to study the relationship between the rate of crimes in a particular category per head of population ($y$) and the other 15 variables including percentage of males aged 14-24 (M and $x_{i1}$), indicator variable for a Southern state (So and $x_{i2}$), mean years of schooling (Ed and $x_{i3}$), police expenditure in 1960 (Po1 and $x_{i4}$), police expenditure in 1959 (Po2 and $x_{i5}$), labour force participation rate (LF and $x_{i6}$), number of males per 1000 females (M.F and $x_{i7}$), state population (Pop and $x_{i8}$), number of non-whites per 1000 people (NW and $x_{i9}$), unemployment rate of urban males aged 14-24 (U1 and $x_{i10}$), unemployment rate of urban males aged 35-39 (U2 and $x_{i11}$), gross domestic product per head (GDP and $x_{i12}$), income inequality (Ineq and $x_{i13}$), probability of imprisonment (Prob and $x_{i14}$) and average time served in state prisons (Time and $x_{i15}$). Figure 5 describes their relationships, where the solid red line is fitted with a linear model, and the dashed blue line represents a smoothed curve fitted with the nonparametric method which uses locally-weighted polynomial regression. In these variables, So is binary with values 0 and 1. Nearly all variables exhibit a linear relationship with $y$, except for the variables So, NW, and LF. We fit a linear regression model using all 15 variables, and find that the R-squared of this linear model is 0.8031 with $p$-value $<$ 0.05. This shows that the interpretability of this model is already good. Excluding some outliers and influence points, and fitting the linear model again, the R-squared increases to 0.9173. Thus, we consider the linear models, and the largest model is given by
		\begin{align}
			y_{i}=\theta_{0}+\theta_{1}x_{i1}+\theta_{2}x_{i2}+\cdots+\theta_{15}x_{i15}+e_{i},\ i=1,\cdots,n, \notag
		\end{align}
		where $y_{i}$ is the rate of crimes in the $i$th state, $x_{ij}$ with $j=1,\cdots,15$ represent all 15 covariates, $\theta_{0}$ is the intercept, $\theta_{1},\theta_{2},\cdots,\theta_{15}$ are the regression coefficients, and $e_{i}$ is the error term. We first prioritize the variables using the Mallows model selection criterion and then consider a sequence of nested candidate models.\\
		\indent We divide the data set into training set and test set. Let $n\in\{33, 36, 39, 42\}$ be the training sample size. We standardize the data and calculate the mean squared prediction errors (MSPEs) of eight methods: MMA, JMA, BTMA, S-AIC, S-BIC, BMS, subsampling (Sub) and bagging (Bag). We choose $m=[0.632n]$ as the resample size of subsampling because the resample size of $m=[n^{2/3}]$ cannot satisfy the full rank condition in some training sets.\\
		\indent The MSPE, based on the prediction for the test sample, is given by
		\begin{align}
			\dfrac{1}{N-n}\sum\limits_{i=1}^{N-n}(y_{i}-\hat{y_{i}})^{2}, \notag
		\end{align}
		where $N=47$ is the full sample size, $N-n$ is the test sample size, and $\hat{y_{i}}$ is the forecast of $y_{i}$. We replicate our calculations 1000 times to obtain the mean and variance of MSPEs, which are displayed in Table 4.\\
		\begin{table}
			\caption{Mean and variance of MSPEs for U.S. crime dataset \label{tab:eighth}}
			\vspace{0em}
			\begin{center}
				\footnotesize
				\begin{tabular}{cccccccccc}
					\hline
					& Training set & MMA & JMA & BTMA & S-AIC & S-BIC & BMS & Sub &Bag\\
					\hline
					Mean  & $n=33$  & 0.4224 & 0.4112 & \textbf{0.3898} & 0.4479 & 0.4092 & 0.4216 & 0.4074 & 0.5748\\
					& $n=36$ & 0.3967 & 0.3872 & \textbf{0.3726} & 0.4059 & 0.3833 & 0.3977 & 0.3872 & 0.5095 \\
					& $n=39$ & 0.4030 & 0.3916 & \textbf{0.3838} & 0.4027 & 0.3888 & 0.4086 & 0.3981 & 0.4977 \\
					& $n=42$ & 0.3830 & 0.3729 & \textbf{0.3723} & 0.3768 & 0.3733 & 0.3962 & 0.3830 & 0.4608 	\\
					\hline
					Variance & $n=33$ & 0.0295 & 0.0285 & \textbf{0.0256} & 0.0425 & 0.0283 & 0.0264 & 0.0260 & 0.0559 \\
					& $n=36$ & 0.0360 & 0.0355 & 0.0327 & 0.0438 & 0.0342 & \textbf{0.0319} & 0.0332 & 0.0522 \\
					& $n=39$ & 0.0513 & 0.0484 & 0.0493 & 0.0486 & 0.0493 & 0.0503 & \textbf{0.0471} & 0.0664 \\
					& $n=42$ & 0.0729 & 0.0705 & 0.0740 & \textbf{0.0638} & 0.0718 & 0.0822 & 0.0676 & 0.0843 \\
					\hline
				\end{tabular}
			\end{center}
		\end{table}
		\indent From Table 4, it is readily seen that BTMA always has the smallest MSPE, followed by JMA and S-BIC. The MSPEs of all methods decrease with the increase of the training sample size $n$. In addition, BTMA has the smallest variance of MSPE when $n$ is relatively small ($n=33$ and 36). Its variance becomes larger as the test sample size $47-n$ decreases. This is reasonable because when $n=42$, the test set has only 5 observations.
		\begin{figure}[!htbp]
			\begin{center}
				\includegraphics[scale=0.8]{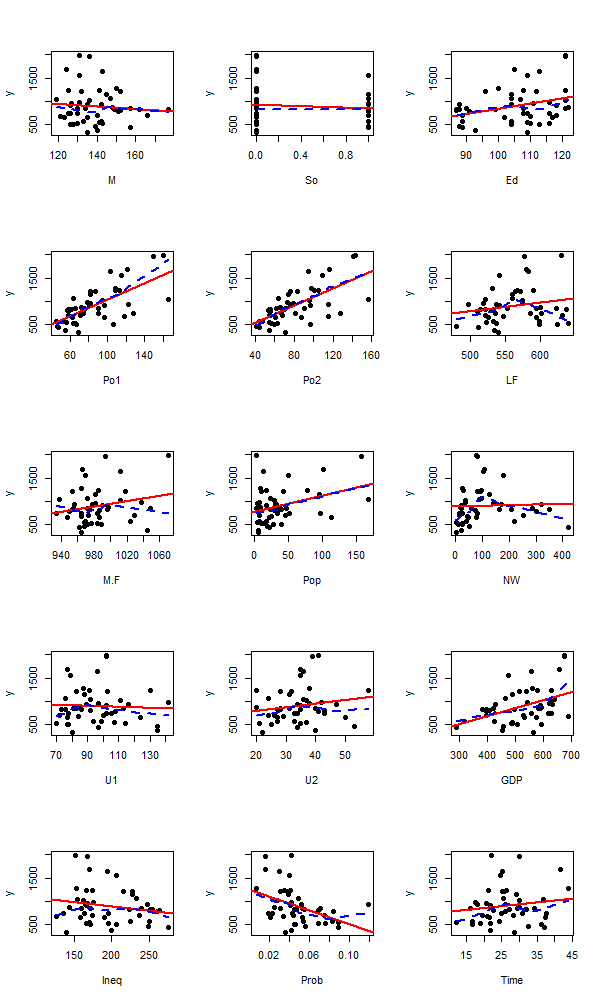}
			\end{center}
			\vspace{-1em}
			\caption{Scatterplot between rate of crimes $y$ and other variables for U.S. crime dataset \label{fig:fifth}}
		\end{figure}%
		\section{Conclusion}
		\label{sec:conc}
		\indent This paper proposed a bootstrap criterion for constructing the model averaging estimator, which encompasses the BMS as a special case. An algorithm is developed to obtain the corresponding BTMA estimator. The asymptotic optimality of the BTMA estimator and the consistency of the bootstrap weight vector are established. The BTMA estimator is proved to have a nonstandard asymptotic distribution when the true model is a linear regression model, and this distribution is utilized to construct confidence intervals. The numerical results presented in the paper demonstrate the superiority of the proposed method, particularly when compared to closely related methods BMS and JMA.\\
		\indent There are still some issues with BTMA that should be addressed in future research. For example, how can we theoretically determine the optimal resample size $m$? What is the corresponding criterion for BTMA in the case of heteroskedastic errors? What are the differences between bootstrapping residuals and bootstrapping pairs in BTMA? It is also interesting to extend BTMA to generalized linear models and nonparametric regression models.
		\section*{Supplementary Material}
		\label{sec:supp}
		\indent The online supplemental file contains the simulation study based on the setting of \cite{shao1996bootstrap}, the empirical analysis of Motor Trend Car data, Lemmas 1-3 and their proofs, and the proofs of Theorems 1-5.
		\setcounter{equation}{1}
		\renewcommand\theequation{A.\arabic{equation}}
		\if1\blind
		{
			\section*{Acknowledgments}
			\label{sec:ackn}
			\addcontentsline{toc}{section}{Acknowledgments}
			Guohua Zou's work was partially supported by the National Natural Science Foundation of China (Grant nos. 11971323 and 12031016) and the Beijing Natural Science Foundation (Grant No. Z210003).
		} \fi
		
		\if0\blind
		{
			
		} \fi
		
		
		
		\bibliographystyle{elsarticle-harv} 
		\bibliography{reference}

\begin{thebibliography}{41}
\expandafter\ifx\csname natexlab\endcsname\relax\def\natexlab#1{#1}\fi
\providecommand{\url}[1]{\texttt{#1}}
\providecommand{\href}[2]{#2}
\providecommand{\path}[1]{#1}
\providecommand{\DOIprefix}{doi:}
\providecommand{\ArXivprefix}{arXiv:}
\providecommand{\URLprefix}{URL: }
\providecommand{\Pubmedprefix}{pmid:}
\providecommand{\doi}[1]{\href{http://dx.doi.org/#1}{\path{#1}}}
\providecommand{\Pubmed}[1]{\href{pmid:#1}{\path{#1}}}
\providecommand{\bibinfo}[2]{#2}
\ifx\xfnm\relax \def\xfnm[#1]{\unskip,\space#1}\fi
\bibitem[{Ando and Li(2014)}]{ando2014model}
\bibinfo{author}{Ando, T.}, \bibinfo{author}{Li, K.C.}, \bibinfo{year}{2014}.
\newblock \bibinfo{title}{A model-averaging approach for high-dimensional
  regression}.
\newblock \bibinfo{journal}{Journal of the American Statistical Association}
  \bibinfo{volume}{109}, \bibinfo{pages}{254--265}.
\bibitem[{Andrews(1991)}]{andrews1991asymptotic}
\bibinfo{author}{Andrews, D.W.}, \bibinfo{year}{1991}.
\newblock \bibinfo{title}{Asymptotic optimality of generalized
  \uppercase{$C_{L}$}, cross-validation, and generalized cross-validation in
  regression with heteroskedastic errors}.
\newblock \bibinfo{journal}{Journal of Econometrics} \bibinfo{volume}{47},
  \bibinfo{pages}{359--377}.
\bibitem[{Bickel and Freedman(1981)}]{bickel1981some}
\bibinfo{author}{Bickel, P.J.}, \bibinfo{author}{Freedman, D.A.},
  \bibinfo{year}{1981}.
\newblock \bibinfo{title}{Some asymptotic theory for the bootstrap}.
\newblock \bibinfo{journal}{The Annals of Statistics} \bibinfo{volume}{9},
  \bibinfo{pages}{1196--1217}.
\bibitem[{Bickel and Sakov(2008)}]{bickel2008choice}
\bibinfo{author}{Bickel, P.J.}, \bibinfo{author}{Sakov, A.},
  \bibinfo{year}{2008}.
\newblock \bibinfo{title}{On the choice of m in the m out of n bootstrap and
  confidence bounds for extrema}.
\newblock \bibinfo{journal}{Statistica Sinica} , \bibinfo{pages}{967--985}.
\bibitem[{Buckland et~al.(1997)Buckland, Burnham and
  Augustin}]{buckland1997model}
\bibinfo{author}{Buckland, S.T.}, \bibinfo{author}{Burnham, K.},
  \bibinfo{author}{Augustin, N.}, \bibinfo{year}{1997}.
\newblock \bibinfo{title}{Model selection: an integral part of inference.}
\newblock \bibinfo{journal}{Biometrics} \bibinfo{volume}{53},
  \bibinfo{pages}{603--618}.
\bibitem[{Bunke and Droge(1984)}]{bunke1984bootstrap}
\bibinfo{author}{Bunke, O.}, \bibinfo{author}{Droge, B.}, \bibinfo{year}{1984}.
\newblock \bibinfo{title}{Bootstrap and cross-validation estimates of the
  prediction error for linear regression models}.
\newblock \bibinfo{journal}{The Annals of Statistics} \bibinfo{volume}{12},
  \bibinfo{pages}{1400--1424}.
\bibitem[{Charkhi and Claeskens(2018)}]{charkhi2018asymptotic}
\bibinfo{author}{Charkhi, A.}, \bibinfo{author}{Claeskens, G.},
  \bibinfo{year}{2018}.
\newblock \bibinfo{title}{Asymptotic post-selection inference for the
  \uppercase{A}kaike information criterion}.
\newblock \bibinfo{journal}{Biometrika} \bibinfo{volume}{105},
  \bibinfo{pages}{645--664}.
\bibitem[{Chen et~al.(2020)Chen, Craiu and Sun}]{chen2020bayesian}
\bibinfo{author}{Chen, B.}, \bibinfo{author}{Craiu, R.V.},
  \bibinfo{author}{Sun, L.}, \bibinfo{year}{2020}.
\newblock \bibinfo{title}{Bayesian model averaging for the
  \uppercase{X}-chromosome inactivation dilemma in genetic association study}.
\newblock \bibinfo{journal}{Biostatistics} \bibinfo{volume}{21},
  \bibinfo{pages}{319--335}.
\bibitem[{Claeskens and Carroll(2007)}]{claeskens2007asymptotic}
\bibinfo{author}{Claeskens, G.}, \bibinfo{author}{Carroll, R.J.},
  \bibinfo{year}{2007}.
\newblock \bibinfo{title}{An asymptotic theory for model selection inference in
  general semiparametric problems}.
\newblock \bibinfo{journal}{Biometrika} \bibinfo{volume}{94},
  \bibinfo{pages}{249--265}.
\bibitem[{Davison et~al.(2003)Davison, Hinkley and Young}]{davison2003recent}
\bibinfo{author}{Davison, A.C.}, \bibinfo{author}{Hinkley, D.V.},
  \bibinfo{author}{Young, G.A.}, \bibinfo{year}{2003}.
\newblock \bibinfo{title}{Recent developments in bootstrap methodology}.
\newblock \bibinfo{journal}{Statistical Science} \bibinfo{volume}{18},
  \bibinfo{pages}{141--157}.
\bibitem[{De~Bin et~al.(2016)De~Bin, Janitza, Sauerbrei and
  Boulesteix}]{de2016subsampling}
\bibinfo{author}{De~Bin, R.}, \bibinfo{author}{Janitza, S.},
  \bibinfo{author}{Sauerbrei, W.}, \bibinfo{author}{Boulesteix, A.L.},
  \bibinfo{year}{2016}.
\newblock \bibinfo{title}{Subsampling versus bootstrapping in resampling-based
  model selection for multivariable regression}.
\newblock \bibinfo{journal}{Biometrics} \bibinfo{volume}{72},
  \bibinfo{pages}{272--280}.
\bibitem[{Efron(1979)}]{efron1979bootstrap}
\bibinfo{author}{Efron, B.}, \bibinfo{year}{1979}.
\newblock \bibinfo{title}{Bootstrap methods: another look at the jackknife}.
\newblock \bibinfo{journal}{The Annals of Statistics} \bibinfo{volume}{7},
  \bibinfo{pages}{1--26}.
\bibitem[{Efron(1982)}]{efron1982jackknife}
\bibinfo{author}{Efron, B.}, \bibinfo{year}{1982}.
\newblock \bibinfo{title}{The Jackknife, the Bootstrap and Other Resampling
  Plans}.
\newblock \bibinfo{publisher}{Society for Industrial and Applied Mathematics,
  Philadelphia}.
\bibitem[{Efron(1983)}]{efron1983estimating}
\bibinfo{author}{Efron, B.}, \bibinfo{year}{1983}.
\newblock \bibinfo{title}{Estimating the error rate of a prediction rule:
  improvement on cross-validation}.
\newblock \bibinfo{journal}{Journal of the American Statistical Association}
  \bibinfo{volume}{78}, \bibinfo{pages}{316--331}.
\bibitem[{Efron(2014)}]{efron2014estimation}
\bibinfo{author}{Efron, B.}, \bibinfo{year}{2014}.
\newblock \bibinfo{title}{Estimation and accuracy after model selection}.
\newblock \bibinfo{journal}{Journal of the American Statistical Association}
  \bibinfo{volume}{109}, \bibinfo{pages}{991--1007}.
\bibitem[{Freedman(1981)}]{freedman1981bootstrapping}
\bibinfo{author}{Freedman, D.A.}, \bibinfo{year}{1981}.
\newblock \bibinfo{title}{Bootstrapping regression models}.
\newblock \bibinfo{journal}{The Annals of Statistics} \bibinfo{volume}{9},
  \bibinfo{pages}{1218--1228}.
\bibitem[{Gao et~al.(2016)Gao, Zhang, Wang and Zou}]{gao2016model}
\bibinfo{author}{Gao, Y.}, \bibinfo{author}{Zhang, X.}, \bibinfo{author}{Wang,
  S.}, \bibinfo{author}{Zou, G.}, \bibinfo{year}{2016}.
\newblock \bibinfo{title}{Model averaging based on leave-subject-out
  cross-validation}.
\newblock \bibinfo{journal}{Journal of Econometrics} \bibinfo{volume}{192},
  \bibinfo{pages}{139--151}.
\bibitem[{Hansen(2007)}]{hansen2007least}
\bibinfo{author}{Hansen, B.E.}, \bibinfo{year}{2007}.
\newblock \bibinfo{title}{Least squares model averaging}.
\newblock \bibinfo{journal}{Econometrica} \bibinfo{volume}{75},
  \bibinfo{pages}{1175--1189}.
\bibitem[{Hansen(2008)}]{hansen2008least}
\bibinfo{author}{Hansen, B.E.}, \bibinfo{year}{2008}.
\newblock \bibinfo{title}{Least-squares forecast averaging}.
\newblock \bibinfo{journal}{Journal of Econometrics} \bibinfo{volume}{146},
  \bibinfo{pages}{342--350}.
\bibitem[{Hansen and Racine(2012)}]{hansen2012jackknife}
\bibinfo{author}{Hansen, B.E.}, \bibinfo{author}{Racine, J.S.},
  \bibinfo{year}{2012}.
\newblock \bibinfo{title}{Jackknife model averaging}.
\newblock \bibinfo{journal}{Journal of Econometrics} \bibinfo{volume}{167},
  \bibinfo{pages}{38--46}.
\bibitem[{Hjort and Claeskens(2003)}]{hjort2003frequentist}
\bibinfo{author}{Hjort, N.L.}, \bibinfo{author}{Claeskens, G.},
  \bibinfo{year}{2003}.
\newblock \bibinfo{title}{Frequentist model average estimators}.
\newblock \bibinfo{journal}{Journal of the American Statistical Association}
  \bibinfo{volume}{98}, \bibinfo{pages}{879--899}.
\bibitem[{Hoeting et~al.(1999)Hoeting, Madigan, Raftery and
  Volinsky}]{hoeting1999bayesian}
\bibinfo{author}{Hoeting, J.A.}, \bibinfo{author}{Madigan, D.},
  \bibinfo{author}{Raftery, A.E.}, \bibinfo{author}{Volinsky, C.T.},
  \bibinfo{year}{1999}.
\newblock \bibinfo{title}{Bayesian model averaging: A tutorial}.
\newblock \bibinfo{journal}{Statistical Science} \bibinfo{volume}{14},
  \bibinfo{pages}{382--417}.
\bibitem[{Li(1987)}]{li1987asymptotic}
\bibinfo{author}{Li, K.C.}, \bibinfo{year}{1987}.
\newblock \bibinfo{title}{Asymptotic optimality for \uppercase{$C_{p}$},
  \uppercase{$C_{L}$}, cross-validation and generalized cross-validation:
  discrete index set}.
\newblock \bibinfo{journal}{The Annals of Statistics} \bibinfo{volume}{15},
  \bibinfo{pages}{958--975}.
\bibitem[{Liang et~al.(2011)Liang, Zou, Wan and Zhang}]{liang2011optimal}
\bibinfo{author}{Liang, H.}, \bibinfo{author}{Zou, G.}, \bibinfo{author}{Wan,
  A.T.}, \bibinfo{author}{Zhang, X.}, \bibinfo{year}{2011}.
\newblock \bibinfo{title}{Optimal weight choice for frequentist model average
  estimators}.
\newblock \bibinfo{journal}{Journal of the American Statistical Association}
  \bibinfo{volume}{106}, \bibinfo{pages}{1053--1066}.
\bibitem[{Liao et~al.(2019)Liao, Zong, Zhang and Zou}]{liao2019model}
\bibinfo{author}{Liao, J.}, \bibinfo{author}{Zong, X.}, \bibinfo{author}{Zhang,
  X.}, \bibinfo{author}{Zou, G.}, \bibinfo{year}{2019}.
\newblock \bibinfo{title}{Model averaging based on leave-subject-out
  cross-validation for vector autoregressions}.
\newblock \bibinfo{journal}{Journal of Econometrics} \bibinfo{volume}{209},
  \bibinfo{pages}{35--60}.
\bibitem[{Liao and Zou(2020)}]{liao2020corrected}
\bibinfo{author}{Liao, J.}, \bibinfo{author}{Zou, G.}, \bibinfo{year}{2020}.
\newblock \bibinfo{title}{Corrected mallows criterion for model averaging}.
\newblock \bibinfo{journal}{Computational Statistics and Data Analysis}
  \bibinfo{volume}{144}, \bibinfo{pages}{106902}.
\bibitem[{Liu(2015)}]{liu2015distribution}
\bibinfo{author}{Liu, C.A.}, \bibinfo{year}{2015}.
\newblock \bibinfo{title}{Distribution theory of the least squares averaging
  estimator}.
\newblock \bibinfo{journal}{Journal of Econometrics} \bibinfo{volume}{186},
  \bibinfo{pages}{142--159}.
\bibitem[{Proietti and Giovannelli(2021)}]{proietti2021nowcasting}
\bibinfo{author}{Proietti, T.}, \bibinfo{author}{Giovannelli, A.},
  \bibinfo{year}{2021}.
\newblock \bibinfo{title}{Nowcasting monthly \uppercase{GDP} with big data: A
  model averaging approach}.
\newblock \bibinfo{journal}{Journal of the Royal Statistical Society Series A}
  \bibinfo{volume}{184}, \bibinfo{pages}{683--706}.
\bibitem[{Rao and Tibshirani(1997)}]{rao1997out}
\bibinfo{author}{Rao, J.S.}, \bibinfo{author}{Tibshirani, R.},
  \bibinfo{year}{1997}.
\newblock \bibinfo{title}{The out-of-bootstrap method for model averaging and
  selection}.
\newblock \bibinfo{type}{Technical Report}. University of Toronto, Department
  of Statistics.
\bibitem[{Shao(1993)}]{shao1993linear}
\bibinfo{author}{Shao, J.}, \bibinfo{year}{1993}.
\newblock \bibinfo{title}{Linear model selection by cross-validation}.
\newblock \bibinfo{journal}{Journal of the American Statistical Association}
  \bibinfo{volume}{88}, \bibinfo{pages}{486--494}.
\bibitem[{Shao(1996)}]{shao1996bootstrap}
\bibinfo{author}{Shao, J.}, \bibinfo{year}{1996}.
\newblock \bibinfo{title}{Bootstrap model selection}.
\newblock \bibinfo{journal}{Journal of the American Statistical Association}
  \bibinfo{volume}{91}, \bibinfo{pages}{655--665}.
\bibitem[{Shao and Tu(2012)}]{shao2012jackknife}
\bibinfo{author}{Shao, J.}, \bibinfo{author}{Tu, D.}, \bibinfo{year}{2012}.
\newblock \bibinfo{title}{The Jackknife and Bootstrap}.
\newblock \bibinfo{publisher}{Springer, New York}.
\bibitem[{Shibata(1981)}]{shibata1981optimal}
\bibinfo{author}{Shibata, R.}, \bibinfo{year}{1981}.
\newblock \bibinfo{title}{An optimal selection of regression variables}.
\newblock \bibinfo{journal}{Biometrika} \bibinfo{volume}{68},
  \bibinfo{pages}{45--54}.
\bibitem[{Wan et~al.(2010)Wan, Zhang and Zou}]{wan2010least}
\bibinfo{author}{Wan, A.T.K.}, \bibinfo{author}{Zhang, X.},
  \bibinfo{author}{Zou, G.}, \bibinfo{year}{2010}.
\newblock \bibinfo{title}{Least squares model averaging by \uppercase{M}allows
  criterion}.
\newblock \bibinfo{journal}{Journal of Econometrics} \bibinfo{volume}{156},
  \bibinfo{pages}{277--283}.
\bibitem[{Wu(1986)}]{wu1986jackknife}
\bibinfo{author}{Wu, C.J.}, \bibinfo{year}{1986}.
\newblock \bibinfo{title}{Jackknife, bootstrap and other resampling methods in
  regression analysis}.
\newblock \bibinfo{journal}{The Annals of Statistics} \bibinfo{volume}{14},
  \bibinfo{pages}{1261--1295}.
\bibitem[{Yu et~al.(2024)Yu, Lian, Sun, Zhang and Hong}]{yu2024post}
\bibinfo{author}{Yu, D.}, \bibinfo{author}{Lian, H.}, \bibinfo{author}{Sun,
  Y.}, \bibinfo{author}{Zhang, X.}, \bibinfo{author}{Hong, Y.},
  \bibinfo{year}{2024}.
\newblock \bibinfo{title}{Post-averaging inference for optimal model averaging
  estimator in generalized linear models}.
\newblock \bibinfo{journal}{Econometric Reviews} \bibinfo{volume}{43},
  \bibinfo{pages}{98--122}.
\bibitem[{Zhang and Liu(2019)}]{zhang2019inference}
\bibinfo{author}{Zhang, X.}, \bibinfo{author}{Liu, C.A.}, \bibinfo{year}{2019}.
\newblock \bibinfo{title}{Inference after model averaging in linear regression
  models}.
\newblock \bibinfo{journal}{Econometric Theory} \bibinfo{volume}{35},
  \bibinfo{pages}{816--841}.
\bibitem[{Zhang et~al.(2013)Zhang, Wan and Zou}]{zhang2013model}
\bibinfo{author}{Zhang, X.}, \bibinfo{author}{Wan, A.T.}, \bibinfo{author}{Zou,
  G.}, \bibinfo{year}{2013}.
\newblock \bibinfo{title}{Model averaging by jackknife criterion in models with
  dependent data}.
\newblock \bibinfo{journal}{Journal of Econometrics} \bibinfo{volume}{174},
  \bibinfo{pages}{82--94}.
\bibitem[{Zhang et~al.(2020)Zhang, Zou, Liang and
  Carroll}]{zhang2020parsimonious}
\bibinfo{author}{Zhang, X.}, \bibinfo{author}{Zou, G.}, \bibinfo{author}{Liang,
  H.}, \bibinfo{author}{Carroll, R.J.}, \bibinfo{year}{2020}.
\newblock \bibinfo{title}{Parsimonious model averaging with a diverging number
  of parameters}.
\newblock \bibinfo{journal}{Journal of the American Statistical Association}
  \bibinfo{volume}{115}, \bibinfo{pages}{972--984}.
\bibitem[{Zhao and Zou(2020)}]{zhao2020average}
\bibinfo{author}{Zhao, Z.}, \bibinfo{author}{Zou, G.}, \bibinfo{year}{2020}.
\newblock \bibinfo{title}{Average estimation of semiparametric models for
  high-dimensional longitudinal data}.
\newblock \bibinfo{journal}{Journal of Systems Science and Complexity}
  \bibinfo{volume}{33}, \bibinfo{pages}{2013--2047}.
\bibitem[{Zhu and Zou(2018)}]{zhu2018asymptotic}
\bibinfo{author}{Zhu, R.}, \bibinfo{author}{Zou, G.}, \bibinfo{year}{2018}.
\newblock \bibinfo{title}{The asymptotic theory for model averaging in general
  semiparametric models}.
\newblock \bibinfo{journal}{Science China Mathematics} \bibinfo{volume}{48},
  \bibinfo{pages}{1019--1052}.

\end{thebibliography}


\begin{thebibliography}{12}
\expandafter\ifx\csname natexlab\endcsname\relax\def\natexlab#1{#1}\fi
\expandafter\ifx\csname url\endcsname\relax
  \def\url#1{\texttt{#1}}\fi
\expandafter\ifx\csname urlprefix\endcsname\relax\def\urlprefix{URL }\fi

\bibitem[{Chen et~al.(2018)Chen, Li, Linton, and Lu}]{chen2018semiparametric}
Chen, J., Li, D., Linton, O., Lu, Z., 2018. Semiparametric ultra-high
  dimensional model averaging of nonlinear dynamic time series. Journal of the
  American Statistical Association 113~(522), 919--932.

\bibitem[{Fan and Peng(2004)}]{fan2004nonconcave}
Fan, J., Peng, H., 2004. Nonconcave penalized likelihood with a diverging
  number of parameters. The Annals of Statistics 32~(3), 928--961.

\bibitem[{Gunst and Mason(1980)}]{gunst2018regression}
Gunst, R.~F., Mason, R.~L., 1980. Regression Analysis and Its Applications.
  Marcel Dekker, New York.

\bibitem[{Hansen(2007)}]{hansen2007least}
Hansen, B.~E., 2007. Least squares model averaging. Econometrica 75~(4),
  1175--1189.

\bibitem[{Kim and Pollard(1990)}]{kim1990cube}
Kim, J., Pollard, D., 1990. Cube root asymptotics. The Annals of Statistics
  18~(1), 191--219.

\bibitem[{Liu(2015)}]{liu2015distribution}
Liu, C.-A., 2015. Distribution theory of the least squares averaging estimator.
  Journal of Econometrics 186~(1), 142--159.

\bibitem[{Shao(1996)}]{shao1996bootstrap}
Shao, J., 1996. Bootstrap model selection. Journal of the American Statistical
  Association 91~(434), 655--665.

\bibitem[{Van Der~Vaart and Wellner(1996)}]{van1996weak}
Van Der~Vaart, A.~W., Wellner, J.~A., 1996. Weak Convergence. Springer, New
  York.

\bibitem[{Wan et~al.(2010)Wan, Zhang, and Zou}]{wan2010least}
Wan, A.~T., Zhang, X., Zou, G., 2010. Least squares model averaging by
  \uppercase{M}allows criterion. Journal of Econometrics 156~(2), 277--283.

\bibitem[{Whittle(1960)}]{whittle1960bounds}
Whittle, P., 1960. Bounds for the moments of linear and quadratic forms in
  independent variables. Theory of Probability and Its Applications 5~(3),
  302--305.

\bibitem[{Yu et~al.(2024)Yu, Lian, Sun, Zhang, and Hong}]{yu2024post}
Yu, D., Lian, H., Sun, Y., Zhang, X., Hong, Y., 2024. Post-averaging inference
  for optimal model averaging estimator in generalized linear models.
  Econometric Reviews 43~(2-4), 98--122.

\bibitem[{Zhang and Liu(2019)}]{zhang2019inference}
Zhang, X., Liu, C.-A., 2019. Inference after model averaging in linear
  regression models. Econometric Theory 35~(4), 816--841.

\end{thebibliography}
			

	\end{document}


\begin{frontmatter}
		
		
		
		\title{\textbf{Supplementary Material for: Bootstrap Model Averaging}}
		
		
		\author[inst1]{Minghui Song} 
		
		\affiliation[inst1]{organization={School of Mathematical Sciences},
			addressline={Capital Normal University}, 
			city={Beijing},
			postcode={100048}, 
			country={China}}
		
		\author[inst1]{Guohua Zou\corref{cor1}}
		\ead{ghzou@amss.ac.cn}
		\cortext[cor1]{Corresponding author}
		
		\author[inst2]{Alan T.K. Wan}
		
		\affiliation[inst2]{organization={Department of Management Sciences},
			addressline={City University of Hong Kong}, 
			city={Kowloon},
			country={Hong Kong}}
		
		\makeatletter
		\renewenvironment{abstract}{\global\setbox\absbox=\vbox\bgroup
			\hsize=\textwidth\def\baselinestretch{1}%
			\noindent\unskip\textbf{Contents}
			\par\medskip\noindent\unskip}
		{\egroup}
		\begin{abstract}
			\noindent\hspace{-0.75em} The online supplemental file contains the simulation study based on the setting of \cite{shao1996bootstrap}, the empirical analysis of Motor Trend Car data, Lemmas 1-3 and their proofs, and the proofs of Theorems 1-5.
		\end{abstract}
		
	\end{frontmatter}




\section{Simulation based on the setting of Shao (1996)}
\indent This simulation has the same setting to that in \cite{shao1996bootstrap}. We consider a linear model
\begin{align}
	y_{i}=x'_{i}\theta+e_{i}=\sum_{j=1}^{5}\theta_{j}x_{ij}+e_{i}, i=1,\cdots,n, \notag
\end{align}
where $x_{i1}=1$ and $x_{ij}$, $j=2,\cdots,5$ are taken from the solid waste data of \cite{gunst2018regression}, $\theta_{j}$, $j=1,\cdots,5$ are the corresponding regression coefficients, and the independent errors $e_{i}$, $i=1,\cdots,n$ are generated from $N(0,1)$. In Tables S1 and S2, the first column presents the covariate setting for the true model, and the second column lists the set of candidate models. For example, Model \{1,2,4\} means that the candidate model contains the first, second and fourth covariates. We use the symbol star to denote the true model. For instance, if $\theta=(\theta_{1},\theta_{2},\theta_{3},\theta_{4},\theta_{5})=(2,0,0,4,0)$ is the true vector of regression coefficients, then Model \{1,4\} is displayed by star. The candidate model set contains the true model in the first four settings but not  in the last two settings. The settings of Tables S3 and S4 are the same as those in Tables S1 and S2. We take $B = 100$, and repeat our simulations 1000 times.\\
\indent Table S1 reports the empirical probabilities of selecting each model using BMS with different resample size $m$. This reproduction result is consistent with the simulation result in \cite{shao1996bootstrap}. Table S2 records the weights obtained by BTMA, which are the means of the weights based on 1000 simulations. From Table S2, it is seen that for the first three settings, the biggest weight is assigned to the true model, which remains true for the fourth setting if the resample size $m$ is large enough. This makes sense as the true model is included for these settings. When the true model is not included in the set of candidate models, the last two settings exhibit different results about the weights. The fifth setting indicates that, apart from the first covariate which is included in all models, if the remaining covariates in a candidate model are all incorrect, then weights are not assigned to that model. The last setting shows that when all models are under-fitted, the weights tend to be assigned to those models with more covariates.\\
\indent Table S3 shows that BMS performs slightly better than BTMA in the third and fourth settings for most resample sizes. This is not unexpected as these two settings imply that the true model is included in  the candidate model set. However, we also observe that for the first and second settings, BTMA has the smaller risk than BMS. This indicates that the averaging estimator can be better than the selecting estimator even if the candidate model set contains the true model. In the fifth and sixth settings where the candidate model set does not contain the true model, the averaging estimator has much smaller risk than the selecting estimator. Table S3 also implies that we can slightly improve BTMA by selecting the appropriate resample size $m$. The risk of BTMA in Table S4 is the average of the risks based on all resample sizes in Table S3. It is observed from Table S4 that BTMA has better performance than JMA in most cases. The comparisons between BTMA and other methods have similar results and we omit them for saving space.
\newpage
\begin{center}
	\footnotesize
	\LTleft=-50pt
	\begin{longtable}{llcccccccccc}
		\caption{\small Selection probabilities based on 1000 simulations \label{tab:first}} \vspace{-0.5em}\\
		\hline
		True $\theta$ & Model & $m=15$ & $m=20$ & $m=25$ & $m=30$ & $m=40$ & $m=50$ & $m=70$ & $m=100$ & Mallows & BIC\\
		\hline
		\endfirsthead
		\multicolumn{12}{c}{\small Table S1 (continued):\ Selection probabilities based on 1000 simulations \label{tab:firstcontinued}} \vspace{1em}\\
		\hline
		True $\theta$ & Model & $m=15$ & $m=20$ & $m=25$ & $m=30$ & $m=40$ & $m=50$ & $m=70$ & $m=100$ & Mallows & BIC\\
		\hline
		\endhead
		\endfoot
		\endlastfoot
		(2, 0, 0, 4, 0) & $1,4^{*}$   & 0.946 & 0.871 & 0.770 & 0.681 & 0.493 & 0.369 & 0.219 & 0.106 & 0.614 & 0.817\\
		& $1,2,4$     & 0.021 & 0.042 & 0.056 & 0.080 & 0.102 & 0.133 & 0.150 & 0.142 & 0.081 & 0.049\\
		& $1,3,4$     & 0.018 & 0.047 & 0.099 & 0.136 & 0.179 & 0.180 & 0.158 & 0.124 & 0.107 & 0.058\\
		& $1,4,5$     & 0.013 & 0.033 & 0.052 & 0.059 & 0.084 & 0.098 & 0.113 & 0.096 & 0.095 & 0.058\\
		& $1,2,3,4$   & 0.002 & 0.004 & 0.012 & 0.019 & 0.051 & 0.078 & 0.119 & 0.143 & 0.033 & 0.005\\
		& $1,2,4,5$   & 0.000 & 0.000 & 0.005 & 0.011 & 0.034 & 0.042 & 0.063 & 0.097 & 0.035 & 0.008\\
		& $1,3,4,5$   & 0.000 & 0.003 & 0.006 & 0.013 & 0.039 & 0.067 & 0.103 & 0.142 & 0.022 & 0.005\\
		& $1,2,3,4,5$ & 0.000 & 0.000 & 0.000 & 0.001 & 0.018 & 0.033 & 0.075 & 0.150 & 0.013 & 0.000\\
		\hline
		(2, 0, 0, 4, 8) & $1,4,5^{*}$ & 0.957 & 0.898 & 0.828 & 0.730 & 0.570 & 0.450 & 0.309 & 0.192 & 0.672 & 0.861\\
		& $1,2,4,5$   & 0.021 & 0.042 & 0.067 & 0.094 & 0.131 & 0.165 & 0.199 & 0.228 & 0.118 & 0.064\\
		& $1,3,4,5$   & 0.022 & 0.057 & 0.095 & 0.148 & 0.228 & 0.283 & 0.310 & 0.299 & 0.158 & 0.065\\
		& $1,2,3,4,5$ & 0.000 & 0.003 & 0.010 & 0.028 & 0.071 & 0.102 & 0.182 & 0.281 & 0.052 & 0.010\\
		\hline
		(2, 9, 0, 4, 8) & $1,4,5$       & 0.016 & 0.002 & 0.002 & 0.002 & 0.000 & 0.000 & 0.000 & 0.000 & 0.000 & 0.000\\
		& $1,2,4,5^{*}$ & 0.969 & 0.949 & 0.912 & 0.844 & 0.743 & 0.650 & 0.547 & 0.438 & 0.822 & 0.924\\
		& $1,3,4,5$     & 0.003 & 0.002 & 0.002 & 0.002 & 0.000 & 0.000 & 0.000 & 0.000 & 0.000 & 0.000\\
		& $1,2,3,4,5$   & 0.012 & 0.047 & 0.084 & 0.152 & 0.257 & 0.350 & 0.453 & 0.562 & 0.178 & 0.076\\
		\hline
		(2, 9, 6, 4, 8) & $1,2,3,5$       & 0.006 & 0.000 & 0.000 & 0.000  & 0.000  & 0 .000 & 0.000   & 0.000  & 0.000 & 0.000\\
		& $1,2,4,5$       & 0.013 & 0.000 & 0.000 & 0.000  & 0.000  & 0.000  & 0.000   & 0.000  & 0.000 & 0.000\\
		& $1,3,4,5$       & 0.073 & 0.021 & 0.008 & 0.001  & 0.000  & 0.000  & 0.000   & 0.000  & 0.000 & 0.002\\
		& $1,2,3,4,5^{*}$ & 0.908 & 0.979 & 0.992 & 0.999  & 1.000  & 1.000  & 1 .000  & 1.000  & 1.000 & 0.998\\
		\hline
		(2, 0, 0, 4, 8) & $1,2$       & 0.000 & 0.000 & 0.000 & 0.000 & 0.000 & 0.000 & 0.000 & 0.000 & 0.000 & 0.000\\
		& $1,3$       & 0.000 & 0.000 & 0.000 & 0.000 & 0.000 & 0.000 & 0.000 & 0.000 & 0.000 & 0.000\\
		& $1,4$       & 0.000 & 0.000 & 0.000 & 0.000 & 0.000 & 0.000 & 0.000 & 0.000 & 0.000 & 0.000\\
		& $1,5$       & 0.002 & 0.000 & 0.000 & 0.000 & 0.000 & 0.000 & 0.000 & 0.000 & 0.000 & 0.000\\
		& $1,2,3$     & 0.000 & 0.000 & 0.000 & 0.000 & 0.000 & 0.000 & 0.000 & 0.000 & 0.000 & 0.000\\
		& $1,2,4$     & 0.000 & 0.000 & 0.000 & 0.000 & 0.000 & 0.000 & 0.000 & 0.000 & 0.000 & 0.000\\
		& $1,2,5$     & 0.156 & 0.074 & 0.040 & 0.029 & 0.020 & 0.014 & 0.003 & 0.002 & 0.000 & 0.018\\
		& $1,3,4$     & 0.000 & 0.000 & 0.000 & 0.000 & 0.000 & 0.000 & 0.000 & 0.000 & 0.000 & 0.000\\
		& $1,3,5$     & 0.815 & 0.881 & 0.896 & 0.895 & 0.860 & 0.803 & 0.668 & 0.536 & 0.490 & 0.921\\
		& $1,2,3,4$   & 0.000 & 0.000 & 0.000 & 0.000 & 0.000 & 0.000 & 0.000 & 0.000 & 0.000 & 0.000\\
		& $1,2,3,5$   & 0.027 & 0.045 & 0.064 & 0.076 & 0.120 & 0.183 & 0.329 & 0.462 & 0.510 & 0.061\\
		\hline
		(2, 9, 6, 4, 8) & $1,2$       & 0.000 & 0.000 & 0.000 & 0.000 & 0.000 & 0.000 & 0.000 & 0.000 & 0.000 & 0.000\\
		& $1,3$       & 0.000 & 0.000 & 0.000 & 0.000 & 0.000 & 0.000 & 0.000 & 0.000 & 0.000 & 0.000\\
		& $1,4$       & 0.000 & 0.000 & 0.000 & 0.000 & 0.000 & 0.000 & 0.000 & 0.000 & 0.000 & 0.000\\
		& $1,5$       & 0.000 & 0.000 & 0.000 & 0.000 & 0.000 & 0.000 & 0.000 & 0.000 & 0.000 & 0.000\\
		& $1,2,3$     & 0.000 & 0.000 & 0.000 & 0.000 & 0.000 & 0.000 & 0.000 & 0.000 & 0.000 & 0.000\\
		& $1,2,4$     & 0.000 & 0.000 & 0.000 & 0.000 & 0.000 & 0.000 & 0.000 & 0.000 & 0.000 & 0.000\\
		& $1,2,5$     & 0.000 & 0.000 & 0.000 & 0.000 & 0.000 & 0.000 & 0.000 & 0.000 & 0.000 & 0.000\\
		& $1,3,4$     & 0.000 & 0.000 & 0.000 & 0.000 & 0.000 & 0.000 & 0.000 & 0.000 & 0.000 & 0.000\\
		& $1,3,5$     & 0.000 & 0.000 & 0.000 & 0.000 & 0.000 & 0.000 & 0.000 & 0.000 & 0.000 & 0.000\\
		& $1,4,5$     & 0.000 & 0.000 & 0.000 & 0.000 & 0.000 & 0.000 & 0.000 & 0.000 & 0.000 & 0.000\\
		& $1,2,3,4$   & 0.000 & 0.000 & 0.000 & 0.000 & 0.000 & 0.000 & 0.000 & 0.000 & 0.000 & 0.000\\
		& $1,2,3,5$   & 0.146 & 0.088 & 0.067 & 0.057 & 0.051 & 0.042 & 0.037 & 0.039 & 0.049 & 0.049\\
		& $1,2,4,5$   & 0.236 & 0.253 & 0.271 & 0.287 & 0.297 & 0.281 & 0.278 & 0.266 & 0.246 & 0.246\\
		& $1,3,4,5$   & 0.618 & 0.659 & 0.662 & 0.656 & 0.652 & 0.677 & 0.685 & 0.695 & 0.705 & 0.705\\
		\hline
	\end{longtable}
\end{center}
\newpage
\begin{center}
	\footnotesize
	\LTleft=-10pt
	\begin{longtable}{llcccccccccc}
		\caption{\small Weight vectors of BTMA based on 1000 simulations \label{tab:second}} \vspace{-2.5em}\\
		\hline
		True $\theta$ & Model & $m=15$ & $m=20$ & $m=25$ & $m=30$ & $m=40$ & $m=50$ & $m=70$ & $m=100$\\
		\hline
		\endfirsthead
		\multicolumn{12}{c}{\small Table S2 (continued):\ Weight vectors of BTMA based on 1000 simulations \label{tab:secondcontinued}} \vspace{1em}\\
		\hline
		True $\theta$ & Model & $m=15$ & $m=20$ & $m=25$ & $m=30$ & $m=40$ & $m=50$ & $m=70$ & $m=100$\\
		\hline
		\endhead
		\endfoot
		\endlastfoot
		(2, 0, 0, 4, 0) & $1,4^{*}$   & 0.7540 & 0.6780 & 0.6078 & 0.5332 & 0.4252 & 0.3483 & 0.2449 & 0.1638\\
		& $1,2,4$     & 0.0659 & 0.0754 & 0.0866 & 0.1022 & 0.1159 & 0.1259 & 0.1285 & 0.1227\\
		& $1,3,4$     & 0.0673 & 0.0980 & 0.1134 & 0.1362 & 0.1580 & 0.1659 & 0.1559 & 0.1333\\
		& $1,4,5$     & 0.0610 & 0.0715 & 0.0836 & 0.0919 & 0.1005 & 0.1083 & 0.1149 & 0.1151\\
		& $1,2,3,4$   & 0.0202 & 0.0298 & 0.0395 & 0.0500 & 0.0684 & 0.0823 & 0.1064 & 0.1207\\
		& $1,2,4,5$   & 0.0129 & 0.0181 & 0.0220 & 0.0270 & 0.0390 & 0.0458 & 0.0661 & 0.0878\\
		& $1,3,4,5$   & 0.0106 & 0.0177 & 0.0285 & 0.0350 & 0.0516 & 0.0647 & 0.0887 & 0.1133\\
		& $1,2,3,4,5$ & 0.0081 & 0.0115 & 0.0185 & 0.0244 & 0.0413 & 0.0588 & 0.0946 & 0.1434\\
		\hline
		(2, 0, 0, 4, 8) & $1,4,5^{*}$ & 0.8114 & 0.7566 & 0.6939 & 0.6329 & 0.5338 & 0.4556 & 0.3496 & 0.2568\\
		& $1,2,4,5$   & 0.0810 & 0.0953 & 0.1144 & 0.1265 & 0.1504 & 0.1680 & 0.1885 & 0.2101\\
		& $1,3,4,5$   & 0.0775 & 0.1071 & 0.1355 & 0.1659 & 0.2090 & 0.2396 & 0.2650 & 0.2721\\
		& $1,2,3,4,5$ & 0.0301 & 0.0410 & 0.0561 & 0.0747 & 0.1068 & 0.1368 & 0.1968 & 0.2609\\
		\hline
		(2, 9, 0, 4, 8) & $1,4,5$       & 0.1390 & 0.0938 & 0.0683 & 0.0555 & 0.0372 & 0.0274 & 0.0162 & 0.0094\\
		& $1,2,4,5^{*}$ & 0.7337 & 0.7511 & 0.7436 & 0.7242 & 0.6820 & 0.6384 & 0.5675 & 0.4947\\
		& $1,3,4,5$     & 0.0642 & 0.0622 & 0.0591 & 0.0540 & 0.0490 & 0.0438 & 0.0335 & 0.0243\\
		& $1,2,3,4,5$   & 0.0631 & 0.0929 & 0.1290 & 0.1663 & 0.2318 & 0.2904 & 0.3828 & 0.4716\\
		\hline
		(2, 9, 6, 4, 8) & $1,2,3,5$       & 0.3071 & 0.2569 & 0.2145 & 0.1803 & 0.1326 & 0.1055 & 0.0741 & 0.0501\\
		& $1,2,4,5$       & 0.2313 & 0.2146 & 0.1961 & 0.1759 & 0.1446 & 0.1215 & 0.0896 & 0.0632\\
		& $1,3,4,5$       & 0.3999 & 0.3676 & 0.3217 & 0.2781 & 0.2122 & 0.1692 & 0.1147 & 0.0752\\
		& $1,2,3,4,5^{*}$ & 0.0618 & 0.1609 & 0.2677 & 0.3657 & 0.5106 & 0.6038 & 0.7216 & 0.8115\\
		\hline
		(2, 0, 0, 4, 8) & $1,2$       & 0.0001 & 0.0000 & 0.0000 & 0.0000 & 0.0000 & 0.0000 & 0.0000 & 0.0000\\
		& $1,3$       & 0.0001 & 0.0000 & 0.0000 & 0.0000 & 0.0000 & 0.0000 & 0.0000 & 0.0000\\
		& $1,4$       & 0.2507 & 0.2435 & 0.2223 & 0.1962 & 0.1401 & 0.0834 & 0.0216 & 0.0019\\
		& $1,5$       & 0.4579 & 0.4753 & 0.4728 & 0.4642 & 0.4368 & 0.4083 & 0.3758 & 0.3634\\
		& $1,2,3$     & 0.0000 & 0.0000 & 0.0000 & 0.0000 & 0.0000 & 0.0000 & 0.0000 & 0.0000\\
		& $1,2,4$     & 0.0000 & 0.0000 & 0.0000 & 0.0000 & 0.0000 & 0.0000 & 0.0000 & 0.0000\\
		& $1,2,5$     & 0.0812 & 0.0669 & 0.0660 & 0.0656 & 0.0677 & 0.0693 & 0.0731 & 0.0759\\
		& $1,3,4$     & 0.0383 & 0.0559 & 0.0802 & 0.1066 & 0.1608 & 0.2139 & 0.2711 & 0.2892\\
		& $1,3,5$     & 0.1569 & 0.1504 & 0.1524 & 0.1615 & 0.1881 & 0.2167 & 0.2455 & 0.2536\\
		& $1,2,3,4$   & 0.0006 & 0.0009 & 0.0009 & 0.0012 & 0.0014 & 0.0016 & 0.0018 & 0.0020\\
		& $1,2,3,5$   & 0.0129 & 0.0068 & 0.0055 & 0.0047 & 0.0052 & 0.0068 & 0.0112 & 0.0140\\
		\hline
		(2, 9, 6, 4, 8) & $1,2$       & 0.0184 & 0.0184 & 0.0177 & 0.0169 & 0.0144 & 0.0115 & 0.0058 & 0.0014\\
		& $1,3$       & 0.0010 & 0.0003 & 0.0001 & 0.0000 & 0.0000 & 0.0000 & 0.0000 & 0.0000\\
		& $1,4$       & 0.0915 & 0.0707 & 0.0513 & 0.0358 & 0.0146 & 0.0055 & 0.0012 & 0.0004\\
		& $1,5$       & 0.0108 & 0.0070 & 0.0035 & 0.0017 & 0.0002 & 0.0000 & 0.0000 & 0.0000\\
		& $1,2,3$     & 0.0023 & 0.0010 & 0.0006 & 0.0002 & 0.0002 & 0.0005 & 0.0020 & 0.0031\\
		& $1,2,4$     & 0.0054 & 0.0033 & 0.0026 & 0.0022 & 0.0018 & 0.0013 & 0.0008 & 0.0013\\
		& $1,2,5$     & 0.1130 & 0.0956 & 0.0856 & 0.0662 & 0.0365 & 0.0173 & 0.0066 & 0.0022\\
		& $1,3,4$     & 0.0150 & 0.0206 & 0.0251 & 0.0233 & 0.0176 & 0.0121 & 0.0079 & 0.0040\\
		& $1,3,5$     & 0.1343 & 0.1069 & 0.0972 & 0.0872 & 0.0728 & 0.0640 & 0.0430 & 0.0207\\
		& $1,4,5$     & 0.0355 & 0.0241 & 0.0165 & 0.0114 & 0.0068 & 0.0046 & 0.0019 & 0.0007\\
		& $1,2,3,4$   & 0.0041 & 0.0090 & 0.0176 & 0.0269 & 0.0432 & 0.0542 & 0.0610 & 0.0651\\
		& $1,2,3,5$   & 0.2021 & 0.2193 & 0.2124 & 0.2129 & 0.2184 & 0.2267 & 0.2448 & 0.2654\\
		& $1,2,4,5$   & 0.0847 & 0.1109 & 0.1398 & 0.1748 & 0.2241 & 0.2531 & 0.2698 & 0.2702\\
		& $1,3,4,5$   & 0.2820 & 0.3127 & 0.3301 & 0.3404 & 0.3494 & 0.3492 & 0.3553 & 0.3655\\
		\hline
	\end{longtable}
\end{center}
\begin{table}
	\caption{\small Simulated mean squared errors of BMS and BTMA  \label{tab:third}}
	\vspace{-1.0em}
	\begin{center}
		\resizebox{\linewidth}{!}{
			\begin{tabular}{cccccccc}
				\hline
				Resample size & Settings & $(2, 0, 0, 4, 0)^{*}$ & $(2, 0, 0, 4, 8)^{*}$ & $(2, 9, 0, 4, 8)^{*}$ & $(2, 9, 6, 4, 8)^{*}$ & (2, 0, 0, 4, 8) & (2, 9, 6, 4, 8)\\
				\hline
				$m=15$ & BMS  & 0.0570 & 0.0778 & \textbf{0.1137} & \textbf{0.1876} & 1.5256 & 0.9182\\
				& BTMA & \textbf{0.0564} & \textbf{0.0774} & 0.1455 & 0.2123 & \textbf{0.3300} & \textbf{0.2162}\\
				\hline
				$m=20$ & BMS  & 0.0657 & 0.0835 & \textbf{0.1066} & \textbf{0.1372} & 1.4763 & 0.8864\\
				& BTMA & \textbf{0.0608} & \textbf{0.0807} & 0.1137 & 0.1920 & \textbf{0.3192} & \textbf{0.1948}\\
				\hline
				$m=25$ & BMS  & 0.0750 & 0.0895 & \textbf{0.1098} & \textbf{0.1297} & 1.4523 & 0.8782\\
				& BTMA & \textbf{0.0661} & \textbf{0.0843} & 0.1256 & 0.1780 & \textbf{0.3102} & \textbf{0.1655}\\
				\hline
				$m=30$ & BMS  & 0.0817 & 0.0971 & \textbf{0.1145} & \textbf{0.1264} & 1.4523 & 0.8782\\
				& BTMA & \textbf{0.0705} & \textbf{0.0876} & 0.1228 & 0.1655 & \textbf{0.3102} & \textbf{0.1655}\\
				\hline
				$m=40$ & BMS  & 0.0952 & 0.1065 & \textbf{0.1172} & \textbf{0.1258} & 1.4473 & 0.8775\\
				& BTMA & \textbf{0.0778} & \textbf{0.0929} & 0.1203 & 0.1523 & \textbf{0.3020} & \textbf{0.1522}\\
				\hline
				$m=50$ & BMS  & 0.1027 & 0.1119 & 0.1200 & \textbf{0.1258} & 1.4430 & 0.8669\\
				& BTMA & \textbf{0.0830} & \textbf{0.0972} & \textbf{0.1197} & 0.1432 & \textbf{0.2944} & \textbf{0.1447}\\
				\hline
				$m=70$ & BMS  & 0.1115 & 0.1173 & 0.1223 & \textbf{0.1258} & 1.4326 & 0.8627\\
				& BTMA & \textbf{0.0916} & \textbf{0.1032} & \textbf{0.1191} & 0.1362 & \textbf{0.2856} & \textbf{0.1394}\\
				\hline
				$m=100$& BMS  & 0.1173 & 0.1212 & 0.1238 & \textbf{0.1258} & 1.4263 & 0.8596\\
				& BTMA & \textbf{0.0989} & \textbf{0.1082} & \textbf{0.1198} & 0.1314 & \textbf{0.2838} & \textbf{0.1376}\\
				\hline
			\end{tabular}
		}
	\end{center}
	\vspace{2cm}
	\caption{\small Simulated mean squared errors of JMA and BTMA  \label{tab:fourth}}
	\vspace{-1em}
	\begin{center}
		\resizebox{\linewidth}{!}{
			\footnotesize
			\begin{tabular}{ccccccc}
				\hline
				Settings & $(2, 0, 0, 4, 0)^{*}$ & $(2, 0, 0, 4, 8)^{*}$ & $(2, 9, 0, 4, 8)^{*}$ & $(2, 9, 6, 4, 8)^{*}$ & (2, 0, 0, 4, 8) & (2, 9, 6, 4, 8)\\
				\hline
				JMA    & 0.0779 & 0.0925 & 0.1262 & \textbf{0.1616} & 0.3286 & 0.1837\\
				BTMA   & \textbf{0.0756} & \textbf{0.0914}  & \textbf{0.1256}  & 0.1639  & \textbf{0.3051} & \textbf{0.1662}\\
				\hline
			\end{tabular}
		}
	\end{center}
\end{table}
\normalsize
\newpage
\section{Empirical example: Motor Trend Car data}
\indent The dataset is the Motor Trend Car data from the dataset package in R. The data was extracted from the 1974 Motor Trend US magazine, and comprises fuel consumption and 10 aspects of automobile design and performance for 32 automobiles. This dataset has been analyzed by Henderson and Velleman (1981). Here we use this data to predict the miles per gallon (mpg and $y_{i}$). The 10 covariates include Number of cylinders (cyl and $x_{i1}$), Displacement (disp and $x_{i2}$), Gross horsepower (hp and $x_{i3}$), Rear axle ratio (drat and $x_{i4}$), Weight (wt and $x_{i5}$), 1/4 mile time (qsec and $x_{i6}$), Engine (vs and $x_{i7}$), Transmission (am and $x_{i8}$), Number of forward gears (gear and $x_{i9}$) and Number of carburetors (carb and $x_{i10}$). Among them, vs and am are categorical variables with values 0 and 1, while cyl, gear, and carb are integer variables. The scatterplot is presented in Figure S1, with the generation method of the solid red line and dashed blue line being the same as in Example 1. We use all 10 variables to fit a linear model, and find that the R-squared of this linear model is 0.8690 with p-value $<$ 0.05, which demonstrates the strong interpretability of the linear model. Therefore, the largest model is set to be
\begin{align}
	y_{i}=\theta_{0}+\theta_{1}x_{i1}+\theta_{2}x_{i2}+\cdots+\theta_{10}x_{i10}+e_{i},\ i=1,\cdots,n, \notag
\end{align}
where $y_{i}$ is the miles per gallon of the $i$th motor car, $x_{ij}$, $j=1,\cdots,10$ are the covariates with associated coefficients $\theta_{j}$, $j=1,\cdots,10$, $\theta_{0}$ is the intercept, and $e_{i}$ is the random error. We also prepare a candidate model set using the same procedure as in Example 1.\\
\indent Similarly, we split the data into training set and test set. The training sample size $n\in\{20,22,24,26\}$. Then we standardize the data and compare the MSPEs of the eight methods: MMA, JMA, BTMA, S-AIC, S-BIC, BMS, subsampling and bagging. In this example, the number of observations is $N=32$. We repeat the calculations 1000 times, and the mean and variance of MSPEs are shown in Table S5.\\
\begin{table}
	\caption{\small Mean and variance of MSPEs for Motor Trend Car data \label{tab:nineth}}
	\vspace{0em}
	\begin{center}
		\footnotesize
		\begin{tabular}{cccccccccc}
			\hline
			& Training set & MMA & JMA & BTMA & S-AIC & S-BIC & BMS  & Sub & Bag\\
			\hline
			Mean  & $n=20$ & 0.2856 & 0.2776 & \textbf{0.2376} & 0.3336 & 0.2687 & 0.2473 & 0.2513 & 0.8729\\
			& $n=22$       & 0.2490 & 0.2469 & \textbf{0.2233} & 0.2663 & 0.2280 & 0.2321 & 0.2354 & 0.5820\\
			& $n=24$       & 0.2435 & 0.2432 & \textbf{0.2234} & 0.2543 & 0.2253 & 0.2337 & 0.2361 & 0.4650\\
			& $n=26$       & 0.2247&  0.2237 & 0.2142 & 0.2209 & \textbf{0.2118} & 0.2268 & 0.2223 & 0.3600\\
			\hline
			Variance & $n=20$ & 0.0543 & 0.0253 & \textbf{0.0081} & 0.1733 & 0.0930 & 0.0090 & 0.0094 & 0.4032\\
			& $n=22$       & 0.0181 & 0.0132 & \textbf{0.0082} & 0.0501 & 0.0105 & 0.0079 & 0.0090 & 0.2074\\
			& $n=24$       & 0.0216 & 0.0174 & \textbf{0.0100} & 0.0599 & 0.0116 & 0.0105 & 0.0107 & 0.1468\\
			& $n=26$       & 0.0149 & 0.0146 & 0.0137 & 0.0149 & \textbf{0.0128} & 0.0137 & 0.0131 & 0.1703\\
			\hline
		\end{tabular}
	\end{center}
\end{table}
\indent From Table S5, we observe that BTMA has the smallest MSPEs at $n=20$, 22, and 24. As $n$ increases, the mean of MSPEs for all methods gradually decreases, indicating that their predictive performance improve. Specifically, the mean of MSPEs about BTMA and S-BIC gradually approaches. We can observe that S-BIC performs marginally better than BTMA at $n=26$. In addition, it is seen from the variance of MSPEs that BTMA always has good stability.
\begin{figure}[!htbp]
	\begin{center}
		\includegraphics[scale=0.8]{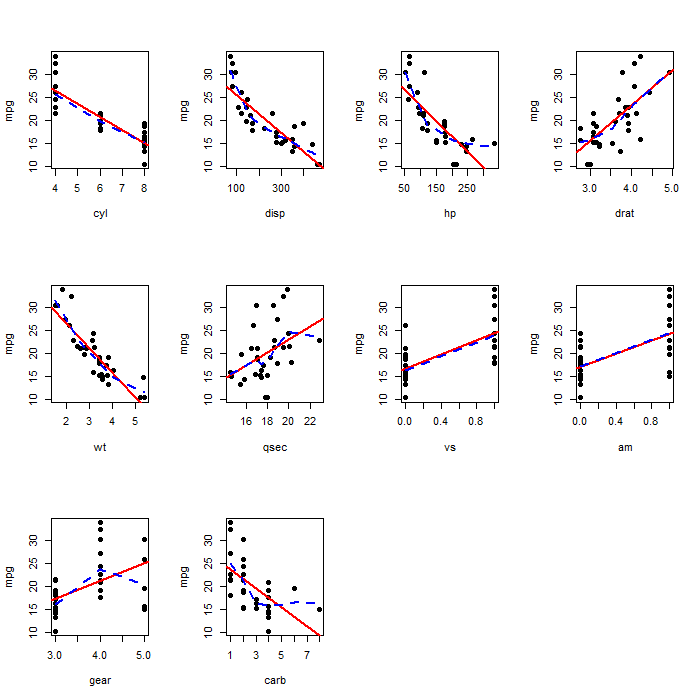}
	\end{center}
	\vspace{-1em}
	\caption{Scatterplot between mpg and other variables for Motor Trend Car data \label{fig:sixth}}
\end{figure}%
\newpage
\section{Proofs of Lemmas and Theorems}
\addcontentsline{toc}{subsection}{Proof of Theorem 1}
\begin{proof}[\textbf{Proof of Theorem 1}]
	From the proof of (14) in \cite{shao1996bootstrap}, if $\xi_{n}^{*}$ is a function of bootstrap sample, then $E_{*}\{\xi_{n}^{*}[1+o_{p_{*}}(1)]\}=E_{*}(\xi_{n}^{*})[1+o_{p}(1)]$, where $o_{p_{*}}(1)$ denotes a quantity $\zeta_{n}^{*}$ satisfying $P_{*}\{|\zeta_{n}^{*}|>\varepsilon\}=o_{p}(1)$ for any $\varepsilon>0$. Since $k_{q}$ is fixed, it is straightforward to find that $(m/n)(X_{(q)}^{*'}X_{(q)}^{*})^{-1}\stackrel{p_{*}}{\rightarrow}(X_{(q)}'X_{(q)})^{-1}$ for $q=1,\cdots,M$. Without loss of generality, for $q,r=1,\cdots,M$, we have
	\begin{align}
		&E_{*}\big(\tilde{\Theta}_{q}^{*}-\hat{\Theta}_{q}\big)\big(\tilde{\Theta}_{r}^{*}-\hat{\Theta}_{r}\big)' \notag\\
		=&E_{*}\bigg[(X_{(q)}^{*'}X_{(q)}^{*})^{-1}\sum\limits_{i=1}^{m}x_{i(q)}^{*}(y_{i}^{*}-x_{i(q)}^{*'}\hat{\Theta}_{q})\bigg]
		\bigg[\sum\limits_{i=1}^{m}(y_{i}^{*'}-\hat{\Theta}_{r}'x_{i(r)}^{*})x_{i(r)}^{*'}(X_{(r)}^{*'}X_{(r)}^{*})^{-1}\bigg] \notag\\
		=&(\dfrac{n}{m})^{2}[I_{k_{q}}+o_{p}(1)](X_{(q)}'X_{(q)})^{-1}E_{*}\bigg[\sum\limits_{i=1}^{n}\pi_{i}x_{i(q)}(y_{i}-x_{i(q)}'\hat{\Theta}_{q})\bigg]
		\bigg[\sum\limits_{i=1}^{n}\pi_{i}(y_{i}'-\hat{\Theta}_{r}'x_{i(r)})x_{i(r)}'\bigg] \notag\\
		&(X_{(r)}'X_{(r)})^{-1}[I_{k_{q}}+o_{p}(1)] \notag\\
		=&(\dfrac{n}{m})^{2}[I_{k_{q}}+o_{p}(1)](X_{(q)}'X_{(q)})^{-1}\bigg[\sum\limits_{i=1}^{n}E_{*}(\pi_{i}^{2})x_{i(q)}x_{i(r)}'(y_{i}-x_{i(q)}'\hat{\Theta}_{q})(y_{i}-x_{i(r)}'\hat{\Theta}_{r}) \notag\\
		&+\underset{i\neq j}{\sum{\sum}}E_{*}(\pi_{i}\pi_{j})x_{i(q)}x_{j(r)}'(y_{i}-x_{i(q)}'\hat{\Theta}_{q})(y_{j}-x_{j(r)}'\hat{\Theta}_{r})\bigg](X_{(r)}'X_{(r)})^{-1}[I_{k_{q}}+o_{p}(1)]\notag\\
		=&(\dfrac{n}{m})^{2}[I_{k_{q}}+o_{p}(1)](X_{(q)}'X_{(q)})^{-1}\bigg[\sum\limits_{i=1}^{n}\dfrac{m}{n}x_{i(q)}x_{i(r)}'(y_{i}-x_{i(q)}'\hat{\Theta}_{q})(y_{i}-x_{i(r)}'\hat{\Theta}_{r}) \notag\\
		&+\dfrac{m^{2}-m}{n^{2}}\bigg(\sum\limits_{i=1}^{n}x_{i(q)}(y_{i}-x_{i(q)}'\hat{\Theta}_{q})\bigg)\bigg(\sum\limits_{j=1}^{n}x_{j(r)}(y_{j}-x_{j(r)}'\hat{\Theta}_{r})\bigg)\bigg](X_{(r)}'X_{(r)})^{-1}[I_{k_{q}}+o_{p}(1)] \notag\\
		=&\dfrac{n}{m}[I_{k_{q}}+o_{p}(1)](X_{(q)}'X_{(q)})^{-1}\sum\limits_{i=1}^{n}x_{i(q)}x_{i(r)}'[e_{i}+o_{p}(1)]^{2}(X_{(r)}'X_{(r)})^{-1}[I_{k_{q}}+o_{p}(1)] \notag\\
		=&\dfrac{1}{m}[I_{k_{q}}+o_{p}(1)](\dfrac{1}{n}X_{(q)}'X_{(q)})^{-1}\dfrac{1}{n}X_{(q)}'X_{(r)}[\sigma^{2}+o_{p}(1)](\dfrac{1}{n}X_{(r)}'X_{(r)})^{-1}[I_{k_{q}}+o_{p}(1)] \notag\\
		=&\dfrac{n\sigma^{2}}{m}(X_{(q)}'X_{(q)})^{-1}X_{(q)}'X_{(r)}(X_{(r)}'X_{(r)})^{-1}+o_{p}(\dfrac{1}{m}). \label{th0.1}
	\end{align}
	It is clear that
	\begin{align}
		\Gamma_{n,m}(\omega)=&\dfrac{||Y-\tilde{\mu}(\omega)||^{2}}{n} \notag\\
		=&\dfrac{||Y-\hat{\mu}(\omega)+\hat{\mu}(\omega)-\tilde{\mu}(\omega)||^{2}}{n} \notag\\
		=&\dfrac{||Y-\hat{\mu}(\omega)||^{2}}{n}+\dfrac{||\hat{\mu}(\omega)-\tilde{\mu}(\omega)||^{2}}{n}+\dfrac{2(Y-\hat{\mu}(\omega))'(\hat{\mu}(\omega)-\tilde{\mu}(\omega))}{n}. \label{th0.2}
	\end{align}
	Now we consider the second and third terms of (\ref{th0.2}). Under the assumptions of Theorem 1, it is seen from (\ref{th0.1}) that
	\begin{align}
		&E_{*}\bigg[\dfrac{||\hat{\mu}(\omega)-\tilde{\mu}(\omega)||^{2}}{n}\bigg] \notag\\
		\triangleq&\dfrac{1}{n}E_{*}
		\begin{bmatrix}
			\sum\limits_{q=1}^{M}\omega_{q}
			\begin{pmatrix}
				\hat{\mu}_{1(q)}-\tilde{\mu}_{1(q)}\\
				\vdots\\
				\hat{\mu}_{n(q)}-\tilde{\mu}_{n(q)}
			\end{pmatrix}
			'
		\end{bmatrix}
		\begin{bmatrix}
			\sum\limits_{q=1}^{M}\omega_{q}
			\begin{pmatrix}
				\hat{\mu}_{1(q)}-\tilde{\mu}_{1(q)}\\
				\vdots\\
				\hat{\mu}_{n(q)}-\tilde{\mu}_{n(q)}
			\end{pmatrix}
		\end{bmatrix} \notag\\
		=&\dfrac{1}{n}E_{*}\bigg[\sum\limits_{q=1}^{M}\omega_{q}^{2}\sum\limits_{i=1}^{n}(\hat{\mu}_{i(q)}-\tilde{\mu}_{i(q)})^{2}
		+2\underset{1<q<r<M}{\sum{\sum}}\omega_{(q)}\omega_{r}\sum\limits_{i=1}^{n}(\hat{\mu}_{i(q)}-\tilde{\mu}_{i(q)})(\hat{\mu}_{i(r)}-\tilde{\mu}_{i(r)})\bigg] \notag\\
		=&\dfrac{1}{n}\bigg[\sum\limits_{q=1}^{M}\omega_{q}^{2}\sum\limits_{i=1}^{n}x_{i(q)}'E_{*}(\hat{\Theta}_{q}-\tilde{\Theta}_{q}^{*})(\hat{\Theta}_{q}-\tilde{\Theta}_{q}^{*})'x_{i(q)}
		+2\underset{1<q<r<M}{\sum{\sum}}\omega_{q}\omega_{r}\sum\limits_{i=1}^{n}x_{i(q)}'E_{*}(\hat{\Theta}_{q} \notag\\
		&-\tilde{\Theta}_{q}^{*})(\hat{\Theta}_{r}-\tilde{\Theta}_{r}^{*})'x_{i(r)}\bigg] \notag\\
		=&\dfrac{1}{n}\bigg\{\sum\limits_{q=1}^{M}\omega_{q}^{2}\sum\limits_{i=1}^{n}x_{i(q)}'\Big[\dfrac{n\sigma^{2}}{m}\big(X_{(q)}'X_{(q)}\big)^{-1}+o_{p}(\dfrac{1}{m})\Big]x_{i(q)}
		+2\underset{1<q<r<M}{\sum{\sum}}\omega_{q}\omega_{r}\sum\limits_{i=1}^{n}x_{i(q)}' \notag\\
		&\Big[\dfrac{n\sigma^{2}}{m}(X_{(q)}'X_{(q)})^{-1}X_{(q)}'X_{(r)}(X_{(r)}'X_{(r)})^{-1}+o_{p}(\dfrac{1}{m})\Big]x_{i(r)}\bigg\} \notag\\
		=&\dfrac{1}{n}\dfrac{n\sigma^{2}}{m}\bigg[\sum\limits_{q=1}^{M}\omega_{q}^{2}\sum\limits_{i=1}^{n}\text{tr}(x_{i(q)}'\big(X_{(q)}'X_{(q)}\big)^{-1}x_{i(q)})
		+2\underset{1<q<r<M}{\sum{\sum}}\omega_{q}\omega_{r}\sum\limits_{i=1}^{n}\text{tr}(x_{i(q)}'(X_{(q)}'X_{(q)})^{-1}X_{(q)}' \notag\\
		&X_{(r)}(X_{(r)}'X_{(r)})^{-1}x_{i(r)})\bigg]+o_{p}(\dfrac{1}{m})  \notag\\
		=&\dfrac{1}{n}\dfrac{n\sigma^{2}}{m}\bigg[\sum\limits_{q=1}^{M}\omega_{q}^{2}\text{tr}(H_{q})+2\underset{1<q<r<M}{\sum{\sum}}\omega_{q}\omega_{r}\text{tr}(H_{q}H_{r})\bigg]+o_{p}(\dfrac{1}{m})  \notag\\
		=&\dfrac{\sigma^{2}}{m}\text{tr}\bigg(\sum\limits_{q=1}^{M}\omega_{q}H_{q}\bigg)^{2}+o_{p}(\dfrac{1}{m}) \notag\\
		=&\dfrac{\sigma^{2}}{m}\text{tr}H^{2}(\omega)+o_{p}(\dfrac{1}{m}), \label{th0.3}
	\end{align}
	and
	\begin{align}
		&E_{*}\big[\dfrac{2(Y-\hat{\mu}(\omega))'(\hat{\mu}(\omega)-\tilde{\mu}(\omega))}{n}\big] \notag\\
		\triangleq&\dfrac{2}{n}E_{*}
		\begin{bmatrix}
			\sum\limits_{q=1}^{M}\omega_{q}
			\begin{pmatrix}
				y_{1}-\hat{\mu}_{1(q)}\\
				\vdots\\
				y_{n}-\hat{\mu}_{n(q)}
			\end{pmatrix}'
		\end{bmatrix}
		\begin{bmatrix}
			\sum\limits_{q=1}^{M}\omega_{q}
			\begin{pmatrix}
				\hat{\mu}_{1(q)}-\tilde{\mu}_{1(q)}\\
				\vdots\\
				\hat{\mu}_{n(q)}-\tilde{\mu}_{n(q)}
			\end{pmatrix}
		\end{bmatrix} \notag\\
		=&\dfrac{2}{n}E_{*}\bigg[\sum\limits_{q=1}^{M}\omega_{q}^{2}\sum\limits_{i=1}^{n}(y_{i}-\hat{\mu}_{i(q)})(\hat{\mu}_{i(q)}-\tilde{\mu}_{i(q)})
		+2\underset{1<q<r<M}{\sum{\sum}}\omega_{q}\omega_{r}\sum\limits_{i=1}^{n}(y_{i}-\hat{\mu}_{i(q)})(\hat{\mu}_{i(r)}-\tilde{\mu}_{i(r)})\bigg] \notag\\
		=&\dfrac{2}{n}\bigg\{\sum\limits_{q=1}^{M}\omega_{q}^{2}\sum\limits_{i=1}^{n}(y_{i}-x_{i(q)}'\hat{\Theta}_{q})[x_{i(q)}'E_{*}(\hat{\Theta}_{q}-\tilde{\Theta}_{q}^{*})]
		+2\underset{1<q<r<M}{\sum{\sum}}\omega_{q}\omega_{r}\sum\limits_{i=1}^{n}(y_{i}-x_{i(q)}'\hat{\Theta}_{q}) \notag\\
		&[x_{i(r)}'E_{*}(\hat{\Theta}_{r}-\tilde{\Theta}_{r}^{*})]\bigg\} \notag\\
		=&4\underset{1<q<r<M}{\sum{\sum}}\omega_{q}\omega_{r}\bigg[\dfrac{1}{n}\sum\limits_{i=1}^{n}(y_{i}-x_{i(q)}'\hat{\Theta}_{q})x_{i(r)}'\bigg]E_{*}(\hat{\Theta}_{r}-\tilde{\Theta}_{r}^{*}) \notag\\
		=&4\underset{1<q<r<M}{\sum{\sum}}\omega_{q}\omega_{r}O_{p}(\dfrac{1}{\sqrt{n}})O_{p}(\dfrac{1}{\sqrt{m}}) \notag\\
		=&O_{p}(\dfrac{1}{\sqrt{nm}}). \label{th0.4}
	\end{align}
	The second to last equality holds because $E||\sum_{i=1}^{n}(y_{i}-x_{i(q)}'\hat{\Theta}_{q})x_{i(r)}'||^{2}
	=E||X_{r}'(I_{n}-H_{q})Y||^{2}=E||X_{r}'(I_{n}-H_{q})e||^{2}\leq\sigma^{2}\text{tr}(X_{r}'X_{r})=O(n)$ and $|E_{*}(\hat{\Theta}_{r}-\tilde{\Theta}_{r}^{*})|\leq[E_{*}(\hat{\Theta}_{r}-\tilde{\Theta}_{r}^{*})^{2}]^{1/2}=O_{p}(1/\sqrt{m})$. Combining (\ref{th0.2})-(\ref{th0.4}), we complete the proof of Theorem 1.
\end{proof}
\begin{lemma}
	Assuming that Conditions (C.1)-(C.6) are satisfied, we have
	\begin{align}
		\max_{1\leq q\leq M}\Big|\Big|(\dfrac{m}{n}X_{(q)}'X_{(q)})(X_{(q)}^{*'}X_{(q)}^{*})^{-1}-I_{k_{q}}\Big|\Big|_{F}\stackrel{p_{*}}{\rightarrow}0. \notag
	\end{align}
\end{lemma}
\begin{proof}
	By Chebyshev's inequality, for all $\varepsilon>0$, we have
	\begin{align}
		&P_{*}\bigg\{\max_{1\leq q\leq M}\Big|\Big|\dfrac{1}{m}X_{(q)}^{*'}X_{(q)}^{*}-\dfrac{1}{n}X_{(q)}'X_{(q)}\Big|\Big|_{F}>\varepsilon\bigg\} \notag\\
		=&P_{*}\bigg\{\max_{1\leq q\leq M}\Big|\Big|\dfrac{1}{m}X_{(q)}^{'}\Pi X_{(q)}-\dfrac{1}{n}X_{(q)}'X_{(q)}\Big|\Big|_{F}>\varepsilon\bigg\} \notag\\
		\leq&\sum\limits_{q=1}^{M}P_{*}\bigg\{\Big|\Big|\dfrac{1}{m}X_{(q)}^{'}\Pi X_{(q)}-\dfrac{1}{n}X_{(q)}'X_{(q)}\Big|\Big|_{F}>\varepsilon\bigg\} \notag\\
		\leq&\sum\limits_{q=1}^{M}\dfrac{1}{\varepsilon^{2}}E_{*}\bigg[\Big|\Big|\dfrac{1}{m}X_{(q)}^{'}\Pi X_{(q)}-\dfrac{1}{n}X_{(q)}'X_{(q)}\Big|\Big|_{F}^{2}\bigg] \notag\\
		=&\dfrac{1}{\varepsilon^{2}}\sum\limits_{q=1}^{M}\sum\limits_{i=1}^{k_{q}}\sum\limits_{j=1}^{k_{q}}E_{*}\bigg(\dfrac{1}{m}\sum\limits_{l=1}^{n}\pi_{l}x_{li(q)}x_{lj(q)}-\dfrac{1}{n}\sum\limits_{l=1}^{n}x_{li(q)}x_{lj(q)}\bigg)^{2} \notag\\
		=&\dfrac{1}{\varepsilon^{2}}\sum\limits_{q=1}^{M}\sum\limits_{i=1}^{k_{q}}\sum\limits_{j=1}^{k_{q}}Var_{*}\bigg(\dfrac{1}{m}\sum\limits_{l=1}^{n}\pi_{l}x_{li(q)}x_{lj(q)}\bigg) \notag\\
		=&\dfrac{1}{m^{2}\varepsilon^{2}}\sum\limits_{q=1}^{M}\sum\limits_{i=1}^{k_{q}}\sum\limits_{j=1}^{k_{q}}\bigg[\sum\limits_{l=1}^{n}Var_{*}(\pi_{l})x_{li(q)}^{2}x_{lj(q)}^{2}+\underset{r\neq s}{\sum\sum}Cov_{*}(\pi_{r},\pi_{s})x_{ri(q)}x_{rj(q)}x_{si(q)}x_{sj(q)}\bigg] \notag\\
		=&\dfrac{1}{m^{2}\varepsilon^{2}}\sum\limits_{q=1}^{M}\bigg\{\dfrac{m(n-1)}{n^{2}}\sum\limits_{i=1}^{k_{q}}\sum\limits_{j=1}^{k_{q}}\sum\limits_{l=1}^{n}x_{li(q)}^{2}x_{lj(q)}^{2}
		-\dfrac{m}{n^{2}}\sum\limits_{i=1}^{k_{q}}\sum\limits_{j=1}^{k_{q}}\underset{r\neq s}{\sum\sum}x_{ri(q)}x_{rj(q)}x_{si(q)}x_{sj(q)}\bigg\} \notag\\
		\leq&\sum\limits_{q=1}^{M}\bigg\{\dfrac{n-1}{mn^{2}\varepsilon^{2}}\sum\limits_{i=1}^{k_{q}}\sum\limits_{j=1}^{k_{q}}\sum\limits_{l=1}^{n}x_{li(q)}^{2}x_{lj(q)}^{2}
		+\dfrac{1}{mn^{2}\varepsilon^{2}}\bigg|\sum\limits_{i=1}^{k_{q}}\sum\limits_{j=1}^{k_{q}}\underset{r\neq s}{\sum\sum}x_{ri(q)}x_{rj(q)}x_{si(q)}x_{sj(q)}\bigg|\bigg\} \notag\\
		\leq&\sum\limits_{q=1}^{M}\bigg\{\dfrac{n-1}{mn^{2}\varepsilon^{2}}\sum\limits_{i=1}^{k_{q}}\sum\limits_{j=1}^{k_{q}}\sum\limits_{l=1}^{n}\max_{i,j,l}\{x_{li(q)}^{2}\}\max_{i,j,l}\{x_{lj(q)}^{2}\} \notag\\
		&+\dfrac{1}{mn^{2}\varepsilon^{2}}\sum\limits_{i=1}^{k_{q}}\sum\limits_{j=1}^{k_{q}}\underset{r,s}{\sum\sum}\max_{i,j,r\neq s}|x_{ri(q)}|\max_{i,j,r,s}|x_{rj(q)}|\max_{i,j,r,s}|x_{si(q)}|\max_{i,j,r,s}|x_{sj(q)}|\bigg\} \notag\\
		=&\dfrac{n-1}{mn^{2}\varepsilon^{2}}O(Mk_{m^{*}}^{2}n)+\dfrac{1}{mn^{2}\varepsilon^{2}}O(Mk_{M^{*}}^{2}n(n-1)) \notag\\
		=&O(\dfrac{n-1}{mn}Mk_{M^{*}}^{2})\rightarrow0. \notag
	\end{align}
	So combining the fact that $\max_{q}\big\{\lambda_{\max}\{(X_{(q)}^{*'}X_{(q)}^{*}/m)^{-1}\}\big\}=1/\min_{q}\big\{\lambda_{\min}\{X_{(q)}^{*'}X_{(q)}^{*}/m\}\big\}
	\leq1/C_{3}$, we obtain,
	\begin{align}
		&\max_{1\leq q\leq M}\Big|\Big|(\dfrac{m}{n}X_{(q)}'X_{(q)})(X_{(q)}^{*'}X_{(q)}^{*})^{-1}-I_{k_{q}}\Big|\Big|_{F} \notag\\
		=&\max_{1\leq q\leq M}\Big|\Big|(\dfrac{1}{n}X_{(q)}'X_{(q)}-\dfrac{1}{m}X_{(q)}^{*'}X_{(q)}^{*})(\dfrac{1}{m}X_{(q)}^{*'}X_{(q)}^{*})^{-1}\Big|\Big|_{F} \notag\\
		\leq&\max_{1\leq q\leq M}\Big|\Big|\dfrac{1}{n}X_{(q)}'X_{(q)}-\dfrac{1}{m}X_{(q)}^{*'}X_{(q)}^{*}\Big|\Big|_{F}\max_{1\leq q\leq M}\Big|\Big|(\dfrac{1}{m}X_{(q)}^{*'}X_{(q)}^{*})^{-1}\Big|\Big|_{2} \notag\\
		=&o_{p_{*}}(1). \notag
	\end{align}
	Thus, we complete the proof of Lemma 1.
\end{proof}
From Lemma 1, we see that for $1\leq q\leq M$,
\begin{align}
	(\dfrac{m}{n}X_{(q)}'X_{(q)})(X_{(q)}^{*'}X_{(q)}^{*})^{-1}-I_{k_{q}}\triangleq\mathfrak{O}_{k_{q}} \notag
\end{align}
is a $k_{q}\times k_{q}$ matrix with all elements being $o_{p_{*}}(1)$. Then we find
\begin{align}
	(X_{(q)}^{*'}X_{(q)}^{*})^{-1}=(n/m)(X_{(q)}'X_{(q)})^{-1}+(n/m)(X_{(q)}'X_{(q)})^{-1}\mathfrak{O}_{k_{q}}. \notag
\end{align}
Under Conditions (C.2) and (C.3), we can also see that
\begin{align}
	\big|\big|(\dfrac{m}{n}X_{(q)}'X_{(q)})(X_{(q)}^{*'}X_{(q)}^{*})^{-1}-I_{k_{q}}\big|\big|_{F}
	\leq\dfrac{C_{2}}{C_{1}+C_{3}}. \label{le1.1}
\end{align}

\begin{lemma}
	Under Conditions (C.1)-(C.4), we have
	\begin{align}
		\Gamma_{n,m}(\omega)\leq\dfrac{||Y-\hat{\mu}(\omega)||^{2}}{n}+C_{M^{*}}\dfrac{\hat{e}_{R^{*}}^{2}}{m}\text{tr}\big\{H(\omega)\big\}+o_{p}(1), \notag
	\end{align}
	where $C_{M^{*}}$ is a positive constant, and $\hat{e}_{R^{*}}^{2}=\max\limits_{1\leq q\leq M,1\leq i\leq n}\{\hat{e}_{q,i}^{2}\}$ indicates the maximum of squared residuals.
\end{lemma}
\begin{proof}
	Recalling Theorem 1, we can write
	\begin{align}
		\Gamma_{n,m}(\omega)
		=&E_{*}\dfrac{||Y-\tilde{\mu}(\omega)||^{2}}{n}\notag\\
		=&\dfrac{||Y-\hat{\mu}(\omega)||^{2}}{n}+E_{*}\dfrac{||\hat{\mu}(\omega)-\tilde{\mu}(\omega)||^{2}}{n}+E_{*}\dfrac{2\langle Y-\hat{\mu}(\omega), \hat{\mu}(\omega)-\tilde{\mu}(\omega)\rangle}{n}. \label{le2.1}
	\end{align}
	So we need only to consider the second term and third terms of (\ref{le2.1}). By Cauchy-Schwarz inequality, we obtain
	\begin{align}
		E_{*}\bigg[\dfrac{||\hat{\mu}(\omega)-\tilde{\mu}(\omega)||^{2}}{n}\bigg]
		=&\dfrac{1}{n}E_{*}\big|\big|X\hat{\Theta}(\omega)-X\tilde{\Theta}(\omega)\big|\big|^{2} \notag\\
		=&\dfrac{1}{n}E_{*}\Big|\Big|\sum\limits_{q=1}^{M}\omega_{q}\big(X_{(q)}\tilde{\Theta}^{*}_{q}-X_{(q)}\hat{\Theta}_{q}\big)\Big|\Big|^{2} \notag\\
		\leq&\dfrac{1}{n}E_{*}\Big(\sum\limits_{q=1}^{M}\omega_{q}\Big|\Big|X_{(q)}\tilde{\Theta}^{*}_{q}-X_{(q)}\hat{\Theta}_{q}\Big|\Big|\Big)^{2} \notag\\
		\leq&\dfrac{1}{n}\sum\limits_{q=1}^{M}\omega_{q}E_{*}\Big|\Big|X_{(q)}\tilde{\Theta}^{*}_{q}-X_{(q)}\hat{\Theta}_{q}\Big|\Big|^{2}. \label{le2.2}
	\end{align}
	From Lemma 1, (\ref{le1.1}) and Conditions (C.2) and (C.3), we have
	\begin{align}
		&E_{*}\Big|\Big|X_{(q)}\tilde{\Theta}^{*}_{q}-X_{(q)}\hat{\Theta}_{q}\Big|\Big|^{2} \notag\\
		=&E_{*}\Big|\Big|X_{(q)}(X_{(q)}^{*'}X_{(q)}^{*})^{-1}X_{(q)}^{*'}\big(Y^{*}-X_{(q)}^{*'}\hat{\Theta}_{q}\big)\Big|\Big|^{2} \notag\\
		=&\dfrac{n^{2}}{m^{2}}E_{*}\Big|\Big|X_{(q)}\big[(X_{(q)}'X_{(q)})^{-1}+(X_{(q)}'X_{(q)})^{-1}\mathfrak{O}_{k_{q}}\big]X_{(q)}'\Pi\big(Y-X_{(q)}\hat{\Theta}_{q}\big)\Big|\Big|^{2} \notag\\
		=&\dfrac{n^{2}}{m^{2}}E_{*}\Big|\Big|X_{(q)}(X_{(q)}'X_{(q)})^{-1}\big[I_{k_{q}}+\mathfrak{O}_{k_{q}}\big]X_{(q)}'X_{(q)}(X_{(q)}'X_{(q)})^{-1}X_{(q)}'\Pi\hat{e}_{q}\Big|\Big|^{2} \notag\\
		\leq&\dfrac{n^{2}}{m^{2}}E_{*}\bigg(\big|\big|H_{q}\Pi\hat{e}_{q}\big|\big|+\Big|\Big|X_{(q)}(X_{(q)}'X_{(q)})^{-1}\mathfrak{O}_{k_{q}}X_{(q)}'H_{q}\Pi\hat{e}_{q}\Big|\Big|\bigg)^{2} \notag\\
		\leq&\dfrac{n^{2}}{m^{2}}\bigg(2E_{*}\big|\big|H_{q}\Pi\hat{e}_{q}\big|\big|_{F}^{2}+2E_{*}\Big|\Big|X_{(q)}(X_{(q)}'X_{(q)})^{-1}\mathfrak{O}_{k_{q}}X_{(q)}'H_{q}\Pi\hat{e}_{q}\Big|\Big|_{F}^{2}\bigg) \notag\\
		\leq&\dfrac{n^{2}}{m^{2}}\bigg(2E_{*}\big|\big|H_{q}\Pi\hat{e}_{q}\big|\big|_{F}^{2}+2E_{*}\Big|\Big|X_{(q)}(X_{(q)}'X_{(q)})^{-1}\big|\big|^{2}_{2}\big|\big|\mathfrak{O}_{k_{q}}\Big|\Big|^{2}_{2}\big|\big|X_{(q)}'\big|\big|^{2}_{2}\big|\big|H_{q}\Pi\hat{e}_{q}\big|\big|_{F}^{2}\bigg) \notag\\
		\leq&2\Big[1+\dfrac{C^{2}_{2}}{C_{1}(C_{1}+C_{3})}\Big]\dfrac{n^{2}}{m^{2}}E_{*}\big|\big|H_{q}\Pi\hat{e}_{q}\big|\big|_{F}^{2} \notag\\
		=&C_{M^{*}}\dfrac{n^{2}}{m^{2}}\text{tr}\Big\{E_{*}\big(\Pi\hat{e}_{q}\hat{e}_{q}'\Pi\big)H_{q}\Big\}. \label{le2.3}
	\end{align}
	Noting that $\hat{e}_{q}=(\hat{e}_{q,1},\cdots,\hat{e}_{q,n})'$, we find
	\begin{align}
		E_{*}\big(\Pi\hat{e}_{q}\hat{e}_{q}'\Pi\big)
		=&
		E_{*}
		\begin{pmatrix}
			\pi_{1}^{2}\hat{e}_{q,1}^{2} &  \pi_{1}\pi_{2}\hat{e}_{q,1}\hat{e}_{q,2} & \cdots &  \pi_{1}\pi_{n}\hat{e}_{q,1}\hat{e}_{q,n}\\
			\pi_{2}\pi_{1}\hat{e}_{q,2}\hat{e}_{q,1} & \pi_{2}^{2}\hat{e}_{q,2}^{2} & \cdots &  \pi_{2}\pi_{n}\hat{e}_{q,2}\hat{e}_{q,n}\\
			\vdots & \vdots & \ddots & \vdots \\
			\pi_{n}\pi_{1}\hat{e}_{q,n}\hat{e}_{q,1} & \pi_{n}\pi_{2}\hat{e}_{q,n}\hat{e}_{q,2} & \cdots & \pi_{n}^{2}\hat{e}_{q,n}^{2}
		\end{pmatrix} \notag\\
		=&
		\begin{pmatrix}
			\frac{m(m+n-1)}{n^{2}}\hat{e}_{q,1}^{2} & \frac{m(m-1)}{n^{2}}\hat{e}_{q,1}\hat{e}_{q,2} & \cdots & \frac{m(m-1)}{n^{2}}\hat{e}_{q,1}\hat{e}_{q,n}\\
			\frac{m(m-1)}{n^{2}}\hat{e}_{q,2}\hat{e}_{q,1} & \frac{m(m+n-1)}{n^{2}}\hat{e}_{q,2}^{2} & \cdots & \frac{m(m-1)}{n^{2}}\hat{e}_{q,2}\hat{e}_{q,n}\\
			\vdots & \vdots & \ddots & \vdots \\
			\frac{m(m-1)}{n^{2}}\hat{e}_{q,n}\hat{e}_{q,1} & \frac{m(m-1)}{n^{2}}\hat{e}_{q,n}\hat{e}_{q,2} & \cdots & \frac{m(m+n-1)}{n^{2}}\hat{e}_{q,n}^{2}
		\end{pmatrix} \notag\\
		=&\dfrac{m}{n}\text{diag}\{\hat{e}_{q,1}^{2}\cdots\hat{e}_{q,n}^{2}\}+\dfrac{m(m-1)}{n^{2}}\hat{e}_{q}\hat{e}_{q}', \notag
	\end{align}
	and
	\begin{align}
		\hat{e}_{q}\hat{e}_{q}'X_{(q)}
		=&A_{q}YY'A_{q}X_{(q)}=\big[I_{n}-X_{(q)}(X_{(q)}'X_{(q)})^{-1}X_{(q)}'\big]YY'\big[I_{n}-X_{(q)}(X_{(q)}'X_{(q)})^{-1}X_{(q)}'\big]X_{(q)}=0. \notag
	\end{align}
	It follows that
	\begin{align}
		\text{tr}\Big\{E_{*}\big(\Pi\hat{e}_{q}\hat{e}_{q}'\Pi\big)H_{q}\Big\}
		=\dfrac{m}{n}\text{tr}\Big\{\text{diag}\big\{\hat{e}_{q,1}^{2}\cdots\hat{e}_{q,n}^{2}\big\}H_{q}\Big\}
		\leq\dfrac{m}{n}\hat{e}_{R^{*}}^{2}\text{tr}\big\{H_{q}\big\}. \notag
	\end{align}
	Combining this and (\ref{le2.3}), we obtain
	\begin{align}
		E_{*}\Big|\Big|X_{(q)}\tilde{\Theta}^{*}_{q}-X_{(q)}\hat{\Theta}_{q}\Big|\Big|^{2}
		\leq&C_{M^{*}}\dfrac{n}{m}\hat{e}_{R^{*}}^{2}\text{tr}\big\{H_{q}\big\}. \notag
	\end{align}
	Thus, by (\ref{le2.2})
	\begin{align}
		E_{*}\bigg[\dfrac{||\hat{\mu}(\omega)-\tilde{\mu}(\omega)||^{2}}{n}\bigg]
		\leq C_{M^{*}}\dfrac{\hat{e}_{R^{*}}^{2}}{m}\text{tr}\big\{H(\omega)\big\}. \label{le2.4}
	\end{align}
	For the third term of (\ref{le2.1}), by Cauchy-Schwarz inequality, we have
	\begin{align}
		&\bigg|E_{*}\bigg[\dfrac{2\langle Y-\hat{\mu}(\omega), \hat{\mu}(\omega)-\tilde{\mu}(\omega)\rangle}{n}\bigg]\bigg| \notag\\
		\leq&\dfrac{2}{n}\bigg|\big[Y-\hat{\mu}(\omega)\big]'E_{*}\big[\hat{\mu}(\omega)-\tilde{\mu}(\omega)\big]\bigg| \notag\\
		=&\dfrac{2}{n}\bigg|\big[Y-\hat{\mu}(\omega)\big]'\sum\limits_{q=1}^{M}\omega_{q}X_{(q)}E_{*}\bigg[(X_{q,m}^{*'}X_{q,m}^{*})^{-1}X_{q,m}^{*'}Y^{*}-(X_{(q)}'X_{(q)})^{-1}X_{(q)}'Y\bigg]\bigg| \notag\\
		=&\dfrac{2}{n}\bigg|\big[Y-\hat{\mu}(\omega)\big]'\sum\limits_{q=1}^{M}\omega_{q}E_{*}\Big[\dfrac{n}{m}X_{(q)}(X_{(q)}'X_{(q)})^{-1}\mathfrak{O}_{k_{q}}X_{(q)}'\Pi Y\Big]\bigg|. \label{le2.5}
	\end{align}
	Noting that $E||e||^{2}=n\sigma^{2}=O(n)$, we have
	\begin{align}
		||Y||^{2}=||\mu+e||^{2}\leq2||\mu||^{2}+2||e||^{2}=O_{p}(n). \label{le2.6}
	\end{align}
	From Lemma 1, we find
	\begin{align}
		&\sum\limits_{q=1}^{M}E_{*}\big|\big|\mathfrak{O}_{k_{q}}\big|\big|_{F}^{2}  \notag\\
		\leq&\sum\limits_{q=1}^{M}E_{*}\big|\big|\dfrac{1}{n}X_{(q)}'X_{(q)}-\dfrac{1}{m}X_{(q)}^{*'}X_{(q)}^{*}\big|\big|_{F}^{2}\big|\big|(\dfrac{1}{m}X_{(q)}^{*'}X_{(q)}^{*})^{-1}\big|\big|_{2}^{2} \notag\\
		\leq&\dfrac{1}{C_{3}}\sum\limits_{q=1}^{M}E_{*}\big|\big|\dfrac{1}{n}X_{(q)}'X_{(q)}-\dfrac{1}{m}X_{(q)}^{*'}X_{(q)}^{*}\big|\big|_{F}^{2}  \notag\\
		=&O(\dfrac{n-1}{mn}Mk_{M^{*}}^{2}). \label{le2.7}
	\end{align}
	So from the expectation inequality, (\ref{le2.6}) and (\ref{le2.7}), we obtain
	\begin{align}
		&\bigg|\bigg|\sum\limits_{q=1}^{M}\omega_{q}E_{*}\Big[\dfrac{n}{m}X_{(q)}(X_{(q)}'X_{(q)})^{-1}\mathfrak{O}_{k_{q}}X_{(q)}'\Pi Y\Big]\bigg|\bigg| \notag\\
		\leq&\dfrac{n}{m}\sum\limits_{q=1}^{M}\omega_{q}E_{*}\Big|\Big|X_{(q)}(X_{(q)}'X_{(q)})^{-1}\mathfrak{O}_{k_{q}}X_{(q)}'\Pi Y\Big|\Big| \notag\\
		\leq&\dfrac{n}{m}\sum\limits_{q=1}^{M}\omega_{q}E_{*}\Big[\big|\big|X_{(q)}(X_{(q)}'X_{(q)})^{-1}\mathfrak{O}_{k_{q}}X_{(q)}'\big|\big|_{F}\bm\cdot\big|\big|\Pi Y\big|\big|\Big] \notag\\
		\leq&\dfrac{n}{m}\sum\limits_{q=1}^{M}\omega_{q}\big|\big|X_{(q)}(X_{(q)}'X_{(q)})^{-1}\big|\big|_{2}\big|\big|X_{(q)}'\big|\big|_{2}E_{*}\Big[\big|\big|\mathfrak{O}_{k_{q}}\big|\big|_{F}\bm\cdot\big|\big|\Pi Y\big|\big|\Big] \notag\\
		\leq&\dfrac{nC_{2}}{mC_{1}}\sum\limits_{q=1}^{M}\omega_{q}E_{*}\Big[\big|\big|\mathfrak{O}_{k_{q}}\big|\big|_{F}\bm\cdot\big|\big|\Pi Y\big|\big|\Big] \notag\\
		\leq&\dfrac{nC_{2}}{mC_{1}}\sum\limits_{q=1}^{M}\omega_{q}\Big(E_{*}\big|\big|\mathfrak{O}_{k_{q}}\big|\big|_{F}^{2}\Big)^{1/2}\Big(E_{*}\big|\big|\Pi Y\big|\big|^{2}\Big)^{1/2} \notag\\
		\leq&\dfrac{nC_{2}}{mC_{1}}\Big(\sum\limits_{q=1}^{M}E_{*}\big|\big|\mathfrak{O}_{k_{q}}\big|\big|_{F}^{2}\Big)^{1/2}\sum\limits_{q=1}^{M}\omega_{q}\Big(\dfrac{m(m+n-1)}{n^{2}}||Y||^{2}\Big)^{1/2} \notag\\
		=&O_{p}\Big(\dfrac{n^{1/2}(m+n)^{1/2}}{m}M^{3/2}k_{M^{*}}\Big). \notag
	\end{align}
	From this, by Cauchy-Schwarz inequality and eigenvalue inequality, it follows that
	\begin{align}
		&\bigg|\big[Y-\hat{\mu}(\omega)\big]'\sum\limits_{q=1}^{M}\omega_{q}E_{*}\Big[\dfrac{n}{m}X_{(q)}(X_{(q)}'X_{(q)})^{-1}\mathfrak{O}_{k_{q}}X_{(q)}'\Pi Y\Big]\bigg| \notag\\
		\leq&\big|\big|A(\omega)Y\big|\big|\bm\cdot\bigg|\bigg|\sum\limits_{q=1}^{M}\omega_{q}E_{*}\Big[\dfrac{n}{m}X_{(q)}(X_{(q)}'X_{(q)})^{-1}\mathfrak{O}_{k_{q}}X_{(q)}'\Pi Y\Big]\bigg|\bigg| \notag\\
		\leq&O_{p}\Big(\dfrac{n(m+n)^{1/2}}{m}M^{3/2}k_{M^{*}}\Big). \label{le2.8}
	\end{align}
	Hence, from (\ref{le2.5}) and (\ref{le2.8}), we have
	\begin{align}
		\bigg|E_{*}\bigg[\dfrac{2\langle Y-\hat{\mu}(\omega), \hat{\mu}(\omega)-\tilde{\mu}(\omega)\rangle}{n}\bigg]\bigg|
		=O_{p}\Big(\dfrac{(m+n)^{1/2}}{m}M^{3/2}k_{M^{*}}\Big)=o_{p}(1). \label{le2.9}
	\end{align}
	Combining (\ref{le2.1}), (\ref{le2.4}) and (\ref{le2.9}), we see that Lemma 2 is true. This completes the proof.
\end{proof}

\addcontentsline{toc}{subsection}{Proof of Theorem 2}
\begin{proof}[\textbf{Proof of Theorem 2}]
	It is clear that
	\begin{align}
		\dfrac{1}{n}||Y-\hat{\mu}(\omega)||^{2}=\dfrac{1}{n}\bigg[L_{n}(\omega)+2\mu'A(\omega)e-2e'H(\omega)e+||e||^{2}\bigg]. \notag
	\end{align}
	To establish the optimality of the bootstrap criterion, from (\ref{le2.1}), (\ref{le2.4}) and (\ref{le2.9}), we note that
	\begin{align}
		\Gamma_{n,m}(\omega)
		=&\dfrac{||Y-\hat{\mu}(\omega)||^{2}}{n}+E_{*}\dfrac{||\hat{\mu}(\omega)-\tilde{\mu}(\omega)||^{2}}{n}+E_{*}\dfrac{2\langle Y-\hat{\mu}(\omega), \hat{\mu}(\omega)-\tilde{\mu}(\omega)\rangle}{n} \notag\\
		\leq&\dfrac{1}{n}\bigg[L_{n}(\omega)+2\mu'A(\omega)e-2e'H(\omega)e+||e||^{2}\bigg]+C_{M^{*}}\dfrac{\hat{e}_{R^{*}}^{2}}{m}\text{tr}\big\{H(\omega)\big\}+O_{p}\Big(\dfrac{(m+n)^{1/2}}{m}M^{3/2}k_{M^{*}}\Big) \notag\\
		\triangleq&\Phi_{1}(\omega)+\dfrac{1}{n}||e||^{2}, \notag
	\end{align}
	and
	\begin{align}
		\Gamma_{n,m}(\omega)
		\geq&E_{*}\dfrac{||Y-\hat{\mu}(\omega)||^{2}}{n}+E_{*}\dfrac{2\langle Y-\hat{\mu}(\omega), \hat{\mu}(\omega)-\tilde{\mu}(\omega)\rangle}{n} \notag\\
		=&\dfrac{1}{n}\bigg[L_{n}(\omega)+2\mu'A(\omega)e-2e'H(\omega)e+||e||^{2}\bigg]+O_{p}\Big(\dfrac{(m+n)^{1/2}}{m}M^{3/2}k_{M^{*}}\Big) \notag\\
		\triangleq&\Phi_{2}(\omega)+\dfrac{1}{n}||e||^{2}. \notag
	\end{align}
	Then, we have
	\begin{align}
		\dfrac{\Phi_{2}(\omega)}{R_{n}(\omega)/n}-1\leq\dfrac{\Gamma_{n,m}(\omega)-||e||^{2}/n}{R_{n}(\omega)/n}-1\leq\dfrac{\Phi_{1}(\omega)}{R_{n}(\omega)/n}-1. \notag
	\end{align}
	Since $||e||^{2}/n$ is unrelated to $\omega$, Theorem 2 is valid if
	\begin{align}
		\sup_{\omega\in\textbf{H}_{n}}\bigg|\dfrac{\Gamma_{n,m}(\omega)-||e||^{2}/n}{R_{n}(\omega)/n}-1\bigg|\stackrel{p}{\longrightarrow}0, \notag
	\end{align}
	which can be verified through
	\begin{align}
		\sup_{\omega\in\textbf{H}_{n}}\bigg|\dfrac{\Phi_{1}(\omega)}{R_{n}(\omega)/n}-1\bigg|\stackrel{p}{\longrightarrow}0\ \text{and}\
		\sup_{\omega\in\textbf{H}_{n}}\bigg|\dfrac{\Phi_{2}(\omega)}{R_{n}(\omega)/n}-1\bigg|\stackrel{p}{\longrightarrow}0. \notag
	\end{align}
	By Condition (C.6), we have $n(m+n)^{1/2}M^{3/2}k_{M^{*}}\big/m\xi_{n}\rightarrow0$. So in order to prove Theorem 2, we need only to verify that,
	\begin{align}
		\sup_{\omega\in\textbf{H}_{n}}&\bigg|\dfrac{\mu'A(\omega)e}{R_{n}(\omega)}\bigg|\stackrel{p}{\longrightarrow}0, \label{th1.1}\\
		\sup_{\omega\in\textbf{H}_{n}}&\bigg|\dfrac{e'H(\omega)e-\sigma^{2}\text{tr}\big\{H(\omega)\big\}}{R_{n}(\omega)}\bigg|\stackrel{p}{\longrightarrow}0, \label{th1.2}\\
		\sup_{\omega\in\textbf{H}_{n}}&\bigg|\dfrac{2\sigma^{2}\text{tr}\{H(\omega)\}-(C_{M^{*}}n/m)\hat{e}_{R^{*}}^{2}\text{tr}\big\{H(\omega)\big\}}{R_{n}(\omega)}\bigg|\stackrel{p}{\longrightarrow}0, \label{th1.3}
	\end{align}
	and
	\begin{align}
		\sup_{\omega\in\textbf{H}_{n}}&\bigg|\dfrac{L_{n}(\omega)}{R_{n}(\omega)}-1\bigg|\stackrel{p}{\longrightarrow}0, \label{th1.4}
	\end{align}
	while (\ref{th1.3}) holds if
	\begin{align}
		\sup_{\omega\in\textbf{H}_{n}}&\bigg|\dfrac{2\sigma^{2}\text{tr}\{H(\omega)\}-(C_{M^{*}}n/m)\sigma^{2}\text{tr}\big\{H(\omega)\big\}}{R_{n}(\omega)}\bigg|\stackrel{p}{\longrightarrow}0, \label{th1.5}
	\end{align}
	and
	\begin{align}
		\sup_{\omega\in\textbf{H}_{n}}&\bigg|\dfrac{(C_{M^{*}}n/m)\sigma^{2}\text{tr}\{H(\omega)\}
			-(C_{M^{*}}n/m)\hat{e}_{R^{*}}^{2}\text{tr}\big\{H(\omega)\big\}}{R_{n}(\omega)}\bigg|\stackrel{p}{\longrightarrow}0. \label{th1.6}
	\end{align}
	We first consider (\ref{th1.1}) and (\ref{th1.2}). Similar to \cite{wan2010least}, using triangle inequality, Bonferroni inequality, Chebyshev's inequality and Theorem 3 of \cite{whittle1960bounds}, we see that, for all $\varepsilon>0$,
	\begin{align}
		&P\bigg\{\sup_{\omega\in\textbf{H}_{n}}\dfrac{|\langle e,\big(I_{n}-H(\omega)\big)\mu\rangle|/n}{R_{n}(\omega)/n}>\varepsilon\bigg\} \notag\\
		\leq&P\bigg\{\sup_{\omega\in\textbf{H}_{n}}\sum\limits_{q=1}^{M}\omega_{q}|e'(I_{n}-H_{q})\mu|>\varepsilon\xi_{n}\bigg\} \notag\\
		=&P\bigg\{\max_{1\leq q\leq M}|e'(I_{n}-H_{q})\mu|>\varepsilon\xi_{n}\bigg\} \notag\\
		\leq&\sum\limits_{q=1}^{M}P\bigg\{|e'(I_{n}-H(\omega_{q}^{0}))\mu|>\varepsilon\xi_{n}\bigg\} \notag\\
		\leq&\sum\limits_{q=1}^{M}E\bigg\{\dfrac{\langle e,(I_{n}-H(\omega_{q}^{0}))\mu\rangle^{2G}}{(\varepsilon\xi_{n})^{2G}}\bigg\} \notag\\
		\leq&C_{2}'(\varepsilon\xi_{n})^{-2G}\sum\limits_{q=1}^{M}||(I_{n}-H(\omega_{q}^{0}))\mu||^{2G} \notag\\
		\leq&C_{2}'(\varepsilon\xi_{n})^{-2G}\sum\limits_{q=1}^{M}(R_{n}(\omega_{q}^{0}))^{G}\rightarrow 0, \notag
	\end{align}
	and
	\begin{align}
		&P\bigg\{\sup_{\omega\in\textbf{H}_{n}}\dfrac{|\sigma^{2}\text{tr}H(\omega)-\langle e,H(\omega)e\rangle|/n}{R_{n}(\omega)/n}>\varepsilon\bigg\} \notag\\
		\leq&\sum\limits_{q=1}^{M}P\bigg\{|\sigma^{2}\text{tr}H(\omega_{q}^{0})-\langle e,H(\omega_{q}^{0})e\rangle|>\varepsilon\xi_{n}\bigg\} \notag\\
		\leq&\sum\limits_{q=1}^{M}E\bigg\{\dfrac{(\sigma^{2}\text{tr}H(\omega_{q}^{0})-\langle e,H(\omega_{q}^{0})e\rangle)^{2G}}{(\varepsilon\xi_{n})^{2G}}\bigg\} \notag\\
		\leq&C_{3}'(\varepsilon\xi_{n})^{-2G}\sum\limits_{q=1}^{M}[\text{tr}H^{2}(\omega_{q}^{0})]^{G} \notag\\
		\leq&C_{3}'(\varepsilon\xi_{n})^{-2G}\sum\limits_{q=1}^{M}(R_{n}(\omega_{q}^{0}))^{G}\rightarrow 0, \notag
	\end{align}
	where $C_{2}'$ and $C_{3}'$ are positive constants. So (\ref{th1.1}) and (\ref{th1.2}) hold. \\
	\indent To prove (\ref{th1.4}), we need only to verify
	\begin{align}
		\sup_{\omega\in\textbf{H}_{n}}&\dfrac{|\mu A(\omega)H(\omega)e|}{R_{n}(\omega)}\stackrel{p}{\longrightarrow}0, \label{th1.7}
	\end{align}
	and
	\begin{align}
		\sup_{\omega\in\textbf{H}_{n}}&\dfrac{\big|||H(\omega)e||^{2}-\sigma^{2}\text{tr}H^{2}(\omega)\big|}{R_{n}(\omega)}\stackrel{p}{\longrightarrow}0. \label{th1.8}
	\end{align}
	It is seen that
	\begin{align}
		&P\bigg\{\sup_{\omega\in\textbf{H}_{n}}\dfrac{|\langle(I_{n}-H(\omega))\mu,H(\omega)e\rangle|}{R_{n}(\omega)}>\varepsilon\bigg\} \notag\\
		\leq&P\bigg\{\sup_{\omega\in\textbf{H}_{n}}\sum\limits_{q=1}^{M}\sum\limits_{r=1}^{M}\omega_{q}\omega_{r}|e'H_{q}(I_{n}-H_{r})\mu|>\varepsilon\xi_{n}\bigg\} \notag\\
		\leq&\sum\limits_{q=1}^{M}\sum\limits_{r=1}^{M}P\bigg\{|e'H(\omega_{q}^{0})(I_{n}-H(\omega_{r}^{0}))\mu|>\varepsilon\xi_{n}\bigg\} \notag\\
		\leq&\sum\limits_{q=1}^{M}\sum\limits_{r=1}^{M}E\bigg\{\dfrac{\langle(I_{n}-H(\omega_{r}^{0}))\mu,H(\omega_{q}^{0})e\rangle^{2G}}{(\varepsilon\xi_{n})^{2G}}\bigg\} \notag\\
		\leq&C_{4}'(\varepsilon\xi_{n})^{-2G}\sum\limits_{q=1}^{M}\sum\limits_{r=1}^{M}||H(\omega_{q}^{0})(I_{n}-H(\omega_{r}^{0}))||^{2G} \notag\\
		\leq&C_{4}'(\varepsilon\xi_{n})^{-2G}\sum\limits_{q=1}^{M}\sum\limits_{r=1}^{M}\lambda^{2G}_{\max}(H(\omega_{q}^{0}))||(I_{n}-H(\omega_{r}^{0}))||^{2G} \notag\\
		\leq&C_{4}'(\varepsilon\xi_{n})^{-2G}\sum\limits_{q=1}^{M}\sum\limits_{r=1}^{M}||(I_{n}-H(\omega_{r}^{0}))||^{2G} \notag\\
		\leq&C_{4}'(\varepsilon\xi_{n})^{-2G}\sum\limits_{r=1}^{M}(R_{n}(\omega_{r}^{0}))^{G}\rightarrow 0, \notag
	\end{align}
	and
	\begin{align}
		&P\bigg\{\sup_{\omega\in\textbf{H}_{n}}\dfrac{\big|||H(\omega)e||^{2}-\sigma^{2}\text{tr}H^{2}(\omega)\big|}{R_{n}(\omega)}>\varepsilon\bigg\} \notag\\
		\leq&\sum\limits_{q=1}^{M}\sum\limits_{r=1}^{M}P\bigg\{|e'H(\omega_{q}^{0})H(\omega_{r}^{0})e-\sigma^{2}\text{tr}\{H_{q}H_{r}\}|>\varepsilon\xi_{n}\bigg\} \notag\\
		\leq&\sum\limits_{q=1}^{M}\sum\limits_{r=1}^{M}E\bigg\{\dfrac{(\langle e,H(\omega_{q}^{0})H(\omega_{r}^{0})e\rangle-\sigma^{2}\text{tr}\{H_{q}H_{r}\})^{2G}}{(\varepsilon\xi_{n})^{2G}}\bigg\} \notag\\
		\leq&C_{5}'(\varepsilon\xi_{n})^{-2G}\sum\limits_{q=1}^{M}\sum\limits_{r=1}^{M}(\text{tr}\{H^{2}(\omega_{q}^{0})H^{2}(\omega_{r}^{0})\})^{G} \notag\\
		\leq&C_{5}'(\varepsilon\xi_{n})^{-2G}\sum\limits_{q=1}^{M}\sum\limits_{r=1}^{M}\lambda^{2G}_{\max}(H(\omega_{q}^{0}))(\text{tr}\{H^{2}(\omega_{r}^{0})\})^{G} \notag\\
		\leq&C_{5}'(\varepsilon\xi_{n})^{-2G}\sum\limits_{q=1}^{M}\sum\limits_{r=1}^{M}(\text{tr}\{H^{2}(\omega_{r}^{0})\})^{G} \notag\\
		\leq&C_{5}'(\varepsilon\xi_{n})^{-2G}\sum\limits_{r=1}^{M}(R_{n}(\omega_{r}^{0}))^{G}\rightarrow 0, \notag
	\end{align}
	where $C_{4}'$ and $C_{5}'$ are constants. Thus, (\ref{th1.7}) and (\ref{th1.8}) are correct, and hence (\ref{th1.4}) is proved.\\
	\indent By Condition (C.6), it is clear that
	\begin{align}
		&\sup_{\omega\in\textbf{H}_{n}}\bigg|\dfrac{2\sigma^{2}\text{tr}\{H(\omega)\}-(C_{M^{*}}n/m)\sigma^{2}\text{tr}\{H(\omega)\}\big\}}{R_{n}(\omega)}\bigg| \notag\\
		\leq&|2m-C_{M^{*}}n|\dfrac{\sigma^{2}\text{tr}\{H(\omega)\}}{m\xi_{n}} \notag\\
		\leq&\dfrac{2\sigma^{2}\text{tr}\{H(\omega)\}}{\xi_{n}}+\dfrac{C_{M^{*}}n\sigma^{2}\text{tr}\{H(\omega)\}}{m\xi_{n}} \notag\\
		\leq&\big(\dfrac{2k_{M^{*}}}{\xi_{n}}+\dfrac{C_{M^{*}}nk_{M^{*}}}{m\xi_{n}}\big)\sigma^{2} \notag\\
		=&o(1). \notag
	\end{align}
	So (\ref{th1.5}) holds. We now consider (\ref{th1.6}). It is clear that
	\begin{align}
		&\sup_{\omega\in\textbf{H}_{n}}\bigg|\dfrac{(C_{M^{*}}n/m)\sigma^{2}\text{tr}\big\{H(\omega)\big\}-(C_{M^{*}}n/m)\hat{e}_{R^{*}}^{2}\text{tr}\big\{H(\omega)\big\}}{R_{n}(\omega)}\bigg| \notag\\
		\leq&C_{M^{*}}\dfrac{n}{m}|\sigma^{2}-\hat{e}_{R^{*}}^{2}|\sup_{\omega\in\textbf{H}}\bigg|\dfrac{\text{tr}\{H(\omega)\}}{R_{n}(\omega)}\bigg|. \label{th1.9}
	\end{align}
	Let $\hat{e}_{q,i}^{2}$ be the $(i,i)$-th element of the diagonal matrix $\text{diag}\{\hat{e}_{q,1}^{2}\cdots\hat{e}_{q,n}^{2}\}$, which can be expanded as,
	\begin{align}
		\hat{e}_{q,i}^{2}
		=&(y_{i}-\hat{\mu}_{q,i})^{2} \notag\\
		=&\Big|\Big|\textbf{1}_{i}^{0'}Y-\textbf{1}_{i}^{0'}X_{(q)}(X_{(q)}'X_{(q)})^{-1}X_{(q)}'Y\Big|\Big|^{2} \notag\\
		=&Y'\Big[I_{n}-X_{(q)}(X_{(q)}'X_{(q)})^{-1}X_{(q)}'\Big]\textbf{1}_{i}^{0}\textbf{1}_{i}^{0'}\Big[I_{n}-X_{(q)}(X_{(q)}'X_{(q)})^{-1}X_{(q)}'\Big]Y \notag\\
		=&\mu'A_{q}D_{i}A_{q}\mu+2\mu'A_{q}D_{i}A_{q}e+e'A_{q}D_{i}A_{q}e, \notag
	\end{align}
	where $\textbf{1}_{i}^{0}$ is an $n\times1$ vector with the $i$th element being 1 and other elements being 0, $A_{q}=I_{n}-X_{(q)}(X_{(q)}'X_{(q)})^{-1}X_{(q)}'$ and $D_{i}=\textbf{1}_{i}^{0}\textbf{1}_{i}^{0'}$, $i=1,\cdots,n$. Hence,
	\begin{align}
		|\sigma^{2}-\hat{e}_{R^{*}}^{2}|
		=&\Big|\max\limits_{1\leq q\leq M,1\leq i\leq n}\Big\{\mu'A_{q}D_{i}A_{q}\mu+2\mu'A_{q}D_{i}A_{q}e+e'A_{q}D_{i}A_{q}e-\sigma^{2}\Big\}\Big| \notag\\
		\leq&\max\limits_{1\leq q\leq M,1\leq i\leq n}\Big\{\Big|\mu'A_{q}D_{i}A_{q}\mu+2\mu'A_{q}D_{i}A_{q}e+e'A_{q}D_{i}A_{q}e-\sigma^{2}\Big|\Big\} \notag\\
		\leq&\max\limits_{1\leq q\leq M,1\leq i\leq n}\Big\{\mu'A_{q}D_{i}A_{q}\mu\Big\}
		+2\max\limits_{1\leq q\leq M,1\leq i\leq n}\Big\{\big|\mu'A_{q}D_{i}A_{q}e\big|\Big\} \notag\\
		&+\max\limits_{1\leq q\leq M,1\leq i\leq n}\Big\{\big|e'A_{q}D_{i}A_{q}e-\sigma^{2}(1-h_{ii(q)})\big|\Big\}
		+\max\limits_{1\leq q\leq M,1\leq i\leq n}\Big\{\sigma^{2}h_{ii(q)}\Big\}. \label{th1.10}
	\end{align}
	From Condition (C.5) and Cauchy-Schwarz inequality, it is seen that
	\begin{align}
		\max_{1\leq q\leq M}\max_{1\leq i, j\leq n}|h_{ij(q)}|
		=&\max_{1\leq q\leq M}\max_{1\leq i, j\leq n}\Big\{\big|x_{i(q)}'(X_{(q)}'X_{(q)})^{-1}x_{j(q)}\big|\Big\} \notag\\
		\leq&\max_{1\leq q\leq M}\max_{1\leq i, j\leq n}\Big\{\big[x_{i(q)}'(X_{(q)}'X_{(q)})^{-1}x_{i(q)}\big]^{1/2}\big[x_{j(q)}'(X_{(q)}'X_{(q)})^{-1}x_{j(q)}\big]^{1/2}\Big\} \notag\\
		=&\Big(\max_{1\leq q\leq M}\max_{1\leq i\leq n}\Big\{ h_{ii(q)}\Big\}\Big)^{1/2}\Big(\max_{1\leq q\leq M}\max_{1\leq j\leq n}\Big\{ h_{jj(q)}\Big\}\Big)^{1/2} \notag\\
		=&O(\dfrac{k_{M^{*}}}{n}). \notag
	\end{align}
	Thus, using Condition (C.4), we have
	\begin{align}
		&\max\limits_{1\leq q\leq M,1\leq i\leq n}\Big\{\big|(1-h_{ii(q)})^{2}\mu_{i}^{2}\big|\Big\} \notag\\
		\leq&\bigg\{1+2\max\limits_{1\leq q\leq M,1\leq i\leq n}\Big\{|h_{ii(q)}|\Big\}+\max\limits_{1\leq q\leq M,1\leq i\leq n}\Big\{h_{ii(q)}^{2}\Big\}\bigg\}\max\limits_{1\leq i\leq n}\Big\{\mu_{i}^{2}\Big\} \notag\\
		=&\Big[1+O(\dfrac{k_{M^{*}}}{n})+O(\dfrac{k_{M^{*}}^{2}}{n^{2}})\Big]O(1) \notag\\
		=&O(1), \notag
	\end{align}
	and
	\begin{align}
		&\max\limits_{1\leq q\leq M,1\leq i\leq n}\Big\{\big|(1-h_{ii(q)})\sum\limits_{j\neq i}^{n}h_{ji(q)}\mu_{i}\mu_{j}\big|\Big\} \notag\\
		\leq&\sum\limits_{j\neq i}^{n}\max\limits_{1\leq q\leq M,1\leq i\leq n}\Big\{\big|1-h_{ii(q)}\big|\Big\}\max\limits_{1\leq q\leq M,1\leq i,j\leq n}\Big\{\big|h_{ji(q)}\big|\Big\}\max\limits_{1\leq i,j\leq n}\Big\{\big|\mu_{i}\mu_{j}\big|\Big\} \notag\\
		=&nO(1)O(\dfrac{k_{M^{*}}}{n})O(1) \notag\\
		=&O(k_{M^{*}}), \notag
	\end{align}
	and
	\begin{align}
		&\max\limits_{1\leq q\leq M,1\leq i\leq n}\Big\{\big|\underset{j\neq l\neq i}{\sum\sum}h_{ji(q)}h_{li(q)}\mu_{j}\mu_{l}\big|\Big\} \notag\\
		\leq&\underset{j\neq l\neq i}{\sum\sum}\max\limits_{1\leq q\leq M,1\leq i,j\leq n}\Big\{\big|h_{ji(q)}\big|\Big\}\max\limits_{1\leq q\leq M,1\leq i,l\leq n}\Big\{\big|h_{li(q)}\big|\Big\}\max\limits_{1\leq j,l\leq n}\Big\{\big|\mu_{j}\mu_{l}\big|\Big\} \notag\\
		\leq&n^{2}O(\dfrac{k_{M^{*}}}{n})O(\dfrac{k_{M^{*}}}{n})O(1) \notag\\
		=&O(k_{M^{*}}^{2}). \notag
	\end{align}
	Noting that,
	\begin{align}
		\mu'A_{q}D_{i}A_{q}\mu
		=&
		\mu'
		\begin{pmatrix}
			h_{1i(q)^{2}} & \cdots & -h_{1i(q)}(1-h_{ii(q)}) & \cdots & h_{1i(q)}h_{ni(q)}\\
			\vdots & \ddots  & \vdots & \ddots & \vdots \\
			-(1-h_{ii(q)})h_{1i(q)} & \cdots & (1-h_{ii(q)})^{2} & \cdots & -(1-h_{ii(q)})h_{ni(q)}\\
			\vdots & \ddots  & \vdots & \ddots & \vdots \\
			h_{ni(q)}h_{1i(q)} & \cdots & -h_{ni(q)}(1-h_{ii(q)}) & \cdots & h_{ni(q)}^{2}
		\end{pmatrix}
		\mu \notag\\
		=&(1-h_{ii(q)})^{2}\mu_{i}^{2}-2(1-h_{ii(q)})\sum\limits_{j\neq i}^{n}h_{ji(q)}\mu_{i}\mu_{j}+\underset{j\neq l\neq i}{\sum\sum}h_{ji(q)}h_{li(q)}\mu_{j}\mu_{l}, \notag
	\end{align}
	we obtain
	\begin{align}
		&\max\limits_{1\leq q\leq M,1\leq i\leq n}\Big\{\mu'A_{q}D_{i}A_{q}\mu\Big\} \notag\\
		\leq&\max\limits_{1\leq q\leq M,1\leq i\leq n}\Big\{(1-h_{ii(q)})^{2}\mu_{i}^{2}\Big\}
		+2\max\limits_{1\leq q\leq M,1\leq i\leq n}\Big\{\big|(1-h_{ii(q)})\sum\limits_{j\neq i}^{n}h_{ji(q)}\mu_{i}\mu_{j}\big|\Big\} \notag\\
		&+\max\limits_{1\leq q\leq M,1\leq i\leq n}\Big\{\big|\underset{j\neq l\neq i}{\sum\sum}h_{ji(q)}h_{li(q)}\mu_{j}\mu_{l}\big|\Big\} \notag\\
		=&O(1)+O(k_{M^{*}})+O(k_{M^{*}}^{2}) \notag\\
		=&O(k_{M^{*}}^{2}). \notag
	\end{align}
	Therefore, by Condition (C.6), it is found that
	\begin{align}
		&C_{M^{*}}\dfrac{n}{m}\max\limits_{1\leq q\leq M,1\leq i\leq n}\Big\{\mu'A_{q}D_{i}A_{q}\mu\Big\}\sup_{\omega\in\textbf{H}_{n}}\bigg|\dfrac{\text{tr}\{H(\omega)\}}{R_{n}}\bigg| \notag\\
		\leq&C_{M^{*}}\dfrac{n}{m}\max\limits_{1\leq q\leq M,1\leq i\leq n}\Big\{\mu'A_{q}D_{i}A_{q}\mu\Big\}\dfrac{k_{M^{*}}}{\xi_{n}} \notag\\
		=&\dfrac{O(k_{M^{*}}^{2})nk_{M^{*}}}{m\xi_{n}} \notag\\
		=&o(1). \label{th1.11}
	\end{align}
	Similarly,
	\begin{align}
		&C_{M^{*}}\dfrac{n}{m}\max\limits_{1\leq q\leq M,1\leq i\leq n}\Big\{\big|\mu'A_{q}D_{i}A_{q}e\big|\Big\}\sup_{\omega\in\textbf{H}_{n}}\bigg|\dfrac{\text{tr}\{H(\omega)\}}{R_{n}}\bigg| \notag\\
		\leq&C_{M^{*}}\dfrac{n}{m}\max\limits_{1\leq q\leq M,1\leq i\leq n}\Big\{\big|\mu'A_{q}D_{i}A_{q}e\big|\Big\}\dfrac{k_{M^{*}}}{\xi_{n}} \notag\\
		\leq&C_{M^{*}}\dfrac{n}{m}\max\limits_{1\leq q\leq M}\Big\{\big|\big|A_{q}e\big|\big|\Big\}\max\limits_{1\leq q\leq M,1\leq i\leq n}\Big\{\big|\big|D_{i}A_{q}\mu\big|\big|\Big\}\cdot\dfrac{k_{M^{*}}}{\xi_{n}} \notag\\
		\leq&C_{M^{*}}\dfrac{n}{m}\big|\big|e\big|\big|\Big[\max\limits_{1\leq q\leq M,1\leq i\leq n}\Big\{\mu'A_{q}D_{i}A_{q}\mu\Big\}\Big]^{1/2}\dfrac{k_{M^{*}}}{\xi_{n}} \notag\\
		=&\dfrac{O_{p}(\sqrt{n})O(k_{M^{*}})nk_{M^{*}}}{m\xi_{n}} \notag\\
		=&o_{p}(1). \label{th1.12}
	\end{align}
	Further, from Theorem 3 of \cite{whittle1960bounds} and $E(e'A_{q}D_{i}A_{q}e)=\sigma^{2}\text{tr}\big\{A_{q}D_{i}A_{q}\big\}=\sigma^{2}(1-h_{ii(q)})$, we see that for all $\varepsilon>0$,
	\begin{align}
		&P\bigg\{\dfrac{\max\limits_{1\leq q\leq M,1\leq i\leq n}\big|e'A_{q}D_{i}A_{q}e-\sigma^{2}(1-h_{ii(q)})\big|}{\xi_{n}^{1/2}}>\varepsilon\bigg\} \notag\\
		\leq&\sum\limits_{q=1}^{M}\sum\limits_{i=1}^{n}P\bigg\{\big|e'A_{q}D_{i}A_{q}e-\sigma^{2}(1-h_{ii(q)})\big|>\varepsilon\xi_{n}^{1/2}\bigg\} \notag\\
		\leq&\dfrac{1}{\varepsilon^{2G}\xi_{n}^{G}}\sum\limits_{q=1}^{M}\sum\limits_{i=1}^{n}E\Big[e'A_{q}D_{i}A_{q}e-\sigma^{2}(1-h_{ii(q)})\Big]^{2G} \notag\\
		\leq&C_{6}'\varepsilon^{-2G}\xi_{n}^{-G}\sum\limits_{q=1}^{M}\sum\limits_{i=1}^{n}\Big[\text{tr}\Big(A_{q}D_{i}A_{q}A_{q}D_{i}A_{q}\Big)\Big]^{G} \notag\\
		=&C_{6}'\varepsilon^{-2G}\xi_{n}^{-G}\sum\limits_{q=1}^{M}\sum\limits_{i=1}^{n}(1-h_{ii(q)})^{G} \notag\\
		\leq&\dfrac{Mn}{\varepsilon^{2G}\xi_{n}^{G}}C_{6}'\rightarrow0. \notag
	\end{align}
	Hence, $\max\limits_{1\leq q\leq M,1\leq i\leq n}\big|e'A_{q}D_{i}A_{q}e-\sigma^{2}(1-h_{ii(q)})\big|=o_{p}(\xi_{n}^{1/2})$. Noting that
	\begin{align}
		\xi_{n}=\inf_{\omega\in\textbf{H}_{n}}R_{n}(\omega)
		\leq&||\big(I_{n}-H(\omega)\big)\mu||^{2}+\sigma^{2}\text{tr}\big\{H^{2}(\omega)\big\} \notag\\
		\leq&\lambda^{2}_{\max}\big\{I_{n}-H(\omega)\big\}||\mu||^{2}+\sigma^{2}\lambda_{\max}\big\{H(\omega)\big\}\text{tr}\big\{H(\omega)\big\} \notag\\
		=&O(n)+O(k_{M}^{*}) \notag\\
		=&O(n), \notag
	\end{align}
	we obtain,
	\begin{align}
		C_{M^{*}}\dfrac{n}{m}\max\limits_{1\leq q\leq M,1\leq i\leq n}\Big\{\big|e'A_{q}D_{i}A_{q}e-\sigma^{2}(1-h_{ii(q)})\big|\Big\}\dfrac{k_{M^{*}}}{\xi_{n}}=\dfrac{o_{p}(\xi_{n}^{1/2})nk_{M^{*}}}{m\xi_{n}}=o_{p}(1). \label{th1.13}
	\end{align}
	It is readily seen that
	\begin{align}
		C_{M^{*}}\dfrac{n}{m}\max\limits_{1\leq q\leq M,1\leq i\leq n}\Big\{\sigma^{2}h_{ii(q)}\Big\}\dfrac{k_{M^{*}}}{\xi_{n}}=O(\dfrac{k_{M^{*}}^{2}}{m\xi_{n}})=o_{p}(1). \label{th1.14}
	\end{align}
	Thus, combining (\ref{th1.9})-(\ref{th1.14}), we see that (\ref{th1.6}) holds. This completes the proof of Theorem 2.
\end{proof}

\addcontentsline{toc}{subsection}{Proof of Theorem 3}
\begin{proof}[\textbf{Proof of Theorem 3}]
	Denote $\varepsilon_{n}=\xi_{n}^{1/2}n^{-1/2+\delta}$. To prove Theorem 3, following \cite{fan2004nonconcave} and \cite{chen2018semiparametric}, we need only to verify that there exists a constant $C_{0}$, such that for the $M\times1$ vector $u=(u_{1},\cdots,u_{M})'$,
	\begin{align}
		\lim\limits_{n\rightarrow\infty}P\bigg\{\inf_{||u||=C_{0},(\omega^{0}+\varepsilon_{n}u)\in\textbf{H}_{n}}\Gamma_{n,m}(\omega^{0}+\varepsilon_{n}u)>\Gamma_{n,m}(\omega^{0})\bigg\}=1, \notag
	\end{align}
	which means that there exists a minimizer $\hat{\omega}^{\text{BTMA}}$ in $\big\{\omega^{0}+\varepsilon_{n}u\big|\ ||u||\leq C_{0},(\omega^{0}+\varepsilon_{n}u)\in\textbf{H}_{n}\big\}$, such that $||\hat{\omega}^{\text{BTMA}}-\omega^{0}||=O_{p}(\varepsilon_{n})$. It is clear that
	\begin{align}
		&\Gamma_{n,m}(\omega^{0}+\varepsilon_{n}u)-\Gamma_{n,m}(\omega^{0}) \notag\\
		=&E_{*}\dfrac{||Y-\tilde{\mu}(\omega^{0}+\varepsilon_{n}u)||^{2}}{n}-E_{*}\dfrac{||Y-\tilde{\mu}(\omega^{0})||^{2}}{n} \notag\\
		=&\dfrac{1}{n}\bigg\{||Y-\hat{\mu}(\omega^{0}+\varepsilon_{n}u)||^{2}-||Y-\hat{\mu}(\omega^{0})||^{2}\bigg\}
		+\dfrac{1}{n}\bigg\{E_{*}||\hat{\mu}(\omega^{0}+\varepsilon_{n}u)-\tilde{\mu}(\omega^{0}+\varepsilon_{n}u)||^{2}-E_{*}||\hat{\mu}(\omega^{0})\notag\\
		&-\tilde{\mu}(\omega^{0})||^{2}\bigg\}+\dfrac{2}{n}\bigg\{E_{*}\langle Y-\hat{\mu}(\omega^{0}+\varepsilon_{n}u), \hat{\mu}(\omega^{0}+\varepsilon_{n}u)-\tilde{\mu}(\omega^{0}+\varepsilon_{n}u)\rangle
		-E_{*}\langle Y-\hat{\mu}(\omega^{0}), \hat{\mu}(\omega^{0}) \notag\\
		&-\tilde{\mu}(\omega^{0})\rangle\bigg\} \notag\\
		\triangleq&\dfrac{1}{n}(\Psi_{1}+\Psi_{2}+2\Psi_{3}). \label{th2.1}
	\end{align}
	First, note that
	\begin{align}
		\Psi_{1}=&||Y-\hat{\mu}(\omega^{0}+\varepsilon_{n}u)||^{2}-||Y-\hat{\mu}(\omega^{0})||^{2} \notag\\
		=&||\mu-\hat{\mu}(\omega^{0}+\varepsilon_{n}u)||^{2}-||\mu-\hat{\mu}(\omega^{0})||^{2}-2e'H(\varepsilon_{n}u)\mu-2e'H(\varepsilon_{n}u)e \notag\\
		=&\varepsilon_{n}^{2}u'\Lambda'\Lambda u-2\varepsilon_{n}\omega^{0'}\Omega'\Lambda u-2e'H(\varepsilon_{n}u)\mu-2e'H(\varepsilon_{n}u)e.\notag
	\end{align}
	We will show that $\varepsilon_{n}^{2}u'\Lambda'\Lambda u$ is the leading term in $\Psi_{1}$ and the remaining terms of $\Psi_{1}$ are asymptotically dominated by it. From Condition (C.10), we have
	\begin{align}
		\varepsilon_{n}^{2}u'\Lambda'\Lambda u>\varepsilon_{n}^{2}\lambda_{\text{min}}\big\{\Lambda'\Lambda\big\}u'u>C_{3}n\varepsilon_{n}^{2}||u||^{2}>0 \label{th2.2}
	\end{align}
	in probability tending to 1. By the definition of $\omega^{0}$, we see that $E||\Omega\omega^{0}||^{2}=E||\mu-\hat{\mu}(\omega^{0})||^{2}=\inf_{\omega\in\textbf{H}_{n}}E||\mu-\hat{\mu}(\omega)||^{2}=\xi_{n}$. Hence,
	\begin{align}
		||\Omega\omega^{0}||=O_{p}(\xi_{n}^{1/2}).\notag
	\end{align}
	According to Condition (C.10), we obtain
	\begin{align}
		|\varepsilon_{n}\omega^{0'}\Omega'\Lambda u|\leq\varepsilon_{n}||\Omega\omega^{0}||\bm\cdot||\Lambda||_{2}\bm\cdot||u||=O_{p}(n^{1/2}\xi_{n}^{1/2}\varepsilon_{n})||u||.\label{th2.3}
	\end{align}
	By Conditions (C.2) and (C.11), it is readily seen that
	\begin{align}
		\max_{1\leq q\leq M}\{e'H_{q}e\}\leq\max_{1\leq q\leq M}\lambda_{\text{max}}\big\{(X_{(q)}'X_{(q)})^{-1}\big\}\max_{1\leq q\leq M}||X_{(q)}'e||^{2}=O_{p}(k_{M^{*}}),\notag
	\end{align}
	which, by Condition (C.4), implies that
	\begin{align}
		\max_{1\leq q\leq M}|e'H_{q}\mu|\leq\max_{1\leq q\leq M}||H_{q}e||\max_{1\leq q\leq M}||\mu||=O_{p}(n^{1/2}k_{M^{*}}^{1/2}).\notag
	\end{align}
	Thus,
	\begin{align}
		|e'H(\varepsilon_{n}u)\mu|
		\leq&\varepsilon_{n}\big(M\max_{1\leq q\leq M}|e'H_{q}\mu|^{2}\big)^{1/2}||u|| \notag\\
		=&O_{p}(n^{1/2}M^{1/2}k_{M^{*}}^{1/2}\varepsilon_{n})||u||.\label{th2.4}
	\end{align}
	Similarly,
	\begin{align}
		|e'H(\varepsilon_{n}u)e|
		\leq&\varepsilon_{n}\big(M\max_{1\leq q\leq M}|e'H_{q}e|^{2}\big)^{1/2}||u|| \notag\\
		=&O_{p}(M^{1/2}k_{M^{*}}\varepsilon_{n})||u||.\label{th2.5}
	\end{align}
	By Conditions (C.12) and (C.13), and (\ref{th2.2})-(\ref{th2.5}), we see that $\varepsilon_{n}\omega^{0'}\Omega'\Lambda u$, $e'H(\varepsilon_{n}u)\mu$ and $e'H(\varepsilon_{n}u)e$ are asymptotically dominated by $\varepsilon_{n}^{2}u'\Lambda'\Lambda u$.\\
	\indent Next, note that
	\begin{align}
		\Psi_{2}=&E_{*}||\hat{\mu}(\omega^{0}+\varepsilon_{n}u)-\tilde{\mu}(\omega^{0}+\varepsilon_{n}u)||^{2}-E_{*}||\hat{\mu}(\omega^{0})-\tilde{\mu}(\omega^{0})||^{2} \notag\\
		=&E_{*}\big[(\omega^{0}+\varepsilon_{n}u)'\tilde{\Omega}'\tilde{\Omega}(\omega^{0}+\varepsilon_{n}u)-\omega^{0'}\tilde{\Omega}'\tilde{\Omega}\omega^{0}\big] \notag\\
		=&E_{*}\big(\varepsilon_{n}^{2}u'\tilde{\Omega}'\tilde{\Omega}u\big)+2E_{*}\big(\varepsilon_{n}\omega^{0'}\tilde{\Omega}'\tilde{\Omega}u\big).\label{th2.6}
	\end{align}
	From Lemma 2, (\ref{th1.11})-(\ref{th1.13}) and observing that $\tilde{\Omega}'\tilde{\Omega}$ is a positive-definite matrix, we have
	\begin{align}
		\lambda_{\text{max}}\big\{E_{*}(\tilde{\Omega}'\tilde{\Omega})\big\}
		\leq&\text{tr}\big\{E_{*}(\tilde{\Omega}'\tilde{\Omega})\big\} \notag\\
		=&\sum\limits_{q=1}^{M}E_{*}||\hat{\mu}_{q}-\tilde{\mu}_{q}||^{2} \notag\\
		\leq&C_{M^{*}}\dfrac{n}{m}\hat{e}_{R^{*}}^{2}\sum\limits_{q=1}^{M}\text{tr}\{H_{q}\} \notag\\
		\leq&C_{M^{*}}\dfrac{n}{m}\hat{e}_{R^{*}}^{2}Mk_{M^{*}} \notag\\
		\leq&C_{M^{*}}\dfrac{nMk_{M^{*}}}{m}\big[O(k_{M^{*}}^{2})+O_{p}(n^{1/2}k_{M^{*}})+o_{p}(\xi_{n}^{1/2})\big].\notag
	\end{align}
	So
	\begin{align}
		E_{*}\big(\varepsilon_{n}^{2}u'\tilde{\Omega}'\tilde{\Omega}u\big)
		\leq&\varepsilon_{n}^{2}\lambda_{\text{max}}\big\{E_{*}(\tilde{\Omega}'\tilde{\Omega})\big\}||u||^{2} \notag\\
		=&O(\dfrac{nMk_{M^{*}}^{3}}{m }\varepsilon_{n}^{2})+O_{p}(\dfrac{n^{3/2}Mk_{M^{*}}^{2}}{m}\varepsilon_{n}^{2})+o_{p}(\dfrac{nMk_{M^{*}}\xi_{n}^{1/2}}{m}\varepsilon_{n}^{2}).\label{th2.7}
	\end{align}
	Further, by Cauchy-Schwarz inequality, we obtain
	\begin{align}
		\bigg|E_{*}\big(\varepsilon_{n}\omega^{0'}\tilde{\Omega}'\tilde{\Omega}u\big)\bigg|
		\leq&\varepsilon_{n}\big(E_{*}||\tilde{\Omega}\omega^{0}||^{2}\big)^{1/2}\big(E_{*}||\tilde{\Omega}u||^{2}\big)^{1/2} \notag\\
		\leq&\varepsilon_{n}\lambda_{\text{max}}\big\{E_{*}(\tilde{\Omega}'\tilde{\Omega})\big\}||\omega^{0}||\bm\cdot||u|| \notag\\
		=&O(\dfrac{nMk_{M^{*}}^{3}}{m}\varepsilon_{n})+O_{p}(\dfrac{n^{3/2}Mk_{M^{*}}^{2}}{m}\varepsilon_{n})+o_{p}(\dfrac{nMk_{M^{*}}\xi_{n}^{1/2}}{m}\varepsilon_{n}).\label{th2.8}
	\end{align}
	Combining Conditions (C.6), (C.12) and (C.13), and (\ref{th2.6})-(\ref{th2.8}), we see that $\Psi_{2}$ is asymptotically dominated by $\varepsilon_{n}^{2}u'\Lambda'\Lambda u$.\\
	\indent Finally, from Lemma 2, it is found that
	\begin{align}
		\Psi_{3}\leq&\Big|E_{*}\langle Y-\hat{\mu}(\omega^{0}+\varepsilon_{n}u), \hat{\mu}(\omega^{0}+\varepsilon_{n}u)-\tilde{\mu}(\omega^{0}+\varepsilon_{n}u)\rangle
		-E_{*}\langle Y-\hat{\mu}(\omega^{0}), \hat{\mu}(\omega^{0})-\tilde{\mu}(\omega^{0})\rangle\Big| \notag\\
		=&O_{p}\big(\dfrac{n(m+n)^{1/2}}{m}M^{3/2}k_{M^{*}}\big) \notag
	\end{align}
	which is also asymptotically dominated by $\varepsilon_{n}^{2}u'\Lambda'\Lambda u$. This completes the proof of Theorem 3.
\end{proof}

\begin{lemma}
	Denote $\Gamma_{n,m}(q)=\dfrac{1}{n}E_{*}||Y-X_{(q)}\tilde{\Theta}^{*}_{q}||^{2}=\dfrac{1}{n}E_{*}||Y-X\tilde{\beta}^{*}_{q}||^{2}$ as the criterion of bootstrap model selection. Suppose that Conditions (C.1)-(C.7) are satisfied. Then when the $q$th candidate model is over-fitted, we have
	\begin{align}
		\Gamma_{n,m}(q)=\dfrac{1}{n}||Y-X\hat{\beta}_{q}||^{2}+\dfrac{\sigma^{2}}{m}k_{q}+o_{p}(\dfrac{1}{m}); \label{le3.1}
	\end{align}
	and when the $q$th candidate model is under-fitted, we have
	\begin{align}
		\Gamma_{n,m}(q)\leq\dfrac{1}{n}||Y-X\hat{\beta}_{q}||^{2}+\dfrac{C_{9}}{m}k_{q}+o_{p}(\dfrac{1}{m}), \label{le3.2}
	\end{align}
	where $C_{9}$ is a positive constant.
\end{lemma}
\begin{proof}
	It is clear that
	\begin{align}
		\Gamma_{n,m}(q)
		=&\dfrac{1}{n}||Y-X\hat{\beta}_{q}||^{2}+\dfrac{1}{n}E_{*}||X\hat{\beta}_{q}-X\tilde{\beta}^{*}_{q}||^{2}+\dfrac{1}{n}E_{*}\langle Y-X\hat{\beta}_{q},X\hat{\beta}_{q}-X\tilde{\beta}^{*}_{q}\rangle. \label{le3.3}
	\end{align}
	Note that
	\begin{align}
		E_{*}\langle Y-X\hat{\beta}_{q},X\hat{\beta}_{q}-X\tilde{\beta}^{*}_{q}\rangle
		=(Y-X\hat{\beta}_{q})'XE_{*}(\hat{\beta}_{q}-\tilde{\beta}^{*}_{q})
		=(Y-X\hat{\beta}_{q})'XS'_{q}E_{*}(\hat{\Theta}_{q}-\tilde{\Theta}^{*}_{q})
		=0. \label{le3.4}
	\end{align}
	Therefore, (\ref{le3.1}) is obtained by Theorem 2 of \cite{shao1996bootstrap}.\\
	\indent In the following, we consider the case where the $q$th candidate model is under-fitted. Note that, by Conditions (C.1) and (C.2), we have
	\begin{align}
		&\sum\limits_{i=1}^{n}x_{i(q)}'(X_{(q)}'X_{(q)})^{-1}E_{*}\Big[\mathfrak{O}_{k_{q}}X'_{(q)}\Pi\hat{e}_{q}\hat{e}'_{q}\Pi X_{(q)}\Big](X_{(q)}'X_{(q)})^{-1}x_{i(q)} \notag\\
		=&E_{*}\bigg[\text{tr}\Big\{X_{(q)}(X_{(q)}'X_{(q)})^{-1}\mathfrak{O}_{k_{q}}X'_{(q)}\Pi\hat{e}_{q}\hat{e}'_{q}\Pi X_{(q)}(X_{(q)}'X_{(q)})^{-1}X'_{(q)}\Big\}\bigg] \notag\\
		\leq&E_{*}\bigg[\big|\big|(X_{(q)}'X_{(q)})^{-1}\mathfrak{O}_{k_{q}}\big|\big|_{2}\text{tr}\Big\{X'_{(q)}\Pi\hat{e}_{q}\hat{e}'_{q}\Pi X_{(q)}\Big\}\bigg] \notag\\
		\leq&\dfrac{C_{1}+C_{3}}{nC_{1}C_{3}}\text{tr}\Big\{X'_{(q)}X_{(q)}(X'_{(q)}X_{(q)})^{-1}\Big(\sum\limits_{i=1}^{n}x_{i(q)}x'_{i(q)}\hat{e}^{2}_{q_{i}}\Big)(X'_{(q)}X_{(q)})^{-1}X'_{(q)}X_{(q)}\Big\} \notag\\
		\leq&\dfrac{(C_{1}+C_{3})C_{2}}{C_{1}C_{3}}\sum\limits_{i=1}^{n}x_{i(q)}'(X_{(q)}'X_{(q)})^{-1}\Big(\sum\limits_{i=1}^{n}x_{i(q)}x_{i(q)}'\hat{e}_{q,i}^{2}\Big)(X_{(q)}'X_{(q)})^{-1}x_{i(q)}. \label{le3.5}
	\end{align}
	So from the inequality (\ref{le3.5}), for $q\in\{1,\cdots,M_{0}\}$, we obtain
	\begin{align}
		&E_{*}||X\tilde{\beta}^{*}_{q}-X\hat{\beta}_{q}||^{2} \notag\\
		=&E_{*}||X_{(q)}\tilde{\Theta}^{*}_{q}-X_{(q)}\hat{\Theta}_{q}||^{2} \notag\\
		=&\sum\limits_{i=1}^{n}x_{i(q)}'E_{*}\big[(\tilde{\Theta}^{*}_{q}-\hat{\Theta}_{q})(\tilde{\Theta}^{*}_{q}-\hat{\Theta}_{q})'\big]x_{i(q)} \notag\\
		=&\dfrac{n^{2}}{m^{2}}\Bigg\{\sum\limits_{i=1}^{n}x_{i(q)}'(X_{(q)}'X_{(q)})^{-1}E_{*}\Big[X'_{(q)}\Pi\hat{e}_{q}\hat{e}_{q}\Pi X_{(q)}
		+\mathfrak{O}_{k_{q}}X'_{(q)}\Pi\hat{e}_{q}\hat{e}_{q}\Pi X_{(q)} \notag\\
		&+X'_{(q)}\Pi\hat{e}_{q}\hat{e}_{q}\Pi X_{(q)}\mathfrak{O}_{k_{q}}
		+\mathfrak{O}_{k_{q}}X'_{(q)}\Pi\hat{e}_{q}\hat{e}_{q}\Pi X_{(q)}\mathfrak{O}_{k_{q}}\Big](X'_{(q)}X'_{(q)})x_{i(q)}\Bigg\} \notag\\
		\leq&\dfrac{C_{7}'n}{m}\sum\limits_{i=1}^{n}x_{i(q)}'(X_{(q)}'X_{(q)})^{-1}\Big(\sum\limits_{i=1}^{n}x_{i(q)}x_{i(q)}'\hat{e}_{q,i}^{2}\Big)(X_{(q)}'X_{(q)})^{-1}x_{i(q)}, \label{le3.6}
	\end{align}
	where $C_{7}'$ is a positive constant. It is readily seen that
	\begin{align}
		\sum\limits_{i=1}^{n}x_{i(q)}x_{i(q)}'\hat{e}_{q,i}^{2}
		=&\sum\limits_{i=1}^{n}x_{i(q)}x_{i(q)}'(y_{i}-x_{i}'\hat{\beta}_{q})^{2} \notag\\
		=&\sum\limits_{i=1}^{n}x_{i(q)}x_{i(q)}'y_{i}^{2}
		+\sum\limits_{i=1}^{n}x_{i(q)}x_{i(q)}'\hat{\beta}_{q}'x_{i}x_{i}'\hat{\beta}_{q}
		-2\sum\limits_{i=1}^{n}x_{i(q)}x_{i(q)}'y_{i}x_{i}'\hat{\beta}_{q}. \label{le3.7}
	\end{align}
	We now consider each term of (\ref{le3.7}). For the first term, by Conditions (C.1), (C.4) and (C.7), we have
	\begin{align}
		&\dfrac{1}{n^{2}}\sum\limits_{i=1}^{n}E||x_{i(q)}x_{i(q)}'y_{i}^{2}||_{F}^{2} \notag\\
		=&\dfrac{1}{n^{2}}\sum\limits_{i=1}^{n}E\Big(\text{tr}\big\{x_{i(q)}x_{i(q)}'x_{i(q)}x_{i(q)}'\big\}y_{i}^{4}\Big) \notag\\
		\leq&\dfrac{1}{n^{2}}\sum\limits_{i=1}^{n}E\big(\max_{1\leq i\leq n}\{x_{i(q)}'x_{i(q)}\}\big)^{2}\max_{1\leq i\leq n}\{E(y_{i}^{4}|x_{i})\} \notag\\
		=&O(\dfrac{1}{n}). \label{le3.8}
	\end{align}
	So by the weak law of large numbers (WLLN) and the dimension $k_{q}$ being fixed, we obtain
	\begin{align}
		\sum\limits_{i=1}^{n}x_{i(q)}x_{i(q)}'y_{i}^{2}
		=&\sum\limits_{i=1}^{n}E(x_{i(q)}x_{i(q)}'y_{i}^{2})+o_{p}(n) \notag\\
		=&\sum\limits_{i=1}^{n}E\big[(x_{i(q)}x_{i(q)}')(\mu_{i}^{2}+\sigma^{2})\big]+o_{p}(n) \notag\\
		\leq&E\bigg[\Big(\sum\limits_{i=1}^{n}x_{i(q)}x_{i(q)}'\Big)\big(\max_{1\leq i\leq n}\{\mu_{i}^{2}\}+\sigma^{2}\big)\bigg]+o_{p}(n) \notag\\
		\leq&C_{8}'E(X_{(q)}'X_{(q)})+o_{p}(n), \label{le3.9}
	\end{align}
	where $C_{8}'$ is a positive constant.\\
	\indent For the second term of (\ref{le3.7}), note that,
	\begin{align}
		\hat{\beta}_{q}=&S'_{q}(X_{(q)}'X_{(q)})^{-1}X_{(q)}'(X_{(q)}\Theta_{q}+X_{(q^{c})}\Theta_{q^{c}}+e) \notag\\
		=&S'_{q}\big[\Theta_{q}+(X_{(q)}'X_{(q)})^{-1}X_{(q)}'X_{(q^{c})}\Theta_{q^{c}}+(X_{(q)}'X_{(q)})^{-1}X_{(q)}'e\big], \notag
	\end{align}
	where $(\Theta'_{q}, \Theta'_{q^{c}})'=\Theta_{t}$, and $(X_{(q)}, X_{(q^{c})})=X_{t}$. By $C_{r}$ inequality, we obtain
	\begin{align}
		||\hat{\beta}_{q}||^{4}
		\leq&C'_{r}\Big(||\beta_{q}||^{4}+||(X_{(q)}'X_{(q)})^{-1}||_{F}^{4}||X_{(q)}'||_{F}^{4}||X_{(q^{c})}||_{F}^{4}||\Theta_{q^{c}}||^{4}+||(X_{(q)}'X_{(q)})^{-1}||_{F}^{4}||X_{(q)}'||_{F}^{4}||e||^{4}\Big) \notag\\
		\leq&C'_{r}\Big(O(1)+O\big(\dfrac{1}{n^{2}}\big)||e||^{4}\Big). \label{le3.10}
	\end{align}
	Similar to (\ref{le3.8}), we can find that
	\begin{align}
		&\dfrac{1}{n^{2}}\sum\limits_{i=1}^{n}E||x_{i(q)}x_{i(q)}'\hat{\beta}_{q}'x_{i}x_{i}'\hat{\beta}_{q}||_{F}^{2} \notag\\
		\leq&\dfrac{1}{n^{2}}\sum\limits_{i=1}^{n}E\big(||x_{i(q)}x_{i(q)}'||^{2}_{F}||x_{i}x_{i}'||^{2}_{F}||\hat{\beta}_{q}||^{4}\big) \notag\\
		\leq&\dfrac{1}{n^{2}}\sum\limits_{i=1}^{n}E\Big[||x_{i(q)}x_{i(q)}'||^{2}_{F}||x_{i}x_{i}'||^{2}_{F}E\big(||\hat{\beta}_{q}||^{4}|X\big)\Big] \notag\\
		=&O(\dfrac{1}{n}), \notag
	\end{align}
	where the last equality holds by (\ref{le3.10}). So from WLLN, it is seen that
	\begin{align}
		&\sum\limits_{i=1}^{n}x_{i(q)}x_{i(q)}'\hat{\beta}_{q}'x_{i}x_{i}'\hat{\beta}_{q} \notag\\
		=&\sum\limits_{i=1}^{n}E(x_{i(q)}x_{i(q)}'\hat{\beta}_{q}'x_{i}x_{i}'\hat{\beta}_{q})+o_{p}(n) \notag\\
		=&\sum\limits_{i=1}^{n}E\Bigg\{x_{i(q)}x_{i(q)}'E\bigg[\big(\beta_{q}+S'_{q}(X_{(q)}'X_{(q)})^{-1}X_{(q)}'X_{(q)^{c}}\Theta_{q^{c}}\big)'x_{i}x_{i}'(\beta_{q}+S'_{q}(X_{(q)}'X_{(q)})^{-1}X_{(q)}'X_{(q)^{c}}\Theta_{q^{c}}\big) \notag\\
		&+\big(S'_{q}(X_{(q)}'X_{(q)})^{-1}X_{(q)}'e\big)'x_{i}x_{i}'\big(S'_{q}(X_{(q)}'X_{(q)})^{-1}X_{(q)}'e\big)\Big|X\bigg]\Bigg\}+o_{p}(n) \notag\\
		\leq&\sum\limits_{i=1}^{n}E\Bigg\{x_{i(q)}x_{i(q)}'\bigg[2\big(x'_{i}\beta_{q}\big)^{2}+2\big(x'_{i}S'_{q}(X_{(q)}'X_{(q)})^{-1}X_{(q)}'X_{(q)^{c}}\Theta_{q^{c}}\big)^{2} \notag\\
		&+\sigma^{2}\text{tr}\Big\{\big(S'_{q}(X_{(q)}'X_{(q)})^{-1}X_{(q)}'\big)'x_{i}x_{i}'\big(S'_{q}(X_{(q)}'X_{(q)})^{-1}X_{(q)}'\big)\Big\}\bigg]\Bigg\}+o_{p}(n) \notag\\
		\leq&C_{9}'E(X_{(q)}'X_{(q)})+o_{p}(n), \label{le3.11}
	\end{align}
	where $C_{9}'$ is a positive constant.\\
	\indent For the third term of (\ref{le3.7}), by (\ref{le3.9}) and (\ref{le3.11}), we obtain
	\begin{align}
		\Big|\sum\limits_{i=1}^{n}x_{i(q)}x_{i(q)}'y_{i}x_{i}'\hat{\beta}_{q}\Big|
		\leq&\dfrac{1}{2}\sum\limits_{i=1}^{n}x_{i(q)}x_{i(q)}'[y_{i}^{2}+(x_{i}'\hat{\beta}_{q})^{2}] \notag\\
		\leq&\dfrac{1}{2}C_{8}'E(X_{(q)}'X_{(q)})+\dfrac{1}{2}C_{9}'E(X_{(q)}'X_{(q)})+o_{p}(n) \notag\\
		\leq&C_{10}'E(X_{(q)}'X_{(q)})+o_{p}(n). \label{le3.12}
	\end{align}
	Now, combining (\ref{le3.6}), (\ref{le3.7}), (\ref{le3.9}), (\ref{le3.11}), (\ref{le3.12}) and Condition (C.2), we have
	\begin{align}
		&E_{*}||X\tilde{\beta}^{*}_{q}-X\hat{\beta}_{q}||^{2} \notag\\
		\leq&\dfrac{C_{7}'n}{m}\sum\limits_{i=1}^{n}x_{i(q)}'(X_{(q)}'X_{(q)})^{-1}\Big[C_{8}E(X_{(q)}'X_{(q)})+o_{p}(n)\Big](X_{(q)}'X_{(q)})^{-1}x_{i(q)} \notag\\
		=&\dfrac{C_{7}'n}{m}\text{tr}\bigg\{X_{(q)}(X_{(q)}'X_{(q)})^{-1}\Big[C_{8}E(X_{(q)}'X_{(q)})+o_{p}(n)\Big](X_{(q)}'X_{(q)})^{-1}X'_{(q)}\bigg\} \notag\\
		=&\dfrac{C_{7}'n}{m}\text{tr}\bigg\{\Big[C_{8}E(X_{(q)}'X_{(q)})+o_{p}(n)\Big](X_{(q)}'X_{(q)})^{-1}\bigg\} \notag\\
		\leq&\bigg[\dfrac{C_{7}'C_{8}n}{m}\text{tr}\Big\{E(X_{(q)}'X_{(q)})\Big\}+\dfrac{C_{7}'n}{m}\text{tr}\big\{o_{p}(n)\big\}\bigg]\lambda_{\max}\big\{(X_{(q)}'X_{(q)})^{-1}\big\} \notag\\
		\leq&\dfrac{C_{9}n}{m}E\Big(\text{tr}\big\{I_{k_{q}}\big\}\lambda_{\max}\big\{X_{(q)}'X_{(q)}\big\}\Big)\lambda_{\max}\big\{(X_{(q)}'X_{(q)})^{-1}\big\}+o_{p}(\dfrac{n}{m}) \notag\\
		=&\dfrac{C_{9}n}{m}k_{q}+o_{p}(\dfrac{n}{m}), \notag
	\end{align}
	where $C_{8}=\max\{C_{8}',C_{9}',C_{10}'\}$ and $C_{9}$ is a positive constant. From this, (\ref{le3.3}) and (\ref{le3.4}), we see that (\ref{le3.2}) holds. This completes the proof of Lemma 3.
\end{proof}

\addcontentsline{toc}{subsection}{Proof of Theorem 4}
\begin{proof}[\textbf{Proof of Theorem 4}]
	Let $a_{q}=Y'(I_{n}-H_{q})Y$ and $\Gamma$ be an $M\times M$ matrix with the $(q,r)$-th element $\Gamma_{qr}=\dfrac{1}{n}a_{\max\{q,r\}}+\dfrac{1}{n}E_{*}\langle X_{(q)}\hat{\beta}_{q}-X_{(q)}\tilde{\beta}^{*}_{q}, X_{(r)}\hat{\beta}_{r}-X_{(r)}\tilde{\beta}^{*}_{r}\rangle$. It is clear that $\Gamma_{n,m}(\omega)=\omega'\Gamma\omega$ for every $\omega\in\textbf{H}_{n}$. Let $q$ be an index belonging to $\{1,\cdots,M_{0}\}$. We define $\tilde{\omega}_{q}=(\hat{\omega}_{1},\cdots,\hat{\omega}_{q-1},0,\hat{\omega}_{q+1},\cdots,\hat{\omega}_{M}+\hat{\omega}_{q})'$. Similar to \cite{zhang2019inference}, by Cauchy-Schwarz inequality, it follows that,
	\begin{align}
		0\leq&\Gamma_{n,m}(\tilde{\omega}_{q})-\Gamma_{n,m}(\hat{\omega}) \notag\\
		=&\tilde{\omega}_{q'}\Gamma\tilde{\omega}_{q}-\hat{\omega}'\Gamma\hat{\omega} \notag\\
		=&(\tilde{\omega}_{q'}+\hat{\omega})\Gamma(\tilde{\omega}_{q'}-\hat{\omega}) \notag\\
		=&\big(2\hat{\omega}'+(0,\cdots,0,-\hat{\omega}_{q},0,\cdots,0,\hat{\omega}_{q})\big)\Gamma(0,\cdots,0,-\hat{\omega}_{q},0,\cdots,0,\hat{\omega}_{q})' \notag\\
		=&\hat{\omega}^{2}_{q}(\Gamma_{qq}-\Gamma_{qM}-\Gamma_{Mq}+\Gamma_{MM})+2\hat{\omega}_{q}\hat{\omega}'(\Gamma_{M1}-\Gamma_{q1},\cdots,\Gamma_{MM}-\Gamma_{qM})' \notag\\
		=&\hat{\omega}^{2}_{q}\bigg[\dfrac{1}{n}(a_{q}-a_{M})+\dfrac{1}{n}\Big(E_{*}||X_{(q)}\hat{\beta}_{q}-X_{(q)}\tilde{\beta}^{*}_{q}||^{2}+E_{*}||X_{M}\hat{\beta}_{M}-X_{M}\tilde{\beta}^{*}_{M}||^{2} \notag\\
		&-2E_{*}\langle X_{(q)}\hat{\beta}_{q}-X_{(q)}\tilde{\beta}^{*}_{q},X_{M}\hat{\beta}_{M}-X_{M}\tilde{\beta}^{*}_{M}\rangle\Big)\bigg] \notag\\
		&+2\hat{\omega}_{q}\sum\limits_{r=1}^{M}\hat{\omega}_{r}\bigg[\dfrac{1}{n}(a_{M}-a_{\max\{q,r\}})+\dfrac{1}{n}\Big(E_{*}\langle X_{M}\hat{\beta}_{M}-X_{M}\tilde{\beta}^{*}_{M},X_{(r)}\hat{\beta}_{r}-X_{(r)}\tilde{\beta}^{*}_{r}\rangle \notag\\
		&-E_{*}\langle X_{(q)}\hat{\beta}_{q}-X_{(q)}\tilde{\beta}^{*}_{q},X_{(r)}\hat{\beta}_{r}-X_{(r)}\tilde{\beta}^{*}_{r}\rangle\Big)\bigg] \notag\\
		\leq&\hat{\omega}^{2}_{q}\bigg\{\dfrac{1}{n}(a_{q}-a_{M})+\dfrac{1}{n}\Big[E_{*}||X_{(q)}\hat{\beta}_{q}-X_{(q)}\tilde{\beta}^{*}_{q}||^{2}+E_{*}||X_{M}\hat{\beta}_{M}-X_{M}\tilde{\beta}^{*}_{M}||^{2} \notag\\
		&+2\big(E_{*}||X_{(q)}\hat{\beta}_{q}-X_{(q)}\tilde{\beta}^{*}_{q}||^{2}\big)^{1/2}\big(E||X_{M}\hat{\beta}_{M}-X_{M}\tilde{\beta}^{*}_{M}||^{2}\big)^{1/2}\Big]\bigg\} \notag\\
		&+2\hat{\omega}_{q}\sum\limits_{r=1}^{M}\hat{\omega}_{r}\bigg\{\dfrac{1}{n}(a_{M}-a_{\max\{q,r\}})+\dfrac{1}{n}\Big[\big(E_{*}||X_{M}\hat{\beta}_{M}-X_{M}\tilde{\beta}^{*}_{M}||^{2}\big)^{1/2}\big(E||X_{(r)}\hat{\beta}_{r}-X_{(r)}\tilde{\beta}^{*}_{r})||^{2}\big)^{1/2} \notag\\
		&+\big(E_{*}||X_{(q)}\hat{\beta}_{q}-X_{(q)}\tilde{\beta}^{*}_{q}||^{2}\big)^{1/2}\big(E||X_{(r)}\hat{\beta}_{r}-X_{(r)}\tilde{\beta}^{*}_{r})||^{2}\big)^{1/2}\Big]\bigg\} \notag\\
		\leq&\dfrac{1}{n}\bigg\{\hat{\omega}^{2}_{q}(a_{q}-a_{M})+2\hat{\omega}^{2}_{q}(a_{M}-a_{q})
		+\hat{\omega}_{q}\dfrac{n}{m}\Big[\sigma^{2}k_{q}+\sigma^{2}k_{M}+o_{p}(1)\Big] \notag\\
		&+2\hat{\omega}_{q}\sum\limits_{r=1}^{M}\hat{\omega}_{r}\dfrac{n}{m}C_{10}\Big\{\big[k_{M}+o_{p}(1)\big]^{1/2}\big[k_{r}+o_{p}(1)\big]^{1/2}+\big[k_{q}+o_{p}(1)\big]^{1/2}\big[k_{r}+o_{p}(1)\big]^{1/2}\Big\}\bigg\} \notag\\
		\leq&\hat{\omega}^{2}_{q}\dfrac{1}{n}(a_{M}-a_{q})+\hat{\omega}_{q}\dfrac{C_{10}}{m}k_{M}[1+o_{p}(1)], \label{th3.1}
	\end{align}
	where the second to last inequality holds because $a_{M}<a_{r}$ for every $r\in\{1,\cdots,M-1\}$, and the last inequality holds by Lemma 3 and $\hat{\omega}_{r}\in[0,1]$. Therefore, from (\ref{th3.1}), when $\hat{\omega}_{q}\neq0$,
	\begin{align}
		\hat{\omega}_{q}\leq\bigg\{\dfrac{C_{10}}{m}k_{M}[1+o_{p}(1)]\bigg\}n(a_{q}-a_{M})^{-1}. \label{th3.2}
	\end{align}
	Note that for any $q\in\{1,\cdots,M_{0}\}$, we have
	\begin{align}
		\dfrac{1}{n}(a_{q}-a_{M})
		=&\dfrac{1}{n}\big[Y(I_{n}-H_{q})Y-Y'(I_{n}-H_{M})Y\big] \notag\\
		=&\dfrac{1}{n}\big[(X_{(q)}\Theta_{q}+X_{(q)^{c}}\Theta_{q^{c}}+e)'(I_{n}-H_{q})(X_{(q)}\Theta_{q}+X_{(q)^{c}}\Theta_{q^{c}}+e)-(X_{t}\Theta_{t}+e)'(I_{n} \notag\\
		&-H_{M})(X_{t}\Theta_{t}+e)\big] \notag\\
		=&\dfrac{1}{n}\Theta_{q^{c}}'X_{(q)^{c}}'(I-H_{q})X_{(q)^{c}}\Theta_{q^{c}}+\dfrac{2}{n}e'(I-H_{q})X_{(q)^{c}}\Theta_{q^{c}}-\dfrac{1}{n}e'(H_{q}-H_{M})e. \label{th3.3}
	\end{align}
	By Conditions (C.2) and (C.15), and using the fact that $k_{q}$ is fixed, we see that for any $q\in\{1,\cdots,M\}$,
	\begin{align}
		\dfrac{1}{n}\big|e'(I-H_{q})X_{(q)^{c}}\Theta_{q^{c}}\big|
		\leq&\dfrac{1}{n}\big|e'X_{(q)^{c}}\Theta_{q^{c}}\big|+\dfrac{1}{n}\big|e'X_{(q)}(X'_{(q)}X_{(q)})^{-1}X'_{(q)}X_{(q)^{c}}\Theta_{q^{c}}\big| \notag\\
		=&O_{p}(\dfrac{1}{\sqrt{n}}), \label{th3.4}
	\end{align}
	and
	\begin{align}
		E(\dfrac{1}{n}e'H_{q}e)=\dfrac{1}{n}\sigma^{2}\text{tr}\{H_{q}\}=\dfrac{1}{n}\sigma^{2}k_{q}=O(\dfrac{1}{n}). \label{th3.5}
	\end{align}
	From Condition (C.15), there exists a positive define matrix $Q_{(q)}$, such that
	\begin{align}
		\dfrac{1}{n}(X_{(q)}, X_{(q)^{c}})'(X_{(q)}, X_{(q)^{c}})
		=\dfrac{1}{n}
		\begin{pmatrix}
			X_{(q)}'X_{(q)} & X_{(q)}'X_{(q)^{c}}\\
			X_{(q)^{c}}'X_{(q)} & X_{(q)^{c}}'X_{(q)^{c}}
		\end{pmatrix}
		\stackrel{a.s.}{\longrightarrow}Q_{(q)}=
		\begin{pmatrix}
			Q_{11(q)} & Q_{12(q)}\\
			Q_{21(q)} & Q_{22(q)}
		\end{pmatrix}. \notag
	\end{align}
	Since $Q_{(q)}$ is a positive define matrix, we see that $Q_{22(q)}-Q_{21(q)}Q_{11(q)}^{-1}Q_{12(q)}$ is a positive define matrix as well. Recalling that $\Theta_{q^{c}}$ is a nonzero vector, we obtain
	\begin{align}
		\dfrac{1}{n}\Theta_{q^{c}}'X_{(q)^{c}}'(I-H_{q})X_{(q)^{c}}\Theta_{q^{c}}
		=&\dfrac{1}{n}\Theta_{q^{c}}'[X_{(q)^{c}}'X_{(q)^{c}}-X_{(q)^{c}}'X_{(q)}(X_{(q)}'X_{(q)})^{-1}X_{(q)}X_{(q)^{c}}]\Theta_{q^{c}} \notag\\
		\stackrel{a.s.}{\longrightarrow}&\Theta_{q^{c}}'(Q_{22(q)}-Q_{21(q)}Q_{11(q)}^{-1}Q_{12(q)})\Theta_{q^{c}}>0. \label{th3.6}
	\end{align}
	Thus, combining (\ref{th3.3})-(\ref{th3.6}), we have $n(a_{q}-a_{M})^{-1}=O_{p}(1)$, which implies from (\ref{th3.2}) that
	\begin{align}
		\hat{\omega}_{q}=O_{p}(\dfrac{1}{m}). \notag
	\end{align}
	This completes the proof of Theorem 4.
\end{proof}

\addcontentsline{toc}{subsection}{Proof of Theorem 5}
\begin{proof}[\textbf{Proof of Theorem 5}]
	Rewrite $\hat{\omega}=(\hat{\omega}^{*'}_{1}, \hat{\omega}^{*'}_{2})'$, where $\hat{\omega}^{*}_{1}=(\hat{\omega}_{1},\cdots,\hat{\omega}_{M_{0}})'$ and $\hat{\omega}^{*}_{2}=(\hat{\omega}_{M_{0}+1},\cdots,\hat{\omega}_{M})'$ represent the weight vectors of under-fitted models and over-fitted models, respectively. Let $\Gamma^{*}=n\Gamma-||\hat{e}_{f}||^{2}\textbf{1}\textbf{1}'$, where $\hat{e}_{f}=\big[I_{n}-X(X'X)^{-1}X'\big]Y\triangleq(I_{n}-H)Y$, and $||\hat{e}_{f}||^{2}\textbf{1}\textbf{1}'$ is unrelated to $\omega$.
	Accordingly, we rewrite the weight vector from bootstrap criterion as
	\begin{align}
		\hat{\omega}=\mathop{\arg\min}_{\omega\in\textbf{H}_{n}}\omega'\Gamma^{*}\omega\triangleq\mathop{\arg\min}_{\omega\in\textbf{H}_{n}}\Gamma^{*}_{n,m}(\omega). \notag
	\end{align}
	Further, the matrix $\Gamma^{*}$ is decomposed as
	\begin{align}
		\Gamma^{*}=
		\begin{pmatrix}
			\Gamma^{11^{*}} & \Gamma^{12^{*}}  \\
			\Gamma^{21^{*}} & \Gamma^{22^{*}}
		\end{pmatrix}. \notag
	\end{align}
	It can be seen that, for $q\in\{1,\cdots,M_{0}\}$, we have
	\begin{align}
		\dfrac{1}{n}\big|a_{q}-||\hat{e}_{f}||^{2}\big|
		\leq\dfrac{1}{n}\big[Y'(I_{n}-H_{q})Y+Y'(I_{n}-H)Y\big]
		\leq\dfrac{2}{n}||Y||^{2}
		=O_{p}(1),  \notag
	\end{align}
	and for $q\in\{M_{0}+1,\cdots,M\}$, by Condition (C.15), we have
	\begin{align}
		\big|a_{q}-||\hat{e}_{f}||^{2}\big|
		=&\big|Y'(I_{n}-H_{q})Y-Y'(I_{n}-H)Y\big| \notag\\
		=&(X_{t}\Theta_{t}+e)'(H-H_{q})(X_{t}\Theta_{t}+e) \notag\\
		=&e'(H-H_{q})e \notag\\
		\leq&\Big[\lambda_{\max}\big\{\dfrac{1}{n}(X'X)^{-1}\big\}+\lambda_{\max}\big\{(\dfrac{1}{n}X_{q}'X_{q})^{-1}\big\}\Big]||\dfrac{1}{\sqrt{n}}X'e||^{2} \notag\\
		=&O_{p}(1). \notag
	\end{align}
	So by Theorem 4 and Cauchy-Schwarz inequality, it is straightforward to show that
	\begin{align}
		\hat{\omega}^{*'}_{1}\Gamma^{11^{*}}\hat{\omega}^{*}_{1}=O_{p}(\dfrac{n}{m^{2}})=o_{p}(1),\quad\hat{\omega}^{*'}_{1}\Gamma^{12^{*}}\hat{\omega}^{*}_{2}=O_{p}(\dfrac{n^{1/2}}{m})=o_{p}(1). \label{th4.1}
	\end{align}
	\indent Denote $\textbf{H}^{0}_{n}=\big\{\omega\in[0,1]^{M}:\sum_{q=1}^{M_{0}}\omega_{q}=0,\ \sum_{q=M_{0}+1}^{M}\omega_{q}=1\big\}$ and $\Gamma_{n,m}^{22^{*}}(\nu)=\nu'\Gamma^{22^{*}}\nu$. It is clear that
	\begin{align}
		\Gamma^{*}_{n,m}(\hat{\omega})
		=\min_{\omega\in\textbf{H}_{n}}\big\{\Gamma^{*}_{n,m}(\omega)\big\}
		\leq\min_{\omega\in\textbf{H}^{0}_{n}}\big\{\Gamma^{*}_{n,m}(\omega)\big\}
		=\min_{\nu\in\textbf{L}_{n}}\big\{\Gamma^{22^{*}}_{n,m}(\nu)\big\}. \label{th4.2}
	\end{align}
	From (\ref{th4.1}), we also have
	\begin{align}
		\Gamma^{*}_{n,m}(\hat{\omega})
		=\Gamma^{22^{*}}_{n,m}(\hat{\omega}^{*}_{2})+\hat{\omega}^{*'}_{1}\Gamma^{11^{*}}\hat{\omega}^{*}_{1}+2\hat{\omega}^{*'}_{1}\Gamma^{12^{*}}\hat{\omega}^{*}_{2}
		=\Gamma^{22^{*}}_{n,m}(\hat{\omega}^{*}_{2})+o_{p}(1). \label{th4.3}
	\end{align}
	Then combining (\ref{th4.2}) and (\ref{th4.3}), we obtain
	\begin{align}
		\Gamma^{22^{*}}_{n,m}(\hat{\omega}^{*}_{2})\leq\min_{\nu\in\textbf{L}_{n}}\big\{\Gamma^{22^{*}}_{n,m}(\nu)\big\}+o_{p}(1). \notag
	\end{align}
	Following Condition (C.15) and continuous mapping theorem, we have
	\begin{align}
		&\big(\dfrac{1}{n}X'X\big)^{-1}-S'_{M_{0}+\max\{r,t\}}\big(S_{M_{0}+\max\{r,t\}}\dfrac{1}{n}X'XS'_{M_{0}+\max\{r,t\}}\big)^{-1}S_{M_{0}+\max\{r,t\}} \notag\\
		\stackrel{p}{\rightarrow}&Q^{-1}-S'_{M_{0}+\max\{r,t\}}Q_{\max\{r,t\}}^{-1}S_{M_{0}+\max\{r,t\}} \notag\\
		=&Q^{-1}-V_{\max\{r,t\}}. \notag
	\end{align}
	Therefore, for the $(r,t)$-th element of $\Gamma^{22^{*}}$, following Slutsky's theorem, we obtain
	\begin{align}
		\Gamma_{rt}^{22^{*}}
		=&a_{\max\{M_{0}+r,M_{0}+t\}}-||\hat{e}_{f}||^{2}+E_{*}\langle X_{M_{0}+r}\hat{\beta}_{M_{0}+r}-X_{M_{0}+r}\tilde{\beta}_{M_{0}+r}, X_{M_{0}+t}\hat{\beta}_{M_{0}+t}-X_{M_{0}+t}\tilde{\beta}_{M_{0}+t}\rangle \notag\\
		=&Y'(I_{n}-H_{r})(I_{n}-H_{t})Y-||\hat{e}_{f}||^{2}+\dfrac{n\sigma^{2}}{m}k_{\min\{r,t\}}+o_{p}(\dfrac{n}{m}) \notag\\
		=&(X\beta+e)'(I_{n}-H_{\max\{r,t\}})(X\beta+e)-||\hat{e}_{f}||^{2}+\dfrac{n\sigma^{2}}{m}k_{\min\{r,t\}}+o_{p}(\dfrac{n}{m}) \notag\\
		=&(X\beta+e)'(H-H_{\max\{r,t\}})(X\beta+e)+\dfrac{n\sigma^{2}}{m}k_{\min\{r,t\}}+o_{p}(\dfrac{n}{m}) \notag\\
		=&C_{7}\sigma^{2}k_{\min\{r,t\}}+Z'_{n}\Big[Q^{-1}_{n}-S'_{M_{0}+\max\{r,t\}}\big(S_{M_{0}+\max\{r,t\}}Q_{n}S'_{M_{0}+\max\{r,t\}}\big)^{-1}S_{M_{0}+\max\{r,t\}}\Big]Z_{n}+o_{p}(1) \notag\\
		\stackrel{d}{\rightarrow}&C_{7}\sigma^{2}k_{\min\{r,t\}}+Z'\big(Q^{-1}-V_{\max\{r,t\}}\big)Z \notag\\
		=&\Delta_{rt}. \notag
	\end{align}
	Hence, by the joint convergence in distribution, we have $\Gamma^{22^{*}}_{n,m}(\nu)=\nu'\Gamma^{22^{*}}\nu\stackrel{d}{\rightarrow}\nu'\Delta\nu$ , where $\nu'\Delta\nu$ is continuous with respect to $\nu$. For every $r<t$, letting $S_{r,t}=(I_{k_{r}},\textbf{0}_{k_{r}\times(k_{t}-k_{r})})$ be a selection matrix of $X_{(t)}$, we have $S_{r}=(I_{k_{r}},\textbf{0}_{k_{r}\times(k-k_{r})})=(I_{k_{r}},\textbf{0}_{k_{r}\times(k_{t}-k_{r})})(I_{k_{t}},\textbf{0}_{k_{t}\times(k-k_{t})})=S_{r,t}S_{t}$ and
	\begin{align}
		V_{r}QV_{t}
		=&S_{r}'\big(S_{r}QS_{r}'\big)^{-1}S_{r}QS_{t}'\big(S_{t}QS_{t}'\big)^{-1}S_{t} \notag\\
		=&S_{r}'\big(S_{r}QS_{r}'\big)^{-1}S_{r,t}S_{t}QS_{t}'\big(S_{t}QS_{t}'\big)^{-1}S_{t} \notag\\
		=&S_{r}'\big(S_{r}QS_{r}'\big)^{-1}S_{r}. \notag
	\end{align}
	Then, it follows that
	\begin{align}
		Z'\big(Q^{-1}-V_{\max\{r,t\}}\big)Z
		=&Z'\big(Q^{-1}-V_{\max\{r,t\}}-V_{\min\{r,t\}}+V_{\min\{r,t\}}QV_{\max\{r,t\}}\big)Z \notag\\
		=&Z'(Q^{-1}-V_{\min\{r,t\}})Q(Q^{-1}-V_{\max\{r,t\}})Z \notag\\
		=&Z'(Q^{-1}-V_{\min\{r,t\}})T'T(Q^{-1}-V_{\max\{r,t\}})Z, \notag
	\end{align}
	where $Q=T'T$ since $Q$ is a positive define matrix. Further, denote the $R\times R$ matrix $K$ as
	\begin{align}
		K=
		\begin{pmatrix}
			k_{M_{0}+1} & k_{M_{0}+1} & k_{M_{0}+1} & \cdots & k_{M_{0}+1} \\
			k_{M_{0}+1} & k_{M_{0}+2} & k_{M_{0}+2} & \cdots & k_{M_{0}+2} \\
			k_{M_{0}+1} & k_{M_{0}+2} & k_{M_{0}+3} & \cdots & k_{M_{0}+3} \\
			\vdots & \vdots & \vdots & \ddots & \vdots\\
			k_{M_{0}+1} & k_{M_{0}+2} & k_{M_{0}+3} & \cdots & k_{M}
		\end{pmatrix}, \notag
	\end{align}
	and the $R\times R$ matrix $A$ as
	\begin{align}
		A=&
		\begin{pmatrix}
			Z'\big(Q^{-1}-V_{M_{0}+1}\big)Z & Z'\big(Q^{-1}-V_{M_{0}+2}\big)Z & \cdots & Z'\big(Q^{-1}-V_{M}\big)Z \\
			Z'\big(Q^{-1}-V_{M_{0}+2}\big)Z & Z'\big(Q^{-1}-V_{M_{0}+2}\big)Z & \cdots & Z'\big(Q^{-1}-V_{M}\big)Z \\
			\vdots & \vdots & \ddots & \vdots\\
			Z'\big(Q^{-1}-V_{M}\big)Z & Z'\big(Q^{-1}-V_{M}\big)Z & \cdots & Z'\big(Q^{-1}-V_{M}\big)Z
		\end{pmatrix} \notag\\
		=&
		\begin{pmatrix}
			Z'(Q^{-1}-V_{M_{0}+1})T'\\
			Z'(Q^{-1}-V_{M_{0}+2})T'\\
			\vdots\\
			Z'(Q^{-1}-V_{M})T'
		\end{pmatrix}
		\begin{pmatrix}T(Q^{-1}-V_{M_{0}+1})Z, T(Q^{-1}-V_{M_{0}+2})Z,\cdots, T(Q^{-1}-V_{M})Z\end{pmatrix}. \notag
	\end{align}
	It is clear that $A\geq0$, and we find $K>0$ from Lemma 2 of \cite{hansen2007least}. Hence, $C_{7}\sigma^{2}K+A$ is a positive define matrix. This implies that $\nu'\Delta\nu$ attains its unique minimum in the convex set $\textbf{L}_{n}$. Also we note that $\hat{\omega}=O_{p}(1)$, then as in the proof of Theorem 4 of \cite{liu2015distribution} or Theorem 3 of \cite{yu2024post}, by Theorem 4.2.2 of \cite{van1996weak} or Theorem 3.7 of \cite{kim1990cube}, it follows that $(\hat{\omega}_{M_{0}+1}, \hat{\omega}_{M_{0}+2}, \cdots, \hat{\omega}_{M})'=\hat{\omega}^{*}_{2}\stackrel{d}{\rightarrow}\tilde{\nu}=\arg\min_{\nu\in\textbf{L}_{n}}\nu'\Delta\nu$. So by Theorem 4, we have
	\begin{align}
		\sqrt{n}\big(\hat{\beta}(\hat{\omega})-\beta\big)
		=&\sum\limits_{q=1}^{M_{0}}\hat{\omega}_{q}\sqrt{n}(\hat{\beta}_{q}-\beta)+\sum\limits_{q=M_{0}+1}^{M}\hat{\omega}_{q}\sqrt{n}(\hat{\beta}_{q}-\beta) \notag\\
		=&O_{p}(\dfrac{1}{m})O_{p}(\sqrt{n})+\sum\limits_{q=M_{0}+1}^{M}\hat{\omega}_{q}\sqrt{n}S_{q}'(X_{(q)}'X_{(q)})^{-1}X_{(q)}'e \notag\\
		=&o_{p}(1)+\sum\limits_{q=M_{0}+1}^{M}\hat{\omega}_{q}S_{q}'(S_{q}Q_{n}S_{q}')^{-1}S_{q}Z_{n} \notag\\
		=&o_{p}(1)+\sum\limits_{r=1}^{R}\hat{\omega}^{*}_{2,r}S_{M_{0}+r}'(S_{qM_{0}+r}Q_{n}S_{M_{0}+r}')^{-1}S_{M_{0}+r}Z_{n} \notag\\
		\stackrel{d}{\rightarrow}&\sum\limits_{r=1}^{R}\tilde{\nu}_{r}V_{r}Z, \notag
	\end{align}
	where $\hat{\omega}^{*}_{2,r}$ is the $r$-th entry of $\hat{\omega}^{*}_{2}$. As $n\rightarrow\infty$, the distribution of $\hat{\omega}^{*}_{2}$ depends on $Z_{n}$ and $Q_{n}$, and that of $\tilde{\nu}$ relies on $Z$ and $Q$. Therefore, the last convergence is valid by the Slutsky's theorem and the joint convergence in distribution of $\hat{\omega}^{*}_{2}$ and $S_{q}'(S_{q}Q_{n}S_{q}')^{-1}S_{q}Z_{n}$, $q=M_{0}+1,\cdots,M$. This completes the proof of Theorem 5.
\end{proof}
\bibliographystyle{elsarticle-harv} 
\bibliography{reference(supplement)}